\PassOptionsToPackage{table,usenames,dvipsnames}{xcolor}
\documentclass[preprint, 12pt, 3p]{elsarticle} 

\usepackage{amssymb}

\usepackage{lineno}
\usepackage{epsfig}
\usepackage[utf8]{inputenc}
\usepackage{times}
\usepackage{graphicx}
\usepackage[figuresright]{rotating}
\usepackage{float}
\usepackage{longtable}
\usepackage{amsmath,amsfonts,amssymb,amsthm}
\usepackage{mathtools}
\usepackage{commath}
\usepackage{listings}

\usepackage{url}
\usepackage{hyperref}
\usepackage{soul}
\usepackage{wrapfig}
\floatname{algorithm}{Algorithm}
\usepackage[usenames]{color}
\usepackage{flushend}

\usepackage{acronym}
\usepackage{multirow}
\usepackage{multirow}

\usepackage{tcolorbox}
\usepackage{tabularx}
\tcbset{tab2/.style={colback=yellow!10!white,colframe=red!50!black,colbacktitle=cyan!40!white,
coltitle=black,center title}}

\usepackage{colortbl}
\makeatletter
\newcommand{\thickhline}{%
    \noalign {\ifnum 0=`}\fi \hrule height 1.5pt
    \futurelet \reserved@a \@xhline
}
\newcolumntype{"}{@{\hskip\tabcolsep\vrule width 1pt\hskip\tabcolsep}}
\makeatother
\newcolumntype{?}{!{\vrule width 1.5pt}}
\usepackage{xcolor} 

\DeclareMathOperator*{\argmax}{arg\,max}
\DeclareMathOperator*{\argmin}{arg\,min}

\newcommand{\bm}[1]{\mbox{\boldmath{$#1$}}}

\usepackage{booktabs}
\usepackage{caption}

\journal{Ad Hoc Networks}

\usepackage{xpatch}
\makeatletter
\expandafter\xpatchcmd\expandafter\acronym\expandafter{\csname begin\endcsname{description}}{\customAClongtable}{}{\GenericError{}{Patching the acronym list failed}{You're stuck with the description style}{}}
\expandafter\xpatchcmd\expandafter\endacronym\expandafter{\csname end\endcsname{description}}{\customAClongtableend}{}{\GenericError{}{Patching the acronym list failed}{You're stuck with the description style}{}}
\newlength\customAClabelwidth
\newlength\customACdescwidth
\newlength\customACpagewidth
\newcommand\customAClongtable
    {%
        \def\item{\\\hline\customAClongtablegrab}%
        \def\customAClongtablegrab[##1]{##1&}%
        \ifAC@withpage
            \patchcmd\AC@@acro{\ifAC@withpage}{&\ifAC@withpage}{}{\GenericError{}{Patching the acro format failed}{Now go and panic}{}}%
            \patchcmd\AC@@acro{\fi\\}{\fi}{}{\GenericError{}{Patching the acro format failed}{Now go and panic}{}}%
            \setlength\customACdescwidth{\dimexpr\textwidth-\customAClabelwidth-\customACpagewidth-6\tabcolsep-4\arrayrulewidth}%
            \begin{longtable}{|p{\customAClabelwidth}|p{\customACdescwidth}|>{\centering\arraybackslash}p{\customACpagewidth}|}
            \caption{Definition of acronyms}\\
            \hline
            \textbf{Acronym} & \textbf{Meaning} & \textbf{Page}
        \else
            \setlength\customACdescwidth{\dimexpr\textwidth-\customAClabelwidth-4\tabcolsep-3\arrayrulewidth}%
            \begin{longtable}{|p{\customAClabelwidth}|p{\customACdescwidth}|}
            \caption{Definition of acronyms}\\
            \hline
            \textbf{Acronym} & \textbf{Meaning}
        \fi
    }
\newcommand\customAClongtableend{\\\hline\end{longtable}}
\makeatother

\begin{document}


\begin{frontmatter}


\title{Machine Learning for Wireless Communications \\in the Internet of Things: A Comprehensive Survey}




\author{Jithin Jagannath$^\dag$ $^\ddagger$, Nicholas Polosky$^\ddagger$, Anu Jagannath$^\ddagger$,  \\ Francesco Restuccia$^\dag$, and Tommaso Melodia$^\dag$}

\address{$^{\ddagger}$ANDRO Advanced Applied Technology, ANDRO Computational Solutions, LLC, Rome, NY, 13440\\
$^\dag$Department of Electrical and Computer Engineering, Northeastern University, Boston, MA, 02115\\
E-mail: \{jjagannath, npolosky, ajagannath\}@androcs.com\\ \{jagannath.j, melodia, frestuc\}@northeastern.edu
}

\begin{abstract}

The Internet of Things (IoT) is expected to require more effective and efficient wireless communications than ever before. For this reason, techniques such as spectrum sharing, dynamic spectrum access, extraction of signal intelligence and optimized routing will soon become essential components of the IoT wireless communication paradigm. In this vision, IoT devices must be able to not only learn to autonomously extract spectrum knowledge on-the-fly from the network but also leverage such knowledge to dynamically change appropriate wireless parameters (\textit{e.g.}, frequency band, symbol modulation, coding rate, route selection, etc.) to reach the network's optimal operating point.~Given that the majority of the IoT will be composed of tiny, mobile, and energy-constrained devices, traditional techniques based on \emph{a priori} network optimization may not be suitable, since (i) an accurate model of the environment may not be readily available in practical scenarios; (ii) the computational requirements of traditional optimization techniques may prove unbearable for IoT devices. To address the above challenges, much research has been devoted to exploring the use of machine learning to address problems in the IoT wireless communications domain. The reason behind machine learning's popularity is that it provides a general framework to solve very complex problems where a model of the phenomenon being learned is too complex to derive or too dynamic to be summarized in mathematical terms.

This work provides a comprehensive survey of the state of the art in the application of machine learning techniques to address key problems in IoT wireless communications with an emphasis on its ad hoc networking aspect.~First, we present extensive background notions of machine learning techniques.~Then, by adopting a bottom-up approach, we examine existing work on machine learning for the IoT at the physical, data-link and network layer of the protocol stack. Thereafter, we discuss directions taken by the community towards hardware implementation to ensure the feasibility of these techniques. Additionally, before concluding, we also provide a brief discussion of the application of machine learning in IoT beyond wireless communication. Finally, each of these discussions is accompanied by a detailed analysis of the related open problems and challenges.

\end{abstract}


\end{frontmatter}



\section{Introduction}

\ac{IoT} – the term first coined by K. Ashton in 1999 \cite{KAshton} has hence emerged to describe a network of interconnected devices – sensors, actuators, mobile phones, among others – which interact and collaborate with each other to attain common objectives. \ac{IoT} will soon become the most pervasive technology worldwide. In the next few years, cars, kitchen appliances, televisions, smartphones, utility meters, intra-body sensors, thermostats, and almost anything we can imagine will be accessible from anywhere on the planet \cite{whitmore2015internet}. The revolution brought by the \ac{IoT} has been compared to the building of roads and railroads during the Industrial Revolution of the 18th to 19th centuries \cite{ForbesIoT} -- and is expected to radically transform the education, health-care, smart home, manufacturing, mining, commerce, transportation, and surveillance fields, just to mention a few \cite{da2014internet}. 

As the \ac{IoT} gains momentum in every aspect of our lives, the demand for wireless resources will accordingly increase in an unprecedented way. According to the latest Ericsson's mobility report, there are now 5.2 billion mobile broadband subscriptions worldwide, generating more than 130 exabytes per month of wireless traffic \citep{EricssonMobility2018}. Moreover, over 50 billion devices are expected to be in the \ac{IoT} by 2020, which will generate a global network of ``things'' of dimensions never seen before \citep{CiscoEstimates}. Given that only a few radio spectrum bands are available to wireless carriers \citep{SpectrumCrunch}, technologies such as \ac{RF} spectrum sharing through beamforming \citep{shokri2016spectrum,vazquez2018hybrid,lv2018cognitive},  \ac{DSA} \citep{jin2018specguard,chiwewe2017fast,Jagannath_TMC_2018, FederatedWireless,agarwal2016edsa} and anti-jamming technologies \citep{zhang2017framework,huang2017anti,chang2017jamming} will become essential in the near future. These technologies usually require coordination among wireless devices to optimize spectrum usage -- often, they need to be implemented in a distributed manner to ensure scalability, reduce overhead and energy consumption. To address this challenge, \ac{ML} has been widely recognized as the technology of choice for solving classification or regression problems for which no well-defined mathematical model exists. 


The recent introduction of \ac{ML} to wireless communications in the \ac{IoT} has in part to do with the new-found pervasiveness of \ac{ML} throughout the scientific community at large, and in part to do with the nature of the problems that arise in \ac{IoT} wireless communications. With the advent of advances in computing power and ability to collect and store massive amounts of data, \ac{ML} techniques have found their way into many different scientific domains in an attempt to put both of the aforementioned to good use. This concept is equally true in wireless communications. Additionally, problems that arise in wireless communication systems are frequently formulated as classification, detection, estimation, and optimization problems; for all of which \ac{ML} techniques can provide elegant and practical solutions. In this context, the application of \ac{ML} to wireless communications seems almost natural and presents a clear motivation \citep{Bkassiny-ieeecommsurtut2013,Jiang-ieeewcomm2017,Chen-arxiv2017}.

\begin{figure}[h!]
    \centering
    \includegraphics[width=6.5 in]{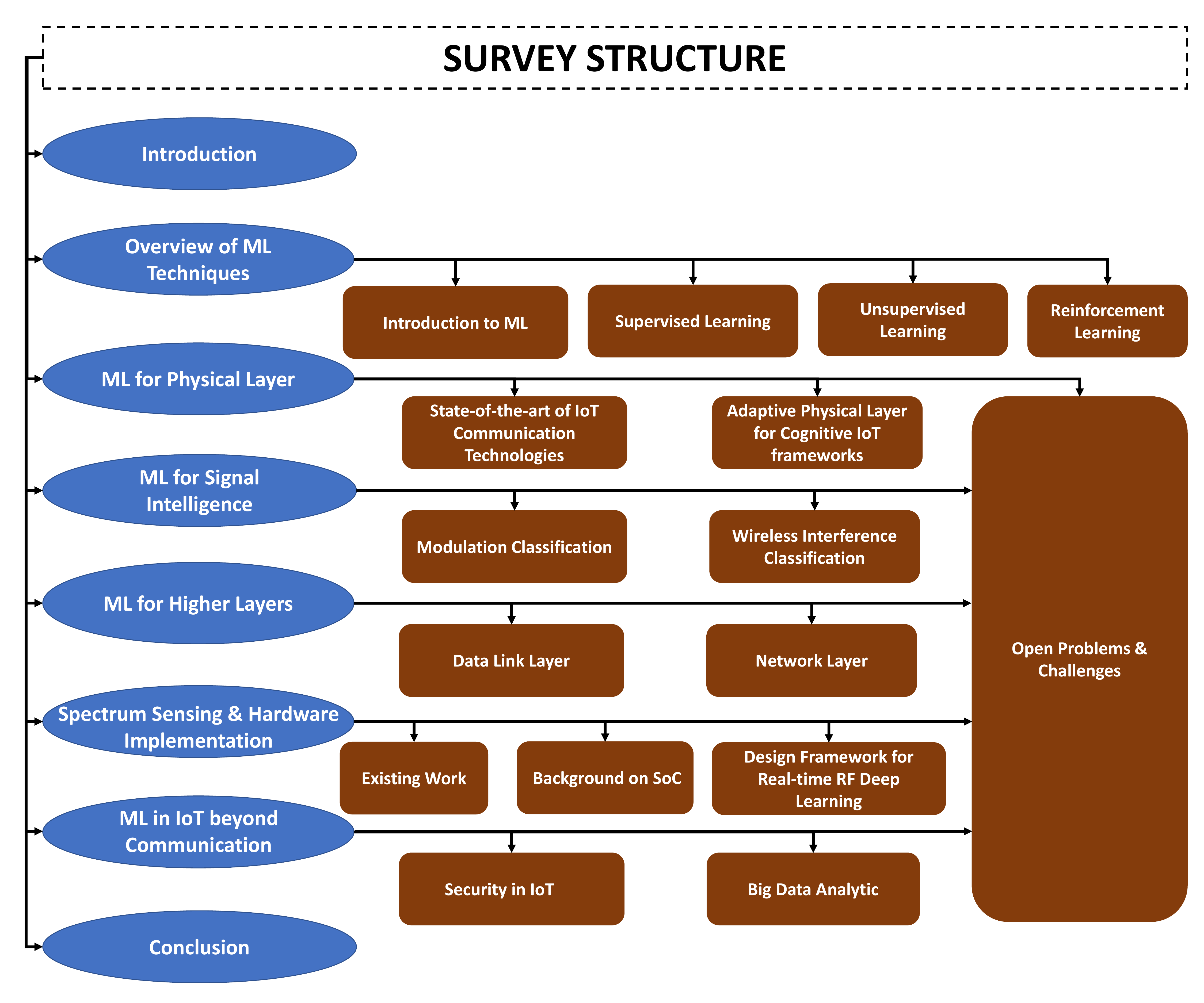}
    \caption{Overall organization of the survey}
    \label{fig:Org}
\end{figure}

The objective of this paper is to provide a detailed insight into the influence \ac{ML} has had on the \ac{IoT} and the broader context of \acp{WANET}. Our hope is to elicit more research in the field to solve some of the key challenges of modern \ac{IoT} communication systems. To begin, we provide an overview of the \ac{ML} techniques in Section \ref{sec:overview}. In Sections \ref{sec:Phy} and \ref{sec:Phy_Int}, we discuss the applications of \ac{ML} to physical layer to improve the communication and acquire signal intelligence respectively. Next, in Section \ref{sec:Higher_Layer}, we discuss how \ac{ML} has been exploited to advance protocol design at the data-link and network layers of the protocol stack. In Section \ref{sec:Hardware}, we discuss the implications of hardware implementations in the context of \ac{ML}. Thereafter, in Section \ref{sec:Beyond_Comm}, we provide a brief discussion on the recent application of \ac{ML} to \ac{IoT} beyond wireless communication. Finally, the conclusion of this paper is provided in Section \ref{sec:conclusion}. The overall structure of the survey paper is depicted in Figure \ref{fig:Org}

\section{Overview of Machine Learning Techniques} \label{sec:overview}

Before we begin, we would like to introduce some standard notations that will be used throughout this paper. We use boldface upper and lower-case letters to denote matrices and column vectors, respectively. For a vector $\mathbf{x}$, $x_i$  denotes the i-th element, $\norm{\mathbf{x}}$ indicates the Euclidean norm, $\mathbf{x}^\intercal$ its transpose, and $\mathbf{x} \cdot \mathbf{y}$ the Euclidean inner product of $\mathbf{x}$ and $\mathbf{y}$. For a matrix $\mathbf{H}$, $H_{ij}$ will indicate the (i,j)-th element of $\mathbf{H}$. The notation $\mathcal{R}$ and $\mathcal{C}$ will indicate the set of real and complex numbers, respectively. The notation $\mathbb{E}_{x\sim p(x)}\left[f(x)\right]$ is used to denote the expected value, or average of the function $f(x)$ where the random variable $x$ is drawn from the distribution $p(x)$. When a probability distribution of a random variable, $x$, is conditioned on a set of parameters, $\boldsymbol{\theta}$, we write $p(x;\boldsymbol{\theta})$ to emphasize the fact that $\boldsymbol{\theta}$ parameterizes the distribution and reserve the typical conditional distribution notation, $p(x|y)$, for the distribution of the random variable $x$ conditioned on the random variable $y$. We use the standard notation for operations on sets where $\cup$ and $\cap$ are the infix operators denoting the union and intersection of two sets, respectively. We use $S_k \subseteq S$ to say that $S_k$ is either a strict subset of or equal to the set $S$ and $x \in S$ to denote that $x$ is an element of the set $S$. $\varnothing$ is used to denote the empty set and $|S|$ the cardinality of a set $S$. Lastly, the convolution operator is denoted as $*$. 

All the notations used in this paper have been summarized in Table. \ref{tab:Notation}. The notations are divided into sections based on where they first appear and if they have been re-defined. Similarly, we also provide all the acronyms used in this paper in Table. \ref{Table:Acro}

\begin{longtable}{|l|l|}
    \caption{Definition of notations\label{tab:Notation}}\\
      \hline\hline \multicolumn{1}{|c|}{\textbf{Notations}} &  \multicolumn{1}{c|}{\textbf{Definitions}} \\ \hline \hline
\endfirsthead
      \multicolumn{2}{|c|}{\textbf{Section \ref{sec:overview}}} \\ \hline\hline
      $x, \mathbf{x}$ & Training example; vector \\ \hline
      $y, \mathbf{y}$ & Training target; vector \\ \hline
      $\hat{y}, \hat{\mathbf{y}}$ & Training target estimate; vector \\ \hline
      $D$ & Set of training data \\ \hline
      $\theta, \boldsymbol{\theta}$ & General model parameter; vector \\ \hline
      $k(\cdot, \cdot)$ & Kernel function \\ \hline
      $G(\cdot)$ & Gini impurity \\ \hline
      $H(\cdot)$ & Entropy function \\ \hline
      $\mathcal{L}(\cdot, \cdot, \cdot)$ & Loss function \\ \hline
      $\mathbf{w}$ & Model weight vector \\ \hline
      $\mathbf{W}, \mathbf{U}, \mathbf{V}$ & Model weight matrix \\ \hline
      $b, w_0$ & Model bias term \\ \hline
      $\mathbf{b}, \mathbf{c}$ & Model bias vector \\ \hline
      $\sigma(\cdot)$ & Sigmoid activation function \\ \hline
      $K$ & Convolution kernel \\ \hline
      $I$ & Input image \\ \hline
      $S(\cdot, \cdot)$ & CNN feature map \\ \hline
      $L$ & Neural network layer \\ \hline
      $C_k$ & Cluster k \\ \hline
      $\boldsymbol{\mu}_k$ & Centroid of a cluster k \\ \hline
      $d_j(\cdot)$ & Discriminant function for a neuron j\\ \hline
      $I(\mathbf{x})$ & Index of minimum occurrence of discriminant function for $\mathbf{x}$ \\ \hline
      $T_{j,I(\mathbf{x})}$ & Topological neighborhood function of $I(\mathbf{x})$ at neuron j\\ \hline
      $S_{i,j}$ & Distance from neuron i to neuron j \\ \hline
      $\eta(t)$ & Learning rate parameter; a function of time \\ \hline
      $\gamma$ & Reward discount parameter \\ \hline
      $\gamma(\cdot)$ & Reward discount parameter \\ \hline
      $S$ & State space \\ \hline
      $A$ & Action space \\ \hline
      $P_a(\cdot, \cdot)$ & State transition function \\ \hline
      $R_a(\cdot, \cdot)$ & Reward function \\ \hline
      $r$ & Observed reward \\ \hline
      $s$ & Observed state \\ \hline
      $a$ & Performed action \\ \hline
      $q_{\pi}(\cdot, \cdot)$ & Action-value function \\ \hline \hline
      \multicolumn{2}{|c|}{\textbf{Section \ref{sec:Phy}}} \\ \hline\hline
      $\mathfrak{s}_i,\mathfrak{s}_{-i}$ & Strategy of player $i$ and strategy of all players except $i$ \\ \hline
      $U_i\left(\mathfrak{s}_i,\mathfrak{s}_{-i}\right)$ & Utility dependent on $\mathfrak{s}_i$ and $\mathfrak{s}_{-i}$\\ \hline
      $\mathbb{P}$ & the set of players\\ \hline
      $\mathfrak{S}_i$ & the set of strategies of player $i$\\ \hline
      $p_i$ & Penalty of player $i$ for inducing interference $\mathcal{I}_i\left(\mathfrak{s}_i,\mathfrak{s}_{-i}\right)$ to other players\\ \hline
      $V_{c,d}$ & Value table for each channel device pair \\ \hline
      $\eta$ & Throughput learning rate of value table\\ \hline
      $\mathcal{C}\left(\epsilon,\omega\right)$ & Collision function which depends on exploration factor\\ & $\epsilon$ and other parameters $\omega$.\\ \hline
      $C^*$ & Collision threshold \\ \hline
      $L\left(\epsilon\right)$ & Loss function \\ \hline
      $s_n$ & System State \\ \hline
      $g_n$  & Channel gain \\ \hline
      $b_n$   & Buffer occupancy \\ \hline
      $n$  & Index of the block  \\ \hline
      $N$   & Maximum number of packets in the buffer \\ \hline
      $B$   & Size of the packet in bits\\ \hline
      $P_a$  & Poisson distribution where $a$ is the number of packets arriving at the buffer\\  \hline
      $\nu$ & Expected number of packets that will arrived in one block \\  \hline
      $p_n$  & Number of packets leaving the buffer in the $n^{th}$ block\\  \hline
      $d_n$  & Number of packets dropped from the buffer in the $n^{th}$ block\\  \hline 
      $M$  & Number of constellation points\\  \hline
      $m_n$  & Bits per symbol in the $n^{th}$ block\\  \hline
      $N_{sym}$  & Number of symbols in a block\\  \hline
      $N_0$  & Noise Spectral Density\\  \hline
      $\epsilon_*$  & Acceptable BER threshold\\  \hline
      $P_n$  & Transmission power in the $n^{th}$ block\\  \hline
      $\Bar{P}$  & Long term average power consumption\\  \hline
      $\mathfrak{T}$  & System Throughput\\  \hline
      $\mathfrak{P}_d$  & Packet drop probability\\  \hline
      $r_n$  & Reward per block\\  \hline
      $\mathfrak{r}\left(t\right), \mathfrak{r}\left(n\right)$  & Continuous and discrete representations of received signal\\  \hline
      $\mathfrak{a}\left(t\right)$  & Modulated amplitude as a function of time $t$\\  \hline
      $\phi\left(t\right)$  & Modulated phase as a function of time $t$\\  \hline
      $g\left(t\right)$  & Additive white Gaussian noise as a function of time $t$\\  \hline
      $\mathcal{A}\left(.\right), \mathcal{P}\left(.\right)$  & Amplitude and Phase distortion functions\\  \hline
      $\alpha_a,\beta_a, \alpha_{\phi}, \beta_{\phi}$  & Scalar values representing channel parameters\\  \hline
      $I_{\rho}$  & Information potential\\  \hline
      $\mathfrak{G}_\sigma(.)$  & Gaussian kernel with standard deviation $\sigma$\\  \hline
      $\rho$  & Entropy order\\  \hline
      $y_i$  & Adaptive system output\\  \hline
      $d_i$  & Desired system output\\  \hline
      $e_i$  & Error measure between actual and desired system output\\  \hline
      $L$  & Mean squared error loss\\  \hline
      $m_i$  & Transmitted symbol\\  \hline
      $\mathfrak{r}_i$  & Received symbol\\  \hline
      $\mu_\chi,var_\chi$  & Mean and variance of mini-batch $\chi$\\  \hline
      $m\left(n\right)$  & Discrete representation of baseband OFDM modulated signal\\  \hline
      $M\left(k\right)$  & Discrete frequency domain representation of $m\left(n\right)$\\  \hline
      $R\left(k\right), H\left(k\right),$  & Discrete frequency domain representation of received signal $\mathfrak{r}\left(n\right)$, \\ $G\left(k\right)$ & channel response $h\left(n\right)$, and additive white Gaussian noise $g\left(n\right)$ \\  \hline
      $y_{i,e=\left(v,c\right)}$ & Output of Neuron $e=\left(v,c\right)$ in the hidden layer $i$\\ \hline
      $z_v$  & Final $v^{th}$ output of the DNN\\  \hline
      $i, d$  & Antenna element and antenna element spacing\\  \hline
      $a_k, \theta_k, \phi_k,$  & Amplitude, incident angle, initial phase, and \\ $ f_0$ & initial frequency of $k^{th}$  incident signal\\  \hline
      $\mathbf{R}(n), R_{mm'}$  & Spatial correlation matrix and its respective diagonal element\\  \hline
      $\mathbf{\Theta}, \mathbf{F}$ & Incident angle matrix and hidden layer matrix\\ \hline \hline
      \multicolumn{2}{|c|}{\textbf{Section \ref{sec:Phy_Int}}} \\ \hline\hline
      $N_s$ & Number of samples \\ \hline
      $\gamma_{max}$ & Maximum value of the power spectral density of the normalized\\ &centered-instantaneous amplitude \\ \hline
      $C_{lk}$ & $l^{th}$ order, $k^{th}$ conjugate cumulant  \\ \hline
      $\delta_0$ & Deviation of normalized amplitude from the unit circle \\ \hline
      $\mathbf{x}^{IQ}_k$ & $k^th$ raw signal training example; I/Q representation \\ \hline
      $\mathbf{x}^{A/\Phi}_k$ & $k^{th}$ raw signal training example; amplitude and phase representation \\ \hline
      $\mathbf{x}^{F}_k$ & $k^{th}$ raw signal training example; frequency domain representation \\ \hline
      $\mathbf{r}_k$ & Received signal, vector form \\ \hline
      $r_{q_n}$ & Received signal quadrature value at index n \\ \hline
      $r_{i_n}$ & Received signal in-phase value at index n \\ \hline
      $x(n)$ & Transmitted signal, function of time \\ \hline
      $y(n)$ & Transmitted signal, function of time \\ \hline
      \hline
      \multicolumn{2}{|c|}{\textbf{Section \ref{sec:Higher_Layer}}} \\ \hline\hline
      $N$ & Total number of nodes in the network \\ \hline
      $N_T$ & Total number of time slots \\ \hline
      $T$ & Set of time slots \\ \hline
      $\mathbf{SA}$ & Slot assignment matrix \\ \hline
      $\mu_{xi}$ & Fuzzy state, a degree that time slot $t_x$ is assigned to node $i$ \\ \hline
      $\mathbf{U}$ & Fuzzy x-partition matrix  \\ \hline
      $\rho$ & Channel utilization  \\ \hline
      $deg(i)$ & Degree of edges incident to $i$  \\ \hline
      $E$ & Energy function \\ \hline
      $\alpha,\; \beta$ & Positive coefficients  \\ \hline
      $f$ & Fuzzification parameter \\ \hline
      $d_{ij}$ & Parameters used to define connectivity between $i$ and $j$  \\ \hline
      $c_r$ & Collision rate \\ \hline
      $P_{req}$ & Packet request rate \\ \hline
      $t_{w}$ & Average packet wait time  \\ \hline
      $p_t$ & Probability of an active DoS attack \\ \hline
      $\Gamma_{th}$ & Chosen threshold  \\ \hline
      $t$ & Time slot \\ \hline
      $h$ & Channel number \\ \hline
      $a_i(t)$ & Node $i$'s action at time slot $t$  \\ \hline
      $R_i$ & Reward for the action  \\ \hline
      $\mathcal{T}$ & Temperature  \\ \hline
      $z(t)$ & Channel observation \\ \hline
      $\mathfrak{h}$ & State history length \\ \hline
      $EX_t$ & Set of experience samples at time $t$  \\ \hline
      $ux$ & Upstream neighbor  \\ \hline
      $\mathcal{K}$ & Set of nodes \\ \hline
      $\mathcal{E}$ & Set of unidirectional wireless link \\ \hline
      $\mathcal{G}(\mathcal{K},\mathcal{E})$ & Directed connective graph \\ \hline
      $\gamma_{ij}$ & Score associated with edge $(i,j)$ \\ \hline
      $l$ & Number of neurons \\ \hline 
      $\tilde{\delta}$ & Normalized advance towards the sink  \\ \hline
      $\tilde{E}$ & Normalized residual energy \\ \hline
      $R_{C}$ & Constant reward if the node is able to reach sink directly  \\ \hline
      $R_{D}$ & Penalty suffered if no next-hop is found   \\ \hline
      $R_{E}$ & Penalty if existing next-hop has residula energy below the threshold \\ \hline
      $\epsilon$ & Probability of exploration \\ \hline
      $P_{ij}^{j}$ & Transition probability  \\ \hline
      $\alpha_1$, $\alpha_2$, $\beta_1$, $\beta_1$ & Tunable weights   \\ \hline
      $c$ & Constant cost associated with consumption of resources like bandwidth, etc. \\ \hline
      $E^{res}_i$ & Residual energy  \\ \hline
      $E^{ini}_i$ & Initial energy \\ \hline
      $E_i$ & Energy cost function associated with $E^{res}_i$ and $E^{ini}_i$   \\ \hline
      $\Bar{E_i}$ & Average residual energy \\ \hline
      $D_i$ & Measure of the energy distribution balance  \\ \hline
      
      $SK$ & Set of sinks  \\ \hline
      $SK_p$ & Subset of sinks  \\ \hline
      $H_{SK_p}^{\mathcal{NB}}$ & Routing information through all neighboring nodes in $\mathcal{NB}$ \\ \hline
\end{longtable}

\begin{acronym}\label{Table:Acro}
\acro{3GPP}{3rd Generation Partnership Project}
\acro{5G}{5th Generation}
\acro{6LOWPAN}{IPv6 over low power wireless personal area networks}
\acro{A3C}{asynchronous advantage actor critic}
\acro{AC}{Actor-Critic}
\acro{ACK}{acknowlegement}
\acro{AM}{amplitude modulation}
\acro{AMC}{automatic modulation classification}
\acro{ANN}{artificial neural network}
\acro{AP}{access point}
\acro{ASIC}{application specific integrated circuit}
\acro{AWGN}{additive white Gaussian noise}
\acro{AXI}{Advanced eXtensible Interface}
\acro{BEP}{belief propagation}
\acro{BER}{bit error rate}
\acro{BLE}{bluetooth low energy}
\acro{BP}{back-propagation}
\acro{BPSK}{binary phase shift keying}
\acro{BPTT}{back-propagation through time}
\acro{BSP}{broadcast scheduling problem}
\acro{BSSID}{basic service set identifier}
\acro{CART}{classification and regression trees}
\acro{CPFSK}{continuous phase frequency shift keying}
\acro{CPU}{central processing unit}
\acro{CR}{cognitive radio}
\acro{CR-IoT}{cognitive radio-based IoT}
\acro{CSMA}{carrier sense multiple access}
\acro{CSMA/CA}{carrier sense multiple access/collision avoidance}
\acro{CDMA}{code division multiple access}
\acro{CE}{cognitive engine}
\acro{CMAC}{cerebellar model articulation controller}
\acro{CNN}{convolutional neural network}
\acro{CR-VANET}{Cognitive Radio-Vehicular Ad Hoc Networks}
\acro{DARPA}{Defense Advanced Research Projects Agency}
\acro{DBN}{deep belief network}
\acro{DBSCAN}{Density-based Spatial Clustering of Applications with Noise}
\acro{DCNN}{deep convolutional neural network}
\acro{DCPC}{distributed constrained power control}
\acro{DMA}{direct memory access}
\acro{DoA}{direction of arrival}
\acro{DoS}{denial of service}
\acro{DRL}{deep reinforcement learning}
\acro{DSA}{dynamic spectrum access}
\acro{DSB}{double-sideband modulation}
\acro{DL}{deep learning}
\acro{DLMA}{deep reinforcement learning multiple access}
\acro{DNN}{deep neural network}
\acro{DP}{dynamic programming}
\acro{DQN}{deep Q-network}
\acro{EAR}{Energy-Aware Routing}
\acro{EM}{Expectation-Maximization}
\acro{FDMA}{frequency division multiple access}
\acro{FHNN}{fuzzy hopfield neural network}
\acro{FIFO}{first-in first-out}
\acro{FPGA}{field-programmable gate array}
\acro{FROMS}{Feedback Routing for Optimizing Multiple Sinks}
\acro{FSK}{frequency shift keying}
\acro{GA}{genetic algorithm}
\acro{GRU}{gated recurrent unit}
\acro{GFSK}{Gaussian frequency shift keying}
\acro{GMM}{Gaussian Mixture Model}
\acro{GMSK}{Gaussian minimum shift keying}
\acro{GPSR}{Greedy Perimeter Stateless Routing}
\acro{HDL}{hardware description language}
\acro{HLS}{high-level synthesis}
\acro{HMFPM}{Hybrid QoS Multicast Routing Framework-Based Protocol for Wireless Mesh Network}
\acro{HNN}{hopfield neural network}
\acro{II}{initiation interval}
\acro{IoT}{Internet of things}
\acro{IPC}{Intelligent Power Control}
\acro{I/Q}{in-phase/quadrature}
\acro{JQP}{join query packet}
\acro{JRP}{join reply packet}
\acro{LATA}{Local Access and Transport Area}
\acro{LANET}{visible light ad hoc network} 
\acro{LMR}{Land Mobile Radio}
\acro{LO}{local oscillator}
\acro{LoRa}{Long Range}
\acro{LoRaWAN}{Long Range Wide Area Network Protocol}
\acro{LoS}{line of sight}
\acro{LS}{least-squares}
\acro{LSTM}{long short term memory}
\acro{LTE}{long term evolution}
\acro{LTE-A}{long term evolution-advanced}
\acro{M2M}{machine-to-machine}
\acro{MAC}{medium access control}
\acro{MAP}{maximum a posteriori}
\acro{MANET}{mobile ad hoc network}
\acro{MIMO}{multiple input multiple output}
\acro{MDP}{markov decision process}
\acro{ML}{machine learning}
\acro{MLP}{multi-layer perceptron}
\acro{MMSE}{minimum mean square error}
\acro{MST}{multi-stage training}
\acro{M-QAM}{M-ary quadrature amplitude modulation}
\acro{MVDR}{minimum variance distortionless response}
\acro{MUSIC}{multiple signal classification}
\acro{NACK}{negative acknowledgement}
\acro{NB-IoT}{narrowband \ac{IoT}}
\acro{NCNN}{noisy chaotic neural network}
\acro{NDP}{node disconnection probability}
\acro{NE}{Nash equilibrium}
\acro{NLP}{natural language processing}
\acro{NOMA}{non-orthogonal multiple access}
\acro{NSG}{non-cooperative strategic game}
\acro{OFDM}{orthogonal frequency-division multiplexing}
\acro{OSPF}{open shortest path first}
\acro{PAM}{pulse-amplitude modulation}
\acro{PCA}{Principal component analysis}
\acro{PL}{programmable logic}
\acro{POMDP}{partially observable markov decision process}
\acro{PS}{processing system}
\acro{PSD}{power spectral density}
\acro{PSK}{phase shift keying}
\acro{PSO}{particle swarm optimization}
\acro{PU}{primary user}
\acro{QARC}{Video Quality Aware Rate Control}
\acro{QAM}{quadrature amplitude modulation}
\acro{QoE}{quality of experience}
\acro{QoS}{quality of service}
\acro{QPSK}{quadrature phase shift keying}
\acro{RAM}{random access memory}
\acro{RBF}{radial basis function}
\acro{RBFNN}{radial basis function neural network}
\acro{RF}{radio frequency}
\acro{RFID}{radio frequency identification}
\acro{RL}{reinforcement learning}
\acro{RLGR}{Reinforcement Learning based Geographic Routing}
\acro{RN}{residual network}
\acro{RNN}{recurrent neural network}
\acro{RSS}{received signal strength}
\acro{RSSI}{received signal strength indication}
\acro{SAG}{smart application gateway}
\acro{SAX}{simple aggregation approximation}
\acro{SC}{smart connectivity}
\acro{SC2}{Spectrum Collaboration Challenge}
\acro{SC-FDE}{single carrier frequency domain equalization}
\acro{SGD}{stochastic gradient descent}
\acro{SIR}{Sensor Intelligence Routing}
\acro{SoC}{system on chip}
\acro{SOM}{self-organizing map}
\acro{SNR}{signal-to-noise-ratio}
\acro{SSB}{single-sideband modulation}
\acro{SVC}{sequential vertex coloring}
\acro{SVM}{support vector machine}
\acro{SVR}{support vector regression}
\acro{SU}{secondary user}
\acro{TDMA}{time division multiple access}
\acro{UAN}{underwater acoustic network}
\acro{UF}{unrolling factor}
\acro{UAV}{unmanned aerial vehicle}
\acro{VANET}{vehicular ad hoc network}
\acro{VQPN}{video quality prediction network}
\acro{VQRL}{video quality reinforcement learning} 
\acro{WANET}{wireless ad hoc network}
\acro{WASN}{wireless ad hoc sensor network}
\acro{WBAN}{wireless body area networks}
\acro{WBFM}{wideband Frequency Modulation}
\acro{WIC}{wireless interference classification}
\acro{WSN}{wireless sensor network} 
\end{acronym}

\subsection{Introduction to Machine Learning} \label{Sec:Inrto}
The primary purpose of this section is to provide a brief overview of the field of \ac{ML} itself as well as provide a fundamental description of the algorithms and techniques presented as solutions to the wireless communications problems introduced in subsequent sections. This section aims to be as rigorous as necessary to allow the reader to understand how the presented algorithms are applied to wireless communications problems but does not aim to give an all-encompassing, comprehensive survey of the field of \ac{ML}. Interested readers are urged to refer to \citep{goodfellow2016deep}, \citep{Murphy:2012:MLP:2380985} and \citep{Bishop:2006:PRM:1162264} for a comprehensive understanding of \ac{ML}. The material presented in this section is given from a probabilistic perspective, as many of the concepts of \ac{ML} are rooted in probability and information theory. The rest of Section \ref{Sec:Inrto} provides a kind of road map for Section \ref{sec:overview} as a whole.

\subsubsection{Taxonomy}
Most introductory texts in \ac{ML} split the field into two subdivisions: supervised learning and unsupervised learning. We follow suit and will make the distinction of which subdivision each presented algorithm falls under. As will be shown in later sections of this paper, many problems in \ac{WANET} can be solved using an approach called \ac{RL}. \ac{RL} in its most fundamental form can be viewed as a third and separate subdivision of \ac{ML}, thus we will denote representative algorithms as such. It is important to note that many advanced \ac{RL} algorithms incorporate techniques from both supervised and unsupervised learning yet we will still denote these as \ac{RL} algorithms.

Another common type of learning discussed in \ac{ML} literature is that of \ac{DL}. We view \ac{DL} techniques not as a separate subdivision of \ac{ML} but as a means to achieve the ends associated with each of the three subdivisions stated above. \ac{DL} typically refers to the use of a \ac{DNN}, which we present with more rigor later in Section \ref{sec:FNN}. Thus the ``Deep" qualifier denotes an algorithm that employs a deep neural network to achieve the task. (ex: A \ac{DRL} algorithm would use a \ac{DNN} in a \ac{RL} framework)

\subsubsection{A Note on Modularity}
The concept of modularity is pervasive throughout engineering disciplines and is certainly prevalent in communications. We adopt this precedent throughout this text and present each of the algorithms using a common learning algorithm framework. This framework is primarily composed of the model, the optimization algorithm, the loss function, and a data set.

At its core, a learning algorithm is any algorithm that learns to accomplish some goal given some data to learn from. A common formalism of this definition is given in \cite{mitchellML}: ``A Computer program is said to learn from experience $E$ with respect to some class of tasks $T$ and performance measure $P$, if its performance at tasks $T$, as measured by $P$, improves with experience $E$." While this definition of a learning algorithm is commonly agreed upon, formal definitions of a task, experience, and performance measure are less endemic within the \ac{ML} community, thus we provide examples of each.

In the context of \ac{ML}, tasks usually define some way of processing an object or data structure. A classification task is the process of assigning a class label to an input object or data structure. While different $\mathbf{examples}$ (objects) within the data set will give rise to different class labels, the task of assigning a given example a label is the same for the entire data set. Other examples of tasks addressed in this text include regression (assigning a real value to an example) and structured output (assigning a separate data structure, with a pre-defined form, to an example).

The performance measure, $P$, essentially defines the criteria by which we evaluate a given learning algorithm. In the case of classification, the performance is typically the accuracy of the algorithm, or how many examples the algorithm assigns the correct class label to divided by the total number of examples. It is common practice to divide the entire available data set into two separate data sets, one used for training the algorithm and one used to test the algorithm. The latter, called the test set, is kept entirely separate from the algorithm while training and is used to evaluate the trained algorithm. The performance measure is often a very important aspect of the learning algorithm as it will define the behavior of the system.

The experience, $E$, that a learning algorithm has while learning essentially characterizes the algorithm into one of the three subdivisions defined earlier. Supervised learning algorithms are provided with a data set that contains examples and their associated labels or targets. An unsupervised learning algorithm experiences data sets containing only examples and attempts to learn the properties of the data set. \ac{RL} algorithms experience examples produced by the environment with which they interact. The environment often provides feedback to the \ac{RL} algorithm along with examples.

\subsection{Supervised Learning}
\subsubsection{Overview}

Recall from the previous discussion that in a supervised learning setting the learning algorithm experiences a data set containing examples and their respective labels or targets. An example will typically be denoted as $x$ and its label, or target, as $y$. Together, we have training examples $(x,y) \in D$ existing in our data set $D$. In supervised learning problems, we attempt to learn to predict the label $y$ from the example $x$, or equivalently, estimate the conditional distribution $p(y|x)$. Taking this approach, we will want to obtain a model of this conditional distribution and we will denote the parameters of such a model as $\boldsymbol{\theta}$. Assuming a set of i.i.d data $D = \{x_1, x_2, ... x_n\}$ drawn from the data generating distribution $p_{data}(x)$, the maximum likelihood estimator of the parameters, $\boldsymbol{\theta}$, of a model of the data generating distribution is given as,

\begin{equation}
    \boldsymbol{\theta}_{ML} = \argmax_{\boldsymbol{\theta}}p_{model}(D;\boldsymbol{\theta}) = \argmax_{\boldsymbol{\theta}}\prod_{i=0}^{n}p_{model}(x_i;\boldsymbol{\theta})
\end{equation}
where $p_{model}$ is a function space of probability distributions over the parameters $\boldsymbol{\theta}$. To make the above more computationally appealing, we can take the logarithm on both sides, as this does not change the optimization problem, which gives us,
\begin{equation}
    \boldsymbol{\theta}_{ML} = \argmax_{\boldsymbol{\theta}}\sum_{i=0}^{n}\log(p_{model}(x_i;\boldsymbol{\theta}))
\end{equation}
Additionally, we can divide the right hand side of the equation by $n$, as this does not change the optimization problem either, and we obtain the expectation of the log-probability of the model over the empirical data generating distribution,
\begin{equation}
    \boldsymbol{\theta}_{ML} = \argmax_{\boldsymbol{\theta}}\mathbb{E}_{x\sim\hat{p}_{data}}\log(p_{model}(x_i;\boldsymbol{\theta}))
\end{equation}
Alternatively, we could formulate the maximum likelihood estimation as the minimization of the KL divergence between the empirical data generating distribution and the model distribution given as,
\begin{equation}
    D_{KL}(\hat{p}_{data} || p_{model}) = \mathbb{E}_{x\sim\hat{p}_{data}} [\log(\hat{p}_{data}(x)) - \log(p_{model}(x))]
\end{equation}
Since the data generating distribution is not a function of the model, we can solve the same minimization problem by minimizing
\begin{equation}
    - \mathbb{E}_{x\sim\hat{p}_{data}} \log(p_{model}(x))
\end{equation}
which is exactly equivalent to the maximization problem stated in the maximum likelihood formulation. The above is referred to as the negative log-likelihood of the model distribution and minimizing it results in the minimization of the cross-entropy between the data generating distribution and the model distribution. The significance of this is two-fold. Firstly, the terms cross entropy and negative log-likelihood are often used in literature to describe the loss functions that are being used to evaluate a given \ac{ML} model and the above minimization problem is what is being referred to. Secondly, this gives rise to the narrative that the model associated with the maximum likelihood estimate is, in fact, the same model that most closely resembles the empirical data distribution. This is important considering what we want our model to do, namely, produce correct labels or targets for data drawn from the data generating distribution that the model has not seen before.

For completeness, the maximum likelihood estimator for the conditional distribution, which provides a label's probability given an example, is given as,

\begin{equation}
    \boldsymbol{\theta}_{ML} = \argmax_{\boldsymbol{\theta}}\sum_{i=0}^{n}\log(p_{model}(y_i | x_i;\boldsymbol{\theta}))
\end{equation}
for i.i.d examples, $x_i$.

Often times, regularization on the parameters of the model is desirable, as regularization can lead to better generalization of the model. This is most frequently seen in the different types of neural network models that will be described later in this section. Building on the maximum likelihood perspective of the loss function, we can show that adding a regularization function to our optimization function can be seen as inducing a prior over the model parameters and subsequently changing our estimator to the \ac{MAP} point estimate. Inducing a prior probability on the model parameter results in the following optimization problem,

\begin{equation}
    \boldsymbol{\theta}_{MAP} = \argmax_{\boldsymbol{\theta}}p(\boldsymbol{\theta} | D) = \argmax_{\boldsymbol{\theta}}\log(p(D;\boldsymbol{\theta})) + \log(p(\theta))
\end{equation}
Here, we have made use of Bayes' Rule, the properties of logarithm, and the fact that the optimization problem does not depend on the data generating distribution. If we wish to put a Gaussian prior on the parameters, $p(\boldsymbol{\theta})\sim \mathcal{N}(0, \frac{1}{\lambda}I^2)$ we obtain a log prior proportional to $\lambda \boldsymbol{\theta}^T\boldsymbol{\theta}$, which yields the popular L2-Regularization scheme. Again. we have made use of the fact that the Gaussian prior does not depend on the data distribution and contains constants that do not affect the optimization problem. Thus, the L2-Regularizer can be seen as a cost associated with the magnitude of the model's parameters as well as the placement of a Gaussian prior on the model parameters.

\subsubsection{Support Vector Machines} \label{Sec:SVM}
 
The \ac{SVM} was initially developed to perform the task of binary classification. Since their introduction into the \ac{ML} community, \acp{SVM} have been successfully extended to perform regression and multi-class classification tasks as well. \acp{SVM} are non-parametric models, meaning that the number of parameters that compose the model is not fixed whilst constructing the model. In contrast, a parametric model would have a fixed number of tunable parameters defined before constructing the model. We will first define the \ac{SVM} in the context of linear regression and then expand upon extensions to the algorithm later in the section. It is important to note here the change in notation of the model parameter vector from $\boldsymbol{\theta}$ to $\mathbf{w}$. Throughout the remaining parts of this section, $\mathbf{w}$ is typically used when the literature surrounding the algorithm refers to the parameter vector as a $\textit{weight}$ vector and $\boldsymbol{\theta}$ for a general parameter vector. The decision to forgo notation uniformity was made in an attempt to keep our notation consistent with each algorithm's original presentation, making the text more accessible to readers who may already be familiar with some of the algorithms.

Linear regression is perhaps one of the most well known and prevalent linear predictive models known throughout the \ac{ML} and statistical community. It is typically formulated as follows,

\begin{equation}
    y_i = \mathbf{w}^T\mathbf{x_i} + w_0
\end{equation}
where $y_i$ are the target values, $\mathbf{x_i}$ are individual training examples and weights, $\mathbf{w}$, are the model parameters. A common approach to solving such a problem is to vectorize the output and input variables and solve the normal equations, giving a closed form solution for the \ac{MMSE}. A typical approach to adapt this algorithm to perform classification tasks is the well known logistic regression given as,

\begin{equation}
    p(y=1|\mathbf{x};\mathbf{w}) = \sigma(\mathbf{w}^T\mathbf{x})
\end{equation}
where $\sigma$ is the logistic \textit{sigmoid} function given as,
\begin{equation}
    \sigma(x) = \frac{1}{1 + e^{-x}}
\end{equation}
One favorable property of logistic regression is that it has a well defined probabilistic interpretation that can be viewed as maximizing the likelihood of the conditional distribution $p(y|\mathbf{x})$.
An alternative formulation for a linear classifier is given in what is known as the perceptron algorithm \cite{rosenblatt1957perceptron}. The perceptron algorithm aims to find a hyperplane in the input space that linearly separates inputs that correspond to different classes. It does so using a zero-one loss function, meaning that the model is penalized equally for every point in the training data that it classifies incorrectly. An obvious shortcoming is that the algorithm converges to any hyperplane that separates the data; it need not be the \textit{optimal} hyperplane.

The linear \ac{SVM} \cite{vapnik1963} attempts to find the hyperplane that best separates the data, where the optimal hyperplane maximizes the margin between the nearest points in each class on either side of the plane. While this solution is better, the true power of \acp{SVM} comes from the kernelization of the linear \ac{SVM}, which allows the model to find nonlinear boundaries between different classes by representing the input data in a higher dimensional space. Kernelization of an algorithm is a process by which the parameters of the model are written in terms of a linear combination of the input vectors, which allows the computation of the inner product between a new input vector and the parameter vector of the model to be written as an inner product of the new input and the training inputs. A kernel function can then be substituted for the inner products between training vectors, which can be intuitively interpreted as a function that returns a real value representing the similarity between two vectors. The kernelization of the \ac{SVM} leads to the kernel \ac{SVM} \cite{CortesVapnikSVM}. The most common kernels used to kernelize \acp{SVM} are the linear, polynomial, and \ac{RBF} kernels, given as,

\begin{align}
    k(\mathbf{x_i},\mathbf{x_j}) &= \mathbf{x_i}^T\mathbf{x_j}, \\
    k(\mathbf{x_i},\mathbf{x_j}) &= (\mathbf{x_i}^T\mathbf{x_j} + 1)^d, \text{ and} \\
    k(\mathbf{x_i},\mathbf{x_j}) &= e^{-\frac{(\mathbf{x_i} - \mathbf{x_j})^2}{\sigma^2}}
\end{align}
respectively, where $\sigma$ is a user defined parameter.

\subsubsection{Decision Trees}
Decision trees can be employed for both the tasks of classification and regression. Decision tree algorithms are similar to nearest neighbor type algorithms in a sense that labels for examples lying near each other in input space should be similar; however, they offer a much lighter weight solution to these problems.

A decision tree is essentially nothing more than an aggregation of if conditions that allow a new example to traverse the tree. The tree is traversed until happening upon a leaf node, which would specify the output label. Decision trees can be constructed in a number of different ways, but a common approach is to create trees that minimize some measure of impurity while splitting the data. There are many such impurity measures but each of them essentially conveys how non-homogeneous the data in either child node would be if a given split of the data were to occur. A child node containing only training examples of the same label is referred to as a pure leaf and decision trees are often constructed to contain only pure leaves.

We now discuss two of the most popular impurity functions used in decision tree construction. We first define the training data as $D = \{(\mathbf{x_1},y_1), ..., (\mathbf{x_n},y_n)\},\; y_i \in \{1, ..., c\}$ where $c$ is the number of classes. Additionally, we have $D_k \subseteq D$ where $D_k = \{(\mathbf{x},y) \in D : y = k \}$ and $D = D_1 \cup ... \cup D_c$. We then define the fraction of inputs in $D$ with label $k$ as,

\begin{equation}
    p_k = \frac{|D_k|}{|D|}
\end{equation}
and the Gini Impurity of a leaf node and a tree, respectively as,

\begin{align}
    G(D) &= \sum_{k=1}^c p_k(1-p_k), \text{ and} \\
    G^T(D) &= \frac{|D_L|}{|D|}G^T(D_L) + \frac{|D_R|}{|D|}G^T(D_R)
\end{align}
where $D = D_L \cup D_R$, $D_L \cap D_R = \varnothing$. The idea is then to choose splits in the tree that minimize this measure of impurity. Another popular impurity function is the entropy function. The entropy of the tree has its derivation in using the KL-divergence between the tree label distribution and the uniform distribution to determine how impure it is. Leaving the derivation to the interested reader, we define,

\begin{align}
    H(D) &= -\sum_k p_k\log(p_k), \\
    H^T(D) &= \frac{|D_L|}{|D|}H^T(D_L) + \frac{|D_R|}{|D|}H^T(D_R)    
\end{align}
as the entropy of a leaf and the tree respectively. While decision trees can be strong classifiers on their own, they often benefit from a technique called bagging. We omit the statistical derivation of the benefits of bagging and simply state the essence of bagging: by training many classifiers and considering the average output of the ensemble we can greatly reduce the variance of the overall ensemble classifier. Bagging is often done with decision trees as decision trees are not very robust to errors due to variance in the input data. 

Perhaps the most popular bagged algorithm is that of the Random Forest. Random forests are bagged decision trees generated by the following procedure,
\begin{itemize}
    \item Sample $m$ datatsets $D_1, ..., D_m$ from $D$ with replacement.
    \item For each $D_i$ train a decision tree classifier $h_i(\cdot)$ to the maximum depth and when splitting the tree only consider a subset of features $k$.
    \item The ensemble classifier is then the mean output decision i.e.\\ 
        $h(\mathbf{x}) = \frac{1}{m}\sum_{i=1}^{m}h_i(\mathbf{x})$
   
\end{itemize}
The number of trees $m$ can be set to any number, provided the computational resources are available. If $d$ is the number of features in each training example, the parameter $k\leq d$ is typically set to $k = \sqrt{d}$.

\subsubsection{Feedforward Neural Networks} \label{sec:FNN}
The original formulation of feedforward neural networks was proposed in \cite{rosenblatt1962principles}. It can be seen as an extension to the previously mentioned perceptron algorithm with an element-wise nonlinear transition function applied to the linear classifier. This nonlinear transition function allows the hyperplane decision boundary to take a nonlinear form, allowing the model to separate training data that is not linearly separable. The formulation for a given layer, $l$, is as follows,

\begin{align}
    \mathbf{z}^l &= {\mathbf{W}^{(l)}}^T\mathbf{a}^{l-1} + \mathbf{b}^l \\
    \mathbf{a}^l &= \sigma(\mathbf{z}^l)
\end{align}
where $\mathbf{a}^{l-1}$ are the outputs from the previous layer and may be referred to as the activation values of the previous layer. In the instance where the layer in question is the input layer, $\mathbf{a}^{l-1}$ would be set as $\mathbf{x}$, the training example input. The current layer's activation values are thus denoted as $\mathbf{a}^{l}$ and in the case of the output layer, these values would be synonymous with $\hat{\mathbf{y}}$. The layer weight matrix, ${\mathbf{W}^{(l)}}^T$, consists of column weight vectors for each neuron in the layer and $\mathbf{b}^l$ is a column vector containing the bias term for each neuron. One common implementation approach to handling the bias term is to add an additional parameter to each of the weight vectors and append a $1$ to the input vector. When a bias term is omitted this formulation can be assumed unless otherwise stated throughout the section.

The nonlinear transition function, $\sigma$, is also referred to as the activation function throughout literature and is often chosen from a handful of commonly used nonlinear functions for different applications. The most widely used activation functions are the following,

\begin{align}
    \sigma(z) &= \frac{1}{1 + e^{-z}},\\
    ReLU(z) &= \max(0, z),\;\text{and} \\
    \tanh(z) &= \frac{e^z - e^{-z}}{e^z + e^{-z}}
\end{align}
    
\begin{figure}[h!]
    \centering
    \includegraphics[width=5.5 in]{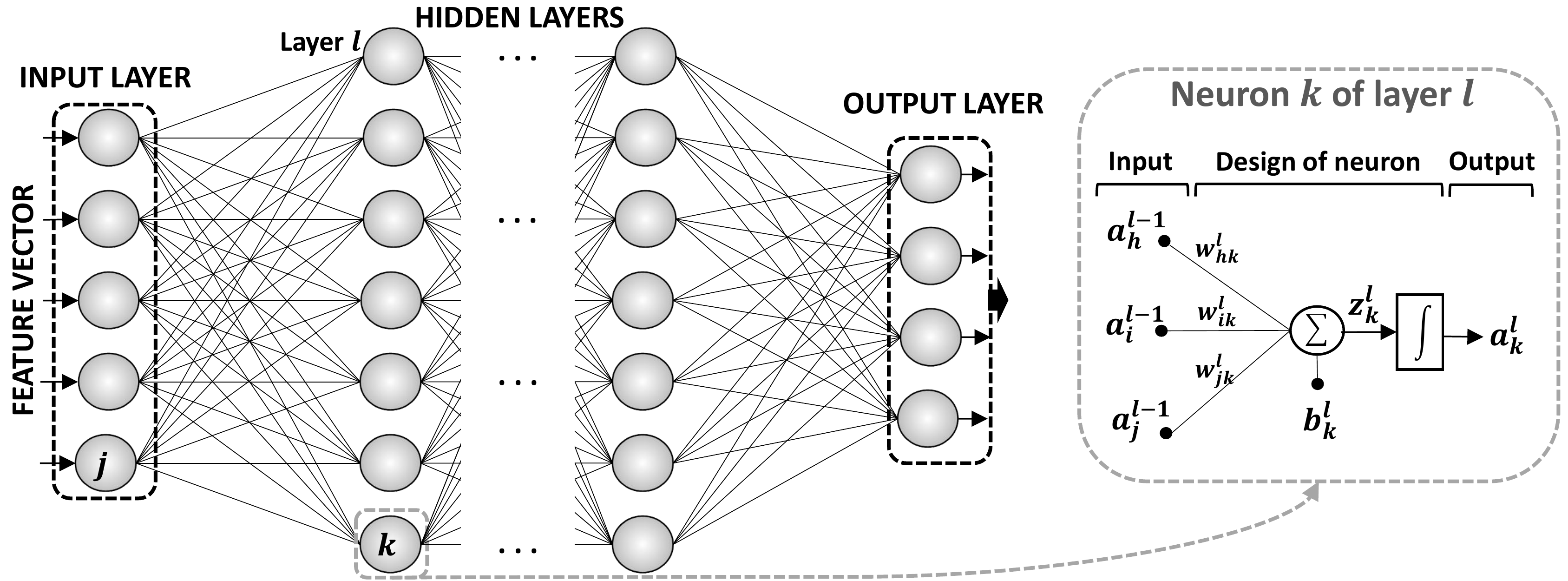}
    \caption{Standard Framework of Feed Forward Neural Network}
    \label{fig:FFNN}
\end{figure}

Additionally, the \ac{RBF} kernel function described earlier in Section \ref{Sec:SVM} can be used as an activation function and doing so give rise to the \ac{RBFNN} \cite{rbfNet}. To increase the complexity of the model, and thus its ability to learn more complex relationships between the input features, network layers can be subsequently added to the model that accept the previous layer's output as input. Doing so results in a \ac{DNN}. The function of the network as a whole $\phi(\mathbf{x})$ thus becomes,

\begin{equation}
    \phi(\mathbf{x}) = \mathbf{W}^{(3)}\sigma(\mathbf{W}^{(2)}\sigma(\mathbf{W}^{(1)}\mathbf{x}))
\end{equation}
where the weight matrices $\mathbf{W}^{(i)}$ are indexed according to the layer they belong to. Intuitively, this allows the first layer to learn linear functions between the input features, the second layer to learn nonlinear combinations of these functions, and the third layer to learn increasingly more complex nonlinear combinations of these functions. This formulation additionally gives rise to a nice graphical interpretation of the model, which is widely used in literature and given in Figure \ref{fig:FFNN}.

This graphical interpretation is also where the feedforward neural network gets its loose biological interpretation. Each solid line in Figure \ref{fig:FFNN} denotes a weighted connection in the graph. The input, output, and hidden layers are denoted as such in the graph and a close up of one node in the graph is provided. This close up calls the single node a neuron, but it can equivalently be referred to simply as a unit in this text and throughout literature. The close up also shows the inputs to the neuron, the weighted connections from the previous layer, the weighted sum of inputs, and the activation value, denoted as $a_i^{l-1}$, $w_{ik}^l$, $z_k^l$, and $a_k^l$, respectively. Occasionally, a neuron employing a given activation function may be referred to as such a unit in this text and throughout literature, i.e. a unit with a \textit{ReLU} activation function may be called a ``\textit{ReLU} unit".

The most common way to train neural networks is by way of the \ac{SGD} optimization algorithm. \ac{SGD} is similar to well-known gradient descent methods with the exception that the true gradient of the loss function with respect to the model parameters is not used to update the parameters. Usually, the gradient is computed using the loss with respect to a single training example or some subset of the entire training set, which is typically referred to as a mini-batch, resulting in mini-batch \ac{SGD}. This results in the updates of the network following a noisy gradient, which in fact, often helps the learning process of the network by being able to avoid convergence on local minima which are prevalent in the non-convex loss landscapes of neural networks. The standard approach to applying \ac{SGD} to the model parameters is through the repeated application of the chain rule of derivation using the famous back-propagation algorithm \cite{rumelharthintonwilliams86b}.

The last layer in any given neural network is called the output layer. The output layer differs from the inner layers in that the choice of the activation function used in the output layer is tightly coupled with the selection of the loss function and the desired structure of the output of the network. Generally, the following discussion of output layers and loss functions applies to all neural networks, including the ones introduced later in this section.

Perhaps the simplest of output unit activation functions is that of the linear output function. It takes the following form,

\begin{equation}
    \hat{\mathbf{y}} = \mathbf{W}^T\mathbf{a} + \mathbf{b}
\end{equation}
where $\mathbf{W}$ is the output layer weight matrix, $\mathbf{a}$ are the latent features given by the activation output from the previous layer, and $\hat{\mathbf{y}}$ are the estimated output targets. Coupling a linear output activation function with a mean squared error loss function results in the maximizing the log-likelihood of the following conditional distribution,

\begin{equation}
    p(\mathbf{y}|\mathbf{x}) = N(\mathbf{y};\hat{\mathbf{y}},I)
\end{equation}

Another task that we have already touched upon in our discussion of \acp{SVM} and perceptrons is that of binary classification. In a binary classification task, the output target assumes one of two values and thus can be characterized by a Bernoulli distribution, $p(y=1|\mathbf{x})$. Since the output of a purely linear layer has a range over the entire real line, we motivate the use of a function that ``squashes" the output to lie in the interval $[0,1]$, thus obtaining a proper probability. We have seen that the logistic \textit{sigmoid} does exactly this and it is in fact the preferred method to obtain a Bernoulli output distribution. Accordingly, the output layer becomes,

\begin{equation}
    \hat{y} = \sigma(\mathbf{w}^T\mathbf{a} + \mathbf{b})
\end{equation}
The negative log-likelihood loss function, used for maximum likelihood estimation, of the above output layer is given as,
\begin{equation}
    \mathcal{L}(\mathbf{y},\mathbf{x},\mathbf{w}) = -\log(p(\mathbf{y}|\mathbf{x};\mathbf{w})) = f((1 - 2\mathbf{y})\mathbf{z}) \label{bernLoss}
\end{equation}
where $f(x) = \log(1 + e^x)$ is called the \textit{softplus} function and $\mathbf{z} = \mathbf{w}^T\mathbf{x} + \mathbf{b}$ is called the activation value. The derivation of (\ref{bernLoss}) is not provided here but can be found in \citep{goodfellow2016deep} for the interested reader.

For a multi-class classification task, the desirable output distribution is that of the Multinoulli distribution. The Multinoulli distribution assigns to each class the probability that a particular example belongs to it, requiring the sum over class probabilities for a single example be equal to 1. The Multinoulli distribution is given as the conditional distribution: $\hat{y}_i = p(y = i | \mathbf{x})$. It is important to note that the output, $\hat{\mathbf{y}}$, is now an $n$-dimensional vector containing the probability that $\mathbf{x}$ belongs to class $i\in[0,n]$ at each index $i$ in the output vector. The targets for such a classification task are often encoded as an $n$-dimensional vector containing $(n-1)$ number of 0's and a single 1, located at an index $j$ which denotes that the associated training example belongs to the class $j$. This type of target vector is commonly referred to as a one-hot vector.
The output function that achieves the Multinoulli distribution in the maximum likelihood setting is called the \textit{softmax} function and is given as,

\begin{equation}
    softmax(\mathbf{z})_i = \frac{e^\mathbf{z}}{\sum_j e^{z_j}} \label{softmax}
\end{equation}
where $z_j$ is the linear activation at an output unit $j$. \textit{Softmax} output units are almost exclusively coupled with a negative log-likelihood loss function. Not only does this give rise to the maximum likelihood estimate for the Multinoulli output distribution but the log in the loss function is able to undo the exponential in the \textit{softmax} which keeps the output units from saturating and allows the gradient to be well-behaved, allowing learning to proceed \citep{goodfellow2016deep}.

\subsubsection{Convolutional Neural Networks}
The \ac{CNN} was originally introduced in \cite{lecun1989generalization} as a means to handle grid-like input data more efficiently. The input of this type could be in the form of a time-series but is more typically found as image-based input. The formulation of \acp{CNN} additionally has biological underpinnings related to the human visual cortex.

\acp{CNN} are very similar to the feedforward networks introduced previously with the exception that they use a convolution operation in place of a matrix multiplication in the computation of a unit's activation value. In this section, we assume the reader is familiar with the concept of the convolution operation on two continuous functions, where one function, the input function, is convolved with the convolution kernel. The primary differences from the aforementioned notion of convolution and convolution in the \ac{CNN} setting are that the convolution operation is discretized (for practical implementation purposes) and that it is often truly the cross-correlation operation that is performed in \acp{CNN} rather than true convolution. This means that the kernel is not typically flipped before convolving it with the input function. This is also primarily done for practical implementation purposes and does not typically affect the efficacy of the \ac{CNN} in practice.

Convolution in the context of \acp{CNN} is thus defined as the following, for an input image $I$,

\begin{equation}
    S(i,j) = (K * I)(i,j) = \sum_m \sum_n I(m,n)K(i-m, j-n)
\end{equation}
where $K$ is the convolution kernel and the output, $S$, is often referred to as the feature map throughout literature. It is important to note that the above formulation is for two-dimensional convolution but can be extended to input data of different dimensions. The entries of $K$ can be seen as analogues of the weight parameters described previously (Section \ref{sec:FNN}) and can be learned in a similar manner using \ac{SGD} and the \ac{BP} algorithm. Intuitively, one can imagine having multiple $K$ kernels in a single \ac{CNN} layer being analogous to having multiple neurons in a single feedforward neural network layer. The output feature maps will be grid-like and subsequent convolutional layers can be applied to these feature maps after the element-wise application of one of the aforementioned nonlinear activation functions.

\begin{figure}[h!]
    \centering
    \includegraphics[width=4 in]{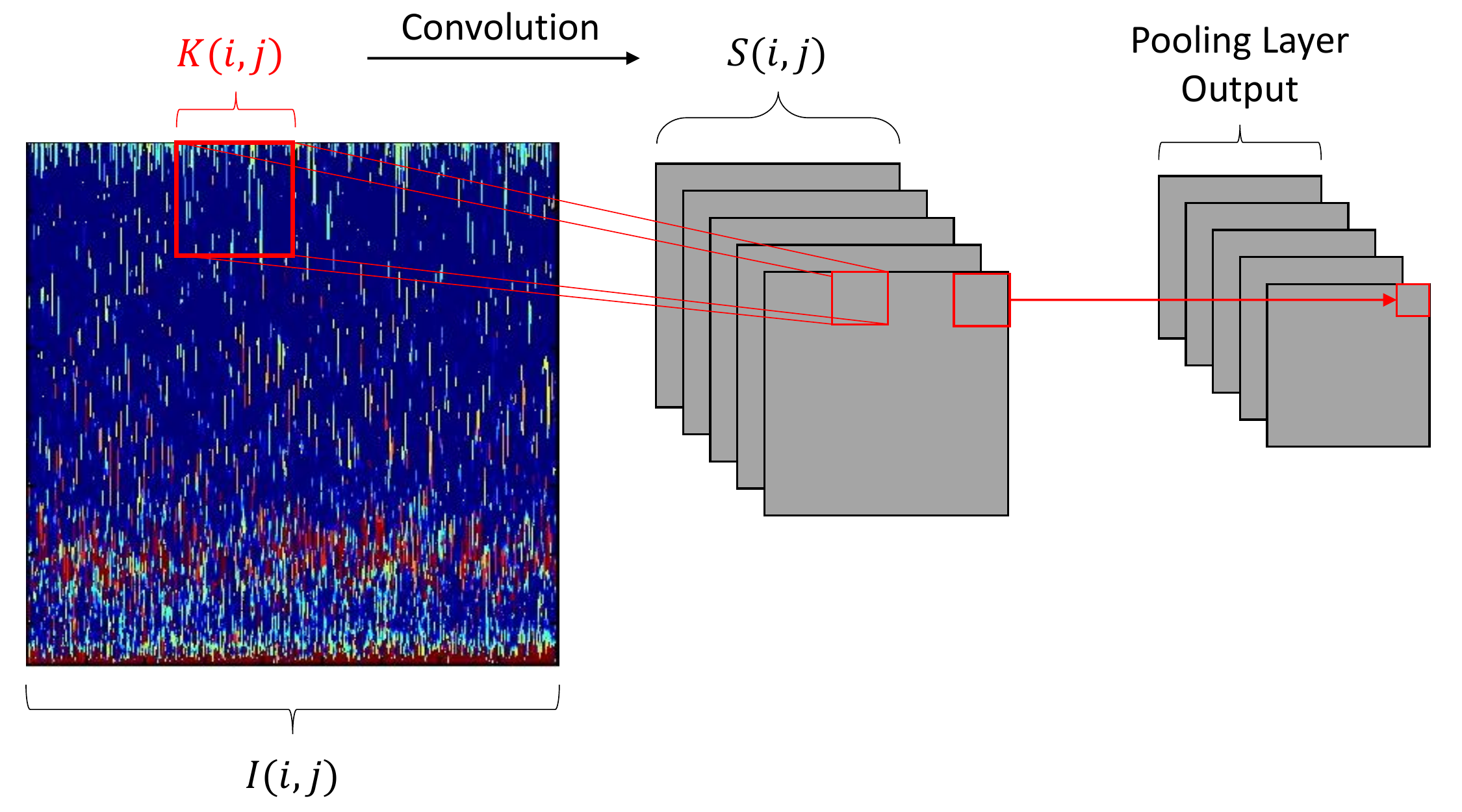}
    \caption{Convolutional and Pooling Layers of a CNN}
    \label{fig:CNN_REP}
\end{figure}

In addition to convolutional layers, \acp{CNN} often employ a separate kind of layer called pooling layers. The primary purpose of a pooling layer is to replace the output of the network at a certain location with a summarization of the outputs within a local neighborhood in the grid. Examples of pooling layers include max pooling \citep{Zhou1988ComputationOO}, average pooling, $L^2$ norm pooling, and distance weighted average pooling. A max pooling layer would summarize some rectangular region of the input image by selecting only the maximum activation value present in the region as output from the pooling layer. Pooling layers improve the efficacy of \acp{CNN} in a few different ways. First, they help make the learned representation of the input invariant to small translations, which is useful when aiming to determine the presence of a feature in the input rather than its location. Second, pooling layers help condense the size of the network since convolutional layers don't inherently do so. A binary classification task taking image data with size $256\times 256\times 3$ will need to reduce the size of the net to a single output neuron to make use of the output layer and cost function pairs described previously in Section \ref{sec:FNN}. Lastly, pooling layers lead to infinitely strong prior distributions making the \ac{CNN} more statistically efficient \citep{goodfellow2016deep}. A pictorial representation of a single convolutional layer followed by a pooling layer is given in Figure \ref{fig:CNN_REP}. The figure depicts a single convolutional layer applied to an input image of a waterfall plot of electroencephalogram data followed by a pooling layer. Subsequent convolutional layers may follow the pooling layer in a \ac{DCNN}, and a nonlinear activation function may be applied to $S(i,j)$ prior to the pooling operation.

Some common adaptations applied to \acp{CNN} come in the form of allowing information flow to skip certain layers within the network. While the following adaptions were demonstrated on \acp{CNN} and \acp{LSTM} (a type of \ac{RNN}), the concepts can be applied to any of the networks presented in this paper. A \ac{RN}, or ResNet \cite{ResNet}, is a neural network which contains a connection from the output of a layer, say $L_{i-2}$, to the input of the layer $L_i$. This connection allows the activation of the $L_{i-2}$ to skip over the layer $L_{i-1}$ such that a ``residual function" is learned from layer $L_{i-2}$ to layer $L_{i}$. A highway neural network \cite{highwayNet} is similar in that it allows a skip connection over layers but additionally applies weights and activation functions to these connections. Lastly, a dense neural network \cite{denseNet} is a network that employs such weighted connections between each layer and all of its subsequent layers. The motivation behind each of these techniques is similar in that they attempt to mitigate learning problems associated with vanishing gradients \cite{vanishGrad}. For each of these networks, the \ac{BP} algorithm used must be augmented to incorporate the flow of error over these connections.

\subsubsection{Recurrent Neural Networks}
The \ac{RNN} was first introduced in \cite{rumelharthintonwilliams86b} as a way to handle the processing of sequential data. These types of neural networks are similar to \acp{CNN} in the sense that they make use of parameter sharing; however, in \acp{RNN}, parameters are shared across time steps or indices in the sequential input. Recurrent nets get their name from the fact that they have recurrent connections between hidden units. We denote this mathematically as follows,

\begin{equation}
    \mathbf{h}^{(t)} = f(\mathbf{h}^{(t-1)}, \mathbf{x}^{(t)} ; \boldsymbol{\theta})
\end{equation}
where the function $f$ could be considered the activation output of a single unit, $\mathbf{h}^{(i)}$ are called the state of the hidden units at a time $i$, $\mathbf{x}^{(i)}$ is the input from the sequence at the index $i$, and $\boldsymbol{\theta}$ are the weight parameters of the network. Note, $\boldsymbol{\theta}$ is not indexed by $i$, signifying that the same network parameters are used to compute the activation at all indices in the the input sequence. Output layers and loss functions appropriate for the desired task are then applied to the hidden unit state $\mathbf{h}$.

\begin{figure}[h!]
    \centering
    \includegraphics[width=4 in]{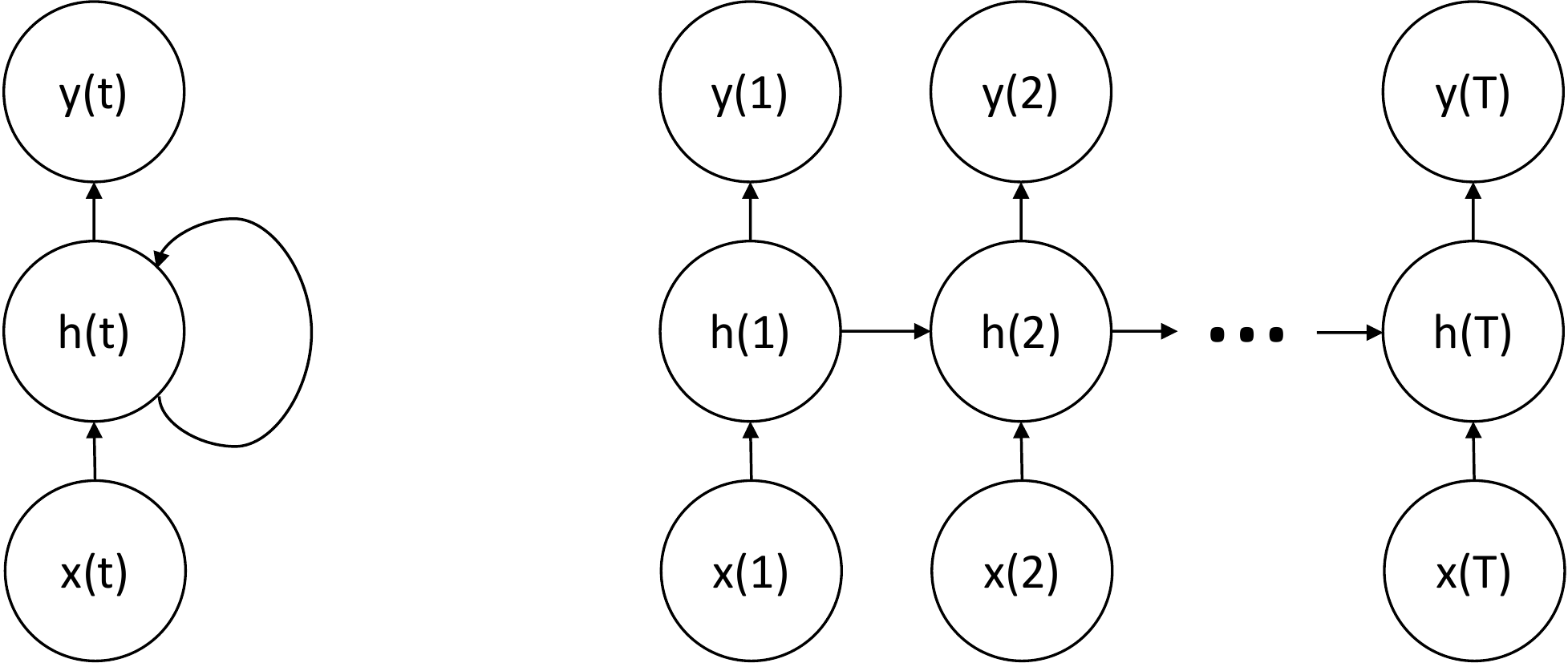}
    \caption{Equivalent graphical formulations for Recurrent Neural Networks}
    \label{fig:RNN_REP}
\end{figure}

Two equivalent graphical representations of \acp{RNN} are provided as reference in Figure \ref{fig:RNN_REP}. The left representation shows the network ``rolled up" with a recurrent connection onto itself. The right representation shows the network ``unrolled" with the recurrent connections now propagating information forward in time. We now provide the forward propagation equations for the hidden unit and use the \textit{softmax} output layer as an example of how the hidden state would be used as input to the output layer. A loss function can then be applied to the \textit{softmax} output as previously discussed in the paper.

\begin{align}
    \mathbf{a}^{(t)} &= \mathbf{W}\mathbf{h}^{(t-1)} + \mathbf{U}\mathbf{x}^{(t)} + \mathbf{b} \\
    \mathbf{h}^{(t)} &= \tanh(\mathbf{a}^{(t)}) \\
    \mathbf{o}^{(t)} &= \mathbf{V}\mathbf{h}^{(t)} + \mathbf{c} \\
    \hat{\mathbf{y}}^{(t)} &= softmax(\mathbf{o}^{(t)})
\end{align}

The matrices $\mathbf{W}$, $\mathbf{U}$, and $\mathbf{V}$ are the weight matrices shared across hidden units. They are used to weight the connections between hidden units from one time step to the next, between the input and hidden state at the current time step, and the hidden state and output at the current time step. The parameters $\mathbf{b}$ and $\mathbf{c}$ are bias term vectors that are shared across time steps.

The loss for a single sequential training example is accumulated over the entire sequence, thus using a negative log-likelihood loss for a sequence $\mathbf{x}^{(t)}$ with output targets $y^{(t)}$ the loss would be,

\begin{equation}
    \mathcal{L}(\{\mathbf{x}^{(1)}, ..., \mathbf{x}^{(\tau)}\}, \{y^{(1)}, ..., y^{(\tau)}\}, \boldsymbol{\theta}) = -\sum_{t} \log(p_{model}(y^{(t)} | \{\mathbf{x}^{(1)}, ..., \mathbf{x}^{(t)}\};\boldsymbol{\theta}))
\end{equation}
The computation for the gradient of the loss with respect to the model parameters is involved and is out of the scope of this paper. For the interested reader, \ac{SGD} is commonly employed to train \acp{RNN}, employing the \ac{BPTT} \cite{werbosBPTT} algorithm to compute the gradients.

Many extensions to the described \ac{RNN} model exist and are worth mentioning. Perhaps the most obvious extension is to add more recurrent layers following the single recurrent layer that was described above, resulting in Deep \acp{RNN} \cite{PascanuGCB13}. This provides similar advantages that were discussed in the motivation for extending feedforward networks to multiple layers. Additionally, more recurrent connections can be added which may skip over time steps, skip over layers, or even move information backward in time resulting in bidirectional \acp{RNN} \citep{SchusterBRNN}. These additional recurrent connections would be weighted and a nonlinear activation function could be applied in the same manner that the basic recurrent connection operates.

The most prevalent extensions to the original \acp{RNN} are those of the \ac{LSTM} and \ac{GRU}, developed originally in \cite{hochreiterLSTM} and \cite{ChoGRU}, respectively. \acp{LSTM} augment the traditional \ac{RNN} framework by adding a self loop on the state of the network. This self loop is coupled with input, output, and forget gates which control whether input values are written to the state, the state values are forgotten within the state, or the state values are written to the output of the network, respectively. These adaptations allow the network to better ``remember" relevant information over longer periods in time. Each of the gates is weighted and have a logistic \textit{sigmoid} activation applied to them, allowing the network to learn how to best use these gates with respect to the task. \acp{GRU} operate in a similar fashion but instead use two gates, namely, the update and reset gates. The update gate controls to what degree the state of the network at the given time step is written back to the state variable as well as what parts of the new state to write to the current state. The reset gates control what parts of the current state to use in the next computation of the new state. Both the \ac{LSTM} and \ac{GRU} have the ability to retain information over longer time periods and aim to mitigate the negative learning mechanics associated with vanishing gradients.

Recurrent networks can also take forms that are significantly different from the models described above. In particular, a \ac{HNN} \cite{Hopfield2554} is a special type of recurrent network formulated to recover corrupted patterns. Specifically, it is a recurrent network where each unit is connected to all other units in the graph except for itself. Additionally, the weight between units is shared and each unit in the network encodes a binary state value, typically either $1$ or $-1$. This formulation aims to mimic the forms of associative memory present in human cognition models and is often trained using a form of Hebbian Learning \cite{Hebb}. The famous summarization of Hebbian learning, ``cells that fire together wire together" drives the idea that when part of the pattern that the \ac{HNN} is trained to recognize is present, all of the units associated with that pattern will ``fire" and the entire pattern will be represented by the network. Another interesting difference from the previously described \ac{RNN} structures is that the \ac{HNN} does not make use of any type of training targets $y$. This makes the \ac{HNN} a type of unsupervised learning algorithm, more of which we discuss in further detail in the next section.

\subsection{Unsupervised Learning}
\subsubsection{Overview}

Unsupervised learning, a separate learning paradigm from the previous described supervised learning, attempts to learn useful properties of the training data rather than learning to map inputs to specific outputs. Examples of unsupervised learning tasks include probability density estimation, denoising, and clustering. Unsupervised learning algorithms only experience the training data examples and are given no target outputs, which are obviously preferable in scenarios when data sets are produced without targets and it would be impractical for a human to go through and label the data set with a target value. Thus, without targets, unsupervised learning algorithms usually try to present the data set in a simpler or easier to understand representation. This simpler representation most commonly manifests itself in the form of lower dimensional representations of data, sparse representations of data, and independent representations of the data.

While some unsupervised learning algorithms draw techniques from previously mentioned supervised learning algorithms, they employ different types of loss functions. Usually, the best types of loss functions to use in unsupervised learning settings will reward the algorithm for preserving information about the input data but penalize the algorithm for not representing the data in one of the three ways discussed in the previous paragraph. The reader may be familiar with the \ac{PCA} algorithm, which is a great example of a linear unsupervised learning algorithm that aims to decorrelate the input data.

\subsubsection{Clustering Algorithms}
\label{sec:clustering}
Clustering algorithms are unsupervised learning algorithms that all share a similar goal of attempting to separate the input data set into some number of partitions, or clusters. The process by which these various algorithms group the data points into clusters is specific to each algorithm but is typically based on a metric which may be a function of distance to other data points, density of the surrounding data points, or fit to a probability distribution, among others. Once a clustering algorithm has grouped the input data into clusters, the algorithm is used to categorized new data points into one of the existing clusters. This categorization is computed using the same metric the algorithm initially used to construct the clusters. The primary shortcomings of clustering algorithms arise from the algorithm having a lack of specification about what similarities the clusters should represent in the data. Thus the algorithm may find some grouping of the input data that the designer did not intend for, rendering the resultant classifier ineffective. Next, a few common clustering algorithms are described in further detail.

\textbf{Lloyd's Algorithm for k-means clustering.} Lloyd's algorithm for \textit{k}-means clustering was initially introduced in \cite{lloyd}, and its presentation has since been proliferated to a multitude of sources. The algorithm itself was developed to obtain a solution to the \textit{k}-means problem, which concerns finding $k$ points (cluster centroids) in the input space which minimize the distance between each training vector and the nearest centroid. Formally the \textit{k}-means problem is as follows. Given a training data set $D = \{\mathbf{x_1}, ..., \mathbf{x_n}\}$, $\mathbf{x_i} \in \mathcal{R}^d$ and an integer $k$, find $k$ points $\boldsymbol{\mu_1}, ..., \boldsymbol{\mu_k} \in \mathcal{R}^d$ which minimize,

\begin{equation}
    f = \sum_{\mathbf{x_i}\in D} \min_{j\in[k]}\norm{\mathbf{x_i} - \boldsymbol{\mu_j}}^2
\end{equation}

Intuitively, minimizing the above expression will attempt to minimize the distance from any given training vector to the nearest cluster centroid. The algorithm developed to find the centroids, the set of $\boldsymbol{\mu_1}, ..., \boldsymbol{\mu_k}$, can be broken out into a two step algorithm that is repeatedly performed until additional iterations no longer further minimize the expression above. We introduce a time parameter $t$ to show how the centroids, and the clusters, $C_1, ..., C_k$ change as the algorithm progresses. For a random initialization of centroids $\boldsymbol{\mu_1}, ..., \boldsymbol{\mu_k}$ the first step, called the assignment step, is given as,

\begin{align}
    C_j^{(t)} = \left\{\mathbf{x_i} : \norm{\mathbf{x_i} - \boldsymbol{\mu_j^{(t)}}}^2 \leq \norm{\mathbf{x_i} - \boldsymbol{\mu_m^{(t)}}}^2 \forall m, 1 \leq m \leq k \right\}, 
    \;\;\text{s.t.}\; C_1 \cap ... \cap C_k = \varnothing
\end{align}

The following step, called the update step, computes the centroids of the newly assigned clusters as follows,

\begin{equation}
    \boldsymbol{\mu_j^{(t+1)}} = \frac{1}{|C_j^{(t)}|}\sum_{\mathbf{x_i} \in C_j^{(t)}} \mathbf{x_i}
\end{equation}

The presented algorithm will converge once there are no further reassignments of any training vectors to new clusters. Once the algorithm is trained, inference is performed by computing the distance from a new input vector, $\mathbf{r}$, and associating it with cluster $j$ according to,

\begin{equation}
    \argmin_j \norm{\mathbf{r} - \boldsymbol{\mu_j}}^2
\end{equation}

\textbf{\acp{GMM}.} Clustering using \acp{GMM} in conjunction with the \ac{EM} \cite{Bishop:2006:PRM:1162264} algorithm is an example of a probability distribution based clustering algorithm and can be seen as an extension to \textit{k}-means clustering algorithms that allow the clusters themselves to take on different shapes other than perfect circles. This ability is realized through modeling each cluster as a Gaussian distribution with parameterized mean and covariance, and the entire clustered data distribution as a weighted linear combination of Gaussian distributions called a Gaussian mixture. Given a training data set $D = \{\mathbf{x_1}, ..., \mathbf{x_N}\}$, $\mathbf{x_i} \in \mathcal{R}^d$ and an integer $K$, model the distribution of a given data point $\mathbf{x}$ as,

\begin{equation}
    p(\mathbf{x}) = \sum_{k=1}^K \pi_k \mathcal{N}(\mathbf{x} | \boldsymbol{\mu_k}, \boldsymbol{\Sigma_k})
\end{equation}

Where $0 \leq \pi_k \leq 1$, $\sum_k \pi_k = 1$, and $\boldsymbol{\mu_k} \in \mathcal{R}^d$, $\boldsymbol{\Sigma_k}\in \mathcal{R}^{d\times d}$ are the mean vector and covariance matrix of the $k$-th Gaussian distribution in the mixture. Following the maximum likelihood approach introduced in the beginning of this section, the maximum likelihood estimate for the \ac{GMM} parameters is given as follows,

\begin{equation}
    \log(p(\mathbf{X} | \boldsymbol{\pi}, \boldsymbol{\mu}, \boldsymbol{\Sigma}) = \sum_{n=1}^N \log \left[ \sum_{k=1}^K \pi_k \mathcal{N}(\mathbf{x_n} | \boldsymbol{\mu_k}, \boldsymbol{\Sigma_k}) \right]
\end{equation}

Where $\mathbf{X}$ is a matrix constructed from the concatenation of the input training vectors. By maximizing the log-likelihood function using the \ac{EM} algorithm, we can obtain the optimal model parameters that give rise to Gaussian distributions that best describe the training input data. To do so we first define,

\begin{equation}
    \gamma(z_k) = p(z_k = 1 | \mathbf{x}) = \frac{\pi_k \mathcal{N}(\mathbf{x} | \boldsymbol{\mu_k}, \boldsymbol{\Sigma_k})}{\sum_{j=1}^K \pi_j \mathcal{N}(\mathbf{x} | \boldsymbol{\mu_j}, \boldsymbol{\Sigma_j})}
    \label{eqn:GMMposterior}
\end{equation}

Where $\mathbf{z} \in \mathcal{R}^K$ is a one-hot vector used to reference any one of the $K$ Gaussian components within the mixture. Thus, $\gamma(z_k)$ as defined above can be interpreted as the probability that the $k$-th component of describes the training vector $\mathbf{x}$ best. This formulation is useful for developing the \ac{EM} algorithm for \ac{GMM}. In order to perform the \ac{EM} algorithm, we must first solve for the maximum likelihood estimates of each of the tunable parameters. Setting the derivatives of $\log(p(\mathbf{X} | \boldsymbol{\pi}, \boldsymbol{\mu}, \boldsymbol{\Sigma})$ equal to 0, we obtain the following equations for each of the \ac{GMM} parameters,

\begin{align}
    \boldsymbol{\mu_k} &= \frac{1}{N_k}\sum_{n=1}^N\gamma(z_{nk})\mathbf(x_n) \label{eqn:maxlikelihoodGMM0} \\
    \boldsymbol{\Sigma_k} &= \frac{1}{N_k}\sum_{n=1}^N\gamma(z_{nk})(\mathbf{x_n} - \mathbf{\mu_k})(\mathbf{x_n} - \mathbf{\mu_k})^T \\
    \pi_k &= \frac{N_k}{N} \quad \text{where,} \\
    N_k &= \sum_{n=1}^N\gamma(z_{nk})
    \label{eqn:maxlikelihoodGMM}
\end{align}

Thus, in the expectation step of the \ac{EM} algorithm, we compute (\ref{eqn:GMMposterior}) with the current model parameters; obtaining probabilities representing which component distribution best describes each input vector. In the maximization step, we compute (\ref{eqn:maxlikelihoodGMM0})-(\ref{eqn:maxlikelihoodGMM}) using the previously computed values of $\gamma(z_{nk})$. Doing so obtains an estimate of the distribution parameters for each component distribution that most likely describe each of the training vectors associated with that component. Iterating through both the expectation and maximization steps yields the \ac{EM} algorithm. E and M steps are typically performed until the log-likelihood of the overall model increases only marginally in any given step.

There are a few well-known difficulties in fitting \acp{GMM} with the \ac{EM} algorithm. Foremost, the log-likelihood function allows for singularities to arise, where one component attempts to describe a single training point. This will send the standard deviation parameter of that component to 0 which will cause the likelihood to tend to infinity. Such a situation can only be avoided by resetting the distribution parameters at fault before restarting the fitting process. The \ac{EM} algorithm is also computationally expensive and typically needs to iterate many times before convergence occurs. To mitigate the computational requirements, the Lloyd's algorithm described earlier can be used to obtain a better initialization for the component distributions.

\textbf{Density-based Clustering.} Density-based clustering algorithms aim to assign clusters to areas in the input training vector space that are particularly dense with respect to the areas around them. Additionally, such algorithms may mark points that lie in a low density area as outliers, not requiring them to belong to any cluster. One of the most popular density-based clustering algorithms is the \ac{DBSCAN} algorithm, originally presented in \cite{EsterDBSCAN}. The \ac{DBSCAN} algorithm provides six definitions, from which the clusters of the training data set, $D = \{\mathbf{x_1}, ..., \mathbf{x_n}\}$,  are built. Two input parameters, $\epsilon$ and $minpts$, and a distance function are required to be provided to the algorithm by the designer. The usages of each are elucidated in the definitions given below:

\begin{itemize}
\item \textbf{Definition 1}: The $\epsilon$-neighborhood, $N_{\epsilon}(\mathbf{x})$, of a training vector $\mathbf{x_i}$ is defined to be the set of all points whose distance from $\mathbf{x_i}$ is less than or equal to $\epsilon$. i.e. $N_{\epsilon}(\mathbf{x_i}) = \{\mathbf{x_j}\in D | \text{dist}(\mathbf{x_i}, \mathbf{x_j}) \leq \epsilon\}$
\item \textbf{Definition 2}: Given $\epsilon$ and $minpts$, $\mathbf{x_j}$ is directly density reachable from $\mathbf{x_i}$ if $\mathbf{x_j}\in N_{\epsilon}(\mathbf{x_i})$ and $|N_{\epsilon}(\mathbf{x_i})|\geq minpts$
\item \textbf{Definition 3}: A training vector $\mathbf{x_j}$ is density reachable from $\mathbf{x_i}$ if $\exists \mathbf{x_i}, ..., \mathbf{x_j}$ such that $\mathbf{x_{k+1}}$ is directly density reachable from $\mathbf{x_{k}}$.
\item \textbf{Definition 4}: $\mathbf{x_{j}}$ is density connected to $\mathbf{x_{i}}$ if $\exists \mathbf{x_{k}}$ such that $\mathbf{x_{j}}$ and $\mathbf{x_{i}}$ are density reachable from $\mathbf{x_{k}}$
\item \textbf{Definition 5}: A set $C$ such that $C\subset D$ and $C\neq \varnothing$, is a cluster if 
    \begin{itemize}
        \item $\forall \mathbf{x_i}, \mathbf{x_j}: $ if $\mathbf{x_i}\in C$ and $\mathbf{x_j}$ is density reachable from $\mathbf{x_i}$ then $\mathbf{x_j}\in C$
        \item $\forall \mathbf{x_i}, \mathbf{x_j} \in C: $ $\mathbf{x_i}$ is density connected to $\mathbf{x_j}$
    \end{itemize}
    \item \textbf{Definition 6}: For clusters $C_1, ..., C_k$ of $D$, $\text{noise} = \{ \mathbf{x_i} \in D | \forall j: \mathbf{x_i} \notin C_j\}$
\end{itemize}

The algorithm for finding clusters within the training data set is as follows. First, an initial random training vector is selected from the training data, $\mathbf{x_i}$, and all points within the $\epsilon$-neighborhood of $\mathbf{x_i}$ are retrieved. If $|N_{\epsilon}(\mathbf{x_i})|$ is less than $minpts$ the vector $\mathbf{x_i}$ is added to the noise set. If $|N_{\epsilon}(\mathbf{x_i})|$ is greater than or equal to $minpts$, (at least $minpts$ number of training examples are directly density reachable from $\mathbf{x_i}$) all points in $|N_{\epsilon}(\mathbf{x_i})|$ are added to the current cluster index set. Using this initial set, all points that are density reachable from $\mathbf{x_i}$ are then retrieved and added to the current cluster index set. The algorithm then increments the cluster index and repeats the preceding process selecting a new initial point in the training set that has not been associated with either the noise set or any cluster set. 

The primary advantage of the \ac{DBSCAN} algorithm is that the number of clusters need not be specified by the designer of the algorithm. Additionally, there are no constraints on the shape of any given cluster, as is the case implicitly with both Lloyd's algorithm and \ac{GMM} clustering. \ac{DBSCAN} also incorporates a noise set, allowing the clusters to be robust to outliers. A disadvantage of the \ac{DBSCAN} algorithm arises when clusters in the data have very different densities, making it difficult to select the appropriate values for $\epsilon$ and $minpts$.

\subsubsection{Autoencoders}

\begin{figure}[h!]
    \centering
    \includegraphics[width=3.7 in]{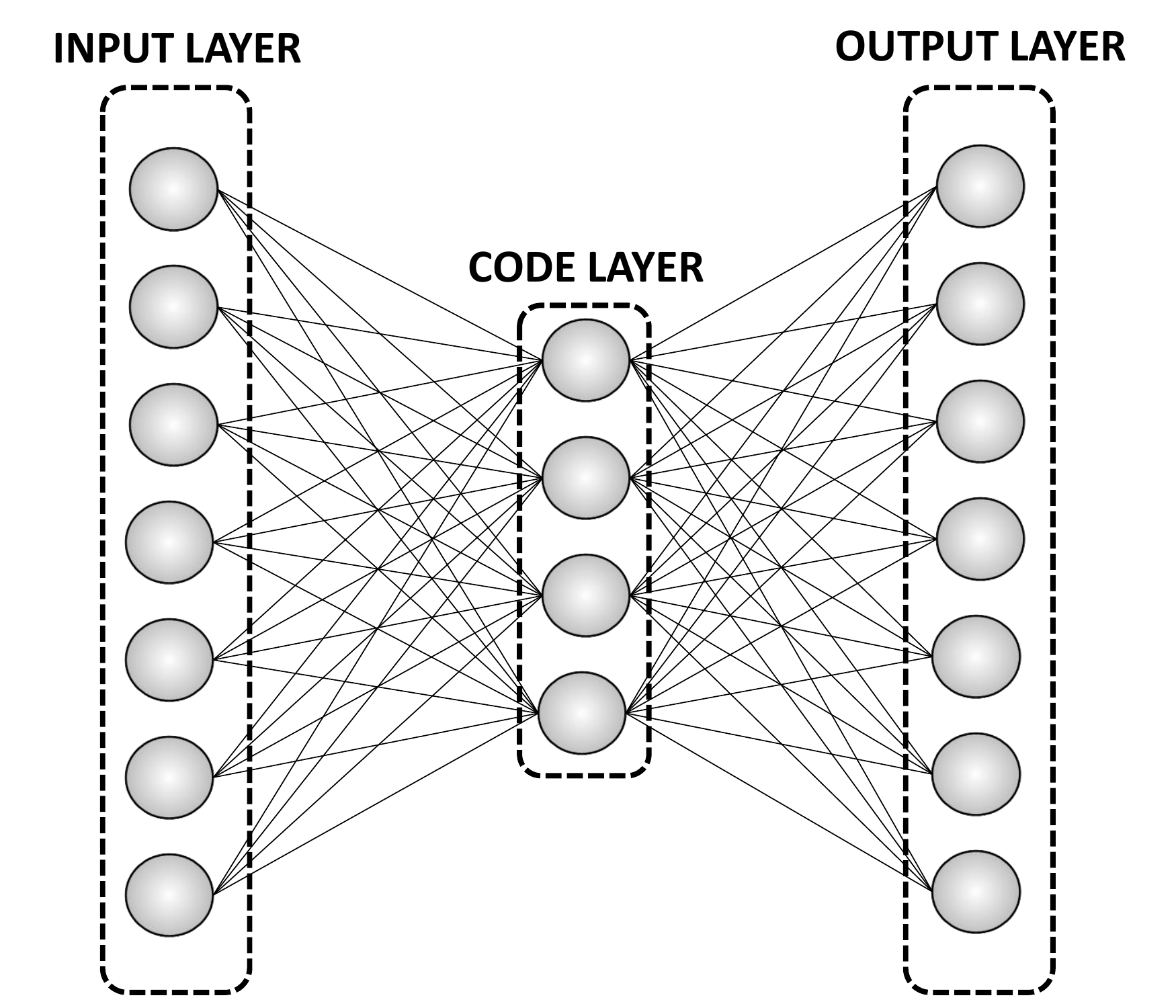}
    \caption{General Structure of an Autoencoder used for Dimensionality Reduction}
    \label{fig:AE_DR}
\end{figure}

Autoencoders were first introduced in \cite{LecunAE} and have a similar structure to \acp{DNN} in that they have an input layer, an output layer, and at least one hidden layer, often called the code layer. Autoencoders, while similar in structure to supervised neural network models, are like other unsupervised learning methods in that they attempt to learn a mapping from the input data to a latent representation that exhibits unique characteristics useful for performing some task. Such latent representations are often learned for the purpose of dimensionality reduction and de-noising; however, in either case, the formulation of the autoencoder splits the model into two parts: the encoder and the decoder. The encoder, usually denoted as $f$, takes the input data and maps it to a latent representation, or code, $\mathbf{h}$, such that $\mathbf{h} = f(\mathbf{x})$. The decoder, $g$, then attempts to reconstruct the original input data from latent representation. The training signal for the autoencoder model is thus computed using a loss function assuming the following form,

\begin{equation}
    \mathcal{L}(\mathbf{x},g(f(\mathbf{x})),\boldsymbol{\theta})
\end{equation}
and may be any function penalizing the dissimilarity between the two arguments. Such a function will force the encoder to learn a latent representation from which the original input data can be reconstructed by the decoder. While the loss function above necessitates the output layer of the decoder to be the same size as the input layer of the encoder, the code layer of the autoencoder is often smaller than the input and output layers. Such is the case of autoencoders used for dimensionality reduction or feature learning; a diagram of such an autoencoder structure is provided in Figure \ref{fig:AE_DR}. This ensures that the code learned by the encoder contains only the most salient information of the data distribution that still allows for reconstruction. In dimensionality reduction and feature learning autoencoders, the decoder becomes inert after the model has been trained and only the encoder portion of the model is used to perform the task.

In denoising autoencoder models, the loss function is augmented such that a corrupted version of the input data is given to the encoder, and the loss is computed using the original input and decoder output. For original input, $\mathbf{x}$, and corrupted version, $\tilde{\mathbf{x}}$, the resulting denoising autoencoder loss function is given as,

\begin{equation}
    \mathcal{L}(\mathbf{x}, g(f(\tilde{\mathbf{x}})),\boldsymbol{\theta})
\end{equation}

The corrupted version of the input data is typically sampled from some corruption process such that each corrupted data is not corrupted in the same way. Unlike dimensionality reduction autoencoders, after the denoising autoencoder model is trained the entire model is kept and used to perform the task. 

\subsubsection{Self Organizing Maps}
The \ac{SOM} \cite{kohonenSOM} was originally introduced as a type of unsupervised learning algorithm with the goal of performing dimensionality reduction and data clustering. The reader may be familiar with the simple clustering algorithm referred to as $k$-means clustering, covered in this text in Section \ref{sec:clustering}, in which each example in the training data is required to belong to one of $k$ different clusters. The obvious pitfall of this algorithm is that the designer of the algorithm must choose the parameter $k$ prior to constructing the model, hence the model's usefulness is contingent on the user's estimate of the appropriate number of clusters. The \ac{SOM} algorithm avoids this by learning the appropriate number of clusters. Additionally, the \ac{SOM} algorithm typically aims to represent the training data as a two-dimensional grid, where examples that are near each other in the input topological space are embedded near each other in the two-dimensional latent representation.

The canonical \ac{SOM} formulation can be viewed as a fully connected single layer feedforward neural network, with units arranged in a two-dimensional grid. As the network sees each input, it computes the similarity between the input vector and each unit in the grid using some discriminant function such as,

\begin{equation}
    d_j(\mathbf{x}) = \sum_{i=1}^N (x_i - w_{ji})^2
\end{equation}
where $d_j(\mathbf{x})$ is the value of the discriminant function at unit $j$, $\mathbf{w_j}$ is the weight vector associated with unit $j$, and $i \in [1,N]$ indexes the $N$ dimensional input and weight vectors. This is often called the competitive process of \acp{SOM} as it is representative of a type of learning called competitive learning.

Once the discriminant function is computed at each unit for a training example the unit with the least value for the discriminant function is selected for what is called the cooperative process of \ac{SOM}. The cooperative process attempts to update the neurons in some local neighborhood around the neuron that provides the closest representation of the input vector (i.e. the neuron with the minimal discriminant function). This creates neighborhoods in the map that will activate similarly for similar input values, thus creating clusters within the map. The topological neighborhood is usually defined as,

\begin{equation}
    T_{j,I(\mathbf{x})} = \exp(\frac{-S^2_{j,I(\mathbf{x})}}{2\sigma^2})
\end{equation}
where $I(\mathbf{x})$ represents the index in the map where the minimal discriminant function occurred and $S_{j,i}$ denotes the distance from a neuron $j$ to a neuron $i$. $\sigma$ is a parameter chosen by the designer and is typically decayed over time using the following schedule for time-dependence,

\begin{equation}
    \sigma_t = \sigma_0 \exp(\frac{-t}{\tau_{\sigma}})
\end{equation}

Once the topological neighborhood is computed, the weight vectors associated with the units in the neighborhood are updated. This is usually referred to as the adaptive process in the context of \acp{SOM}. The change applied to the weight vectors is given as,

\begin{equation}
    \delta w_{ji} = \eta(t) T_{j,I(\mathbf{x})}(t)(x_i - w_{ji})
\end{equation}
where $\eta(t)$ is the learning rate parameter and is also decayed over time using a similar schedule to that of the $\sigma$ parameter,

\begin{equation}
    \eta_t = \eta_0 \exp(\frac{-t}{\tau_{\eta}})
\end{equation}
This process is repeated many times for each training example in the training data set, resulting in the \ac{SOM}.

\subsection{Reinforcement Learning}
\subsubsection{Overview}
\ac{RL} is a learning paradigm that can be considered separate from supervised and unsupervised learning. That being said, \ac{RL} techniques often use ideas and algorithms from both unsupervised and supervised learning. We first describe the problem formulation for \ac{RL} and then present a solution and how it can be extended to include concepts from other learning paradigms \citep{SuttonBarto98}.

\ac{RL} is built on the idea of an agent performing actions within an environment, based on its observations of the environment. The agent generally carries out actions according to a policy, which defines how the agent behaves at a given time. The agent receives reward signals, which define the ultimate goal of the algorithm, from the environment which indicates how well off the agent is at the time step the reward is given. The agent then aims to maximize its cumulative reward by observing its environment and the reward signal received, and then performing actions based on these inputs. The maximization of the cumulative reward is typically defined in terms of a value function. The value function differs from the reward signal in that the reward represents what is a desirable immediate setting and the value function represents how much reward the agent can obtain in the future given the agent's current state. Additionally, \ac{RL} problems typically define a model of the environment. The model is estimated by the agent to determine the dynamics of the environment and is then subsequently used by the agent to devise some sort of plan about how to act.

\ac{RL} problems, as described above, are usually formalized mathematically using finite \ac{MDP}. The tuple $(S, A, P_a(\cdot, \cdot), R_a(\cdot, \cdot))$ defines the dynamics of the \ac{MDP} as well as the state and action spaces, $S$ and $A$. At a given time step, an agent observes a state $s$, chooses an action $a$, receives a reward $r$, and transitions to a new state $s'$. The functions $P_a(\cdot, \cdot)$ and $R_a(\cdot, \cdot)$ define the transition probabilities between states and reward received from the environment when transitioning to a new state. The transition probability function takes the current state, $s$, and a possible new state, $s'$ and outputs the probability of transitioning to that new state, conditioned on an action, $a$. i.e.,

\begin{equation}
    P_a(s, s') = Pr(S_{t+1} = s' | S_t = s, A_t = a)
\end{equation}
$R$ is reward function such that it gives the reward obtained directly after transitioning to state $s'$ from state $s$ via action $a$ and is defined as,

\begin{equation}
    R_a(s, s') = E\left[ R_{t+1} | S_t = s, A_t = a\right]
\end{equation}

A policy is a function which defines how the agent will act given the state it is currently in. The policy is usually denoted as $\pi(a|s)$. Using such a policy, the agent moves about the environment and can start to construct a value function and action-value function based on the return they observe. The action-value function, $q$, for a policy, $\pi$ is given as,

\begin{equation}
    q_{\pi}(s,a) = E_{\pi}\left[ \sum_{k=0}^{\infty} \gamma^k R_{t+k+1} | S_t = s, A_t = a \right]
\end{equation}
where $R_t$ are the observed returns over time and $\gamma$ is a scaling parameter that is used to weight future returns less heavily than immediate returns. The action-value function can be plainly stated as the expected return starting in a state $s$, taking the action $a$, and subsequently following the policy $\pi$. Obtaining values for state action pairs allows for the agent to plan how to act in its environment. Equipped with the optimal action-value function, the solution to the \ac{MDP} is merely choosing the action with the greatest action value.

\subsubsection{Q-Learning}

The method of Q-Learning was introduced in \cite{Watkins89} and is what is called an off-policy control algorithm. The off-policy qualifier simply denotes that the algorithm does not depend on the policy the agent uses to navigate the environment. The Q-Learning is algorithm is defined by the following update rule,

\begin{equation}
    Q(S_t, A_t) \xleftarrow{}  Q(S_t, A_t) + \alpha \left[R_{t+1} + \gamma \max_a Q(S_{t+1}, a) - Q(S_t, A_t) \right]
\end{equation}
Using such an update scheme for action-value pairs will lead to the approximation of the optimal action-value function independent of the policy being followed.

As one would imagine, the state and action spaces of some \ac{RL} problem become extremely vast, making the storage of the action-value function for all state-action pairs impractical. One way to overcome this issue is to introduce a function approximation method which learns to provide values for state-action pairs. This function approximator could be one of the different types of \ac{ML} algorithms discussed previously in this section. Such is the case in the famous \ac{DQN} \cite{mnihDQN}, where a deep convolutional neural network was used to approximate the action-value function when learning to play Atari games.

\subsubsection{REINFORCE}
In the previously presented Q-Learning algorithm, the agent moves about the environment according to some predetermined policy so that it may learn to accurately approximate the action-value function for the state and action spaces belonging to the environment. After learning the action-value function, the agent then navigates through the environment by selecting the actions that map to the greatest action-value function in the given state. The REINFORCE algorithm is inherently different, in that it attempts to learn the optimal policy directly and is thus characterized as a policy gradient method. This distinction specifies that the training signal is in fact a gradient with respect to the parameterized policy function and that the algorithm makes use of the policy gradient theorem \cite{SuttonBarto98}, given as,

\begin{equation}
    \nabla J(\boldsymbol{\theta}) \propto \sum_{s}\mu(s) \sum_{a}q_{\pi}(s,a) \nabla \pi(a|s;\boldsymbol{\theta})
\end{equation}
where $J(\boldsymbol{\theta})$ is a performance measure usually defined as some function of cumulative reward or reward rate, $\boldsymbol{\theta}$ is the policy parameterization vector, and $\mu(s)$ is a distribution over states which denotes the probability of being in any given state.

From the policy gradient theorem, we wish to obtain an expression that specifies exactly how the policy parameters are updated. To do so, we need an expression that provides information about how the performance measure is affected by performing a specific action, $A_t$, in a specific state, $S_t$. Augmenting the policy gradient theorem to allow the parameter update to be computed for every action taken at every state requires the distribution weighted sum to be replaced by the expectation under the policy $\pi$ of the gradient \cite{Williams1992}. Doing so results in,

\begin{align}
    \nabla J(\boldsymbol{\theta}) &= E_{\pi}\left[ \sum_{a} q_{\pi}(S_t,a) \nabla\pi(a|S_t;\boldsymbol{\theta}) \right] \\
    &= E_{\pi}\left[ \sum_{a} \pi(a|S_t;\boldsymbol{\theta}) q_{\pi}(S_t,a) \frac{\nabla\pi(a|S_t;\boldsymbol{\theta})}{\pi(a|S_t;\boldsymbol{\theta})} \right] \\
    &= E_{\pi}\left[ q_{\pi}(S_t,A_t) \frac{\nabla\pi(A_t|S_t;\boldsymbol{\theta})}{\pi(A_t|S_t;\boldsymbol{\theta})} \right] \\
    &= E_{\pi}\left[ G_t \frac{\nabla\pi(A_t|S_t;\boldsymbol{\theta})}{\pi(A_t|S_t;\boldsymbol{\theta})} \right]
\end{align}
Where $G_t$ is the cumulative reward at time $t$. This gives rise to the policy parameter update,

\begin{equation}
    \boldsymbol{\theta}_{t+1} \doteq \boldsymbol{\theta}_t + \alpha G_t \frac{\nabla\pi(A_t|S_t;\boldsymbol{\theta})}{\pi(A_t|S_t;\boldsymbol{\theta})}
\end{equation}
Where $\alpha$ is a step size parameter. Intuitively, such an update moves the parameter vector in a direction that increases the probability of taking action $A_t$ in state $S_t$, proportional to the return received for doing so, normalized by the probability of choosing that action. The above algorithm formulation allows for the policy function to be any differentiable function approximator, which is often one of the neural network structures previously describes in this section. Coupling a neural network with a $softmax$ output function additionally provides the output as a distribution, which is desirable as the policy at a given state should be a distribution over actions.

\subsubsection{Actor-Critic Methods}
Actor-Critic methods \cite{SuttonBarto98} are policy gradient methods that learn a state-value function in addition to the learned policy. The actor, a differentiable function approximator for the policy, learns the optimal policy in a similar fashion described in the REINFORCE algorithm with exception that an eligibility trace is used to update the policy parameters allowing for online learning. The critic should be a differentiable state-value function approximator and also learns using eligibility traces, thus allowing the entire algorithm to learn online.

An eligibility trace vector, $\mathbf{z}$, is a simple way of accumulating parameters that need updating over some time. For an actor policy, $\pi(A|S;\boldsymbol{\theta})$, parameterized by $\boldsymbol{\theta}$, and a critic action-value function, $\hat{v}(S;\mathbf{w})$, parameterized by $\mathbf{w}$, the respective eligibility trace vector updates for the parameters are given as,

\begin{align}
        \mathbf{z}^{\mathbf{w}} &\leftarrow \gamma\lambda^{\mathbf{w}}\mathbf{z}^{\mathbf{w}} + \nabla \hat{v}(S;\mathbf{w}) \\
        \mathbf{z}^{\boldsymbol{\theta}} &\leftarrow \gamma\lambda^{\boldsymbol{\theta}}\mathbf{z}^{\boldsymbol{\theta}} + \nabla \ln(A|S;\boldsymbol{\theta}) 
\end{align}
Where $\lambda^{\mathbf{w}}$ and $\lambda^{\boldsymbol{\theta}}$ are trace decay parameters, and $\gamma$ is discounting parameter. For episodic actor-critic methods, eligibility trace vectors should be initialized to a zero-vector at the start of each episode. Accordingly, for each time step in the episode an action $A_t$ is sampled from the policy approximator and taken in state $S_t$, the agent moves to a new state, $S'_t$, and is given reward $R_t$. Thus, the parameter updates for the episodic actor-critic algorithm are given for each time step as follows,

\begin{align}
        \delta &\leftarrow R_t + \gamma\hat{v}(S'_t;\mathbf{w}) - \hat{v}(S_t;\mathbf{w}) \\
        \mathbf{w} &\leftarrow \mathbf{w} + \alpha^{\mathbf{w}}\delta\mathbf{z}^{\mathbf{w}} \\
        \boldsymbol{\theta} &\leftarrow \boldsymbol{\theta} + \alpha^{\boldsymbol{\theta}}\delta\mathbf{z}^{\boldsymbol{\theta}}
\end{align}
Where $\alpha^{\mathbf{w}}$ and $\alpha^{\boldsymbol{\theta}}$ are parameter space step sizes for each function approximator. Again, both function approximators for actor-critic methods can be implemented with any differentiable model described within this section and is often some neural network structure. In \cite{SuttonBarto98} pseudocode for actor-critic methods can be found along with their extensions to continuous \ac{RL} problems and problems with continuous action spaces.
\section{Machine Learning For Physical Layer}\label{sec:Phy}
\subsection{State-of-the-art of IoT Communication Technologies}

\begin{figure}[h!]
    \centering
    \includegraphics[width=6 in]{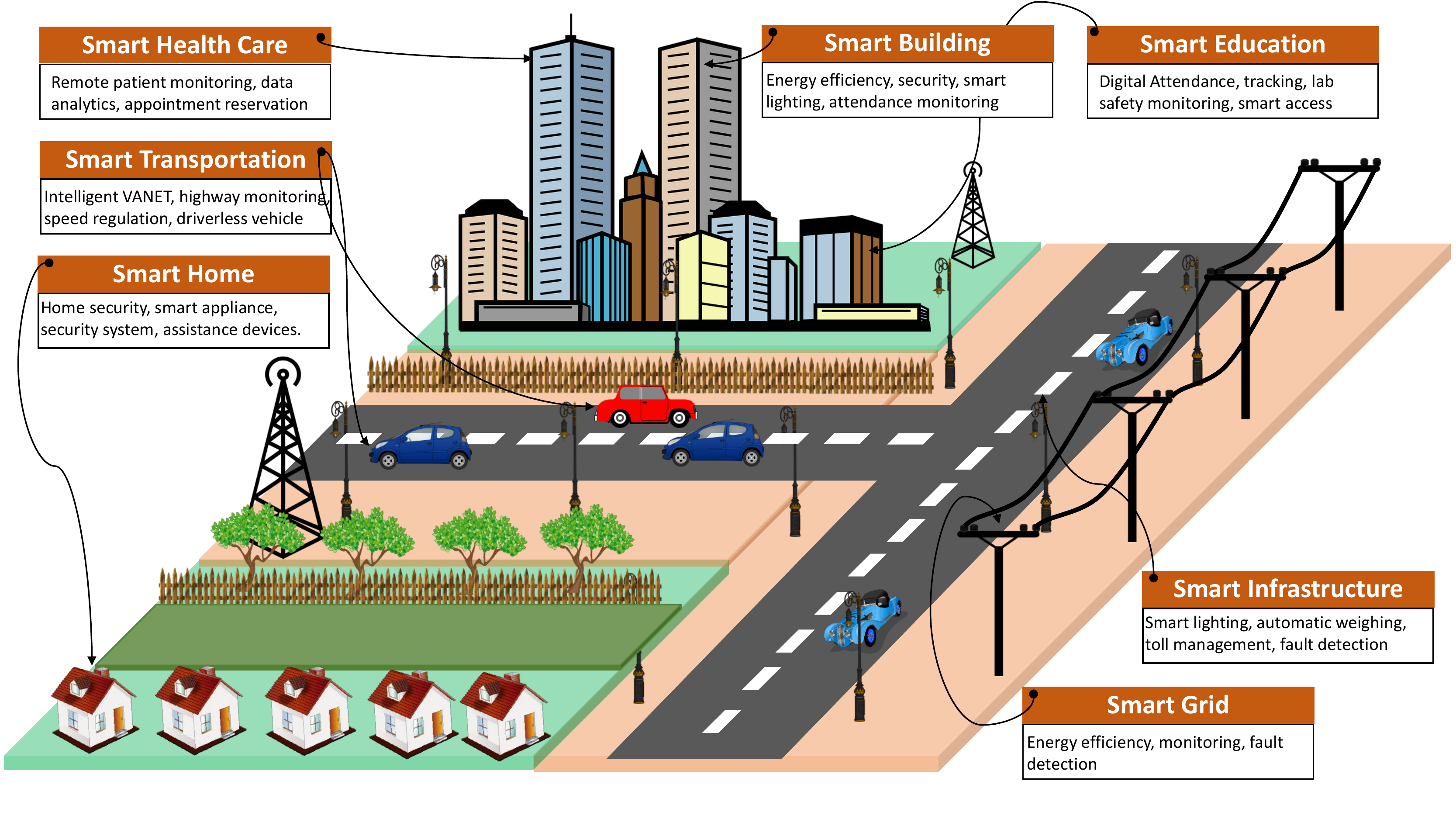}
    \caption{\ac{IoT} network enabling smart city}
    \label{fig:IoTArch}
\end{figure}

\begin{figure}[h!]
    \centering
    \includegraphics[width=5 in]{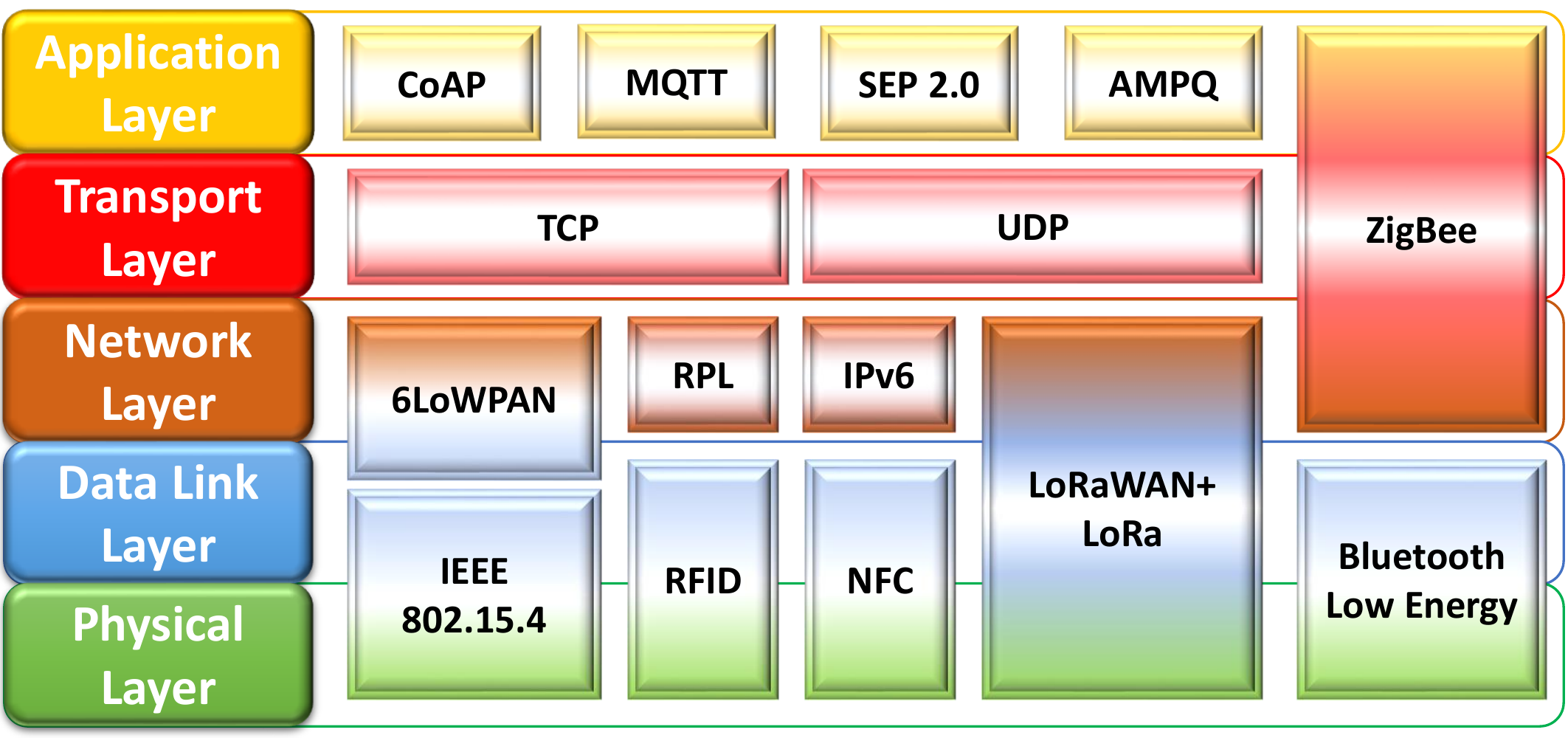}
    \caption{IoT Protocol Stack}
    \label{fig:protocol}
\end{figure}

\ac{IoT} is a broad, emerging trend and hence applies to several key modern concepts that employ several technologies.  In fact, \ac{5G} and \ac{IoT} complement each other in that \ac{5G} wireless networks will catalyze the growth of future \ac{IoT} systems. Achieving the \ac{IoT} vision has been a subject of extensive research to identify and standardize the communication protocols,  ubiquitous connectivity, data storage, computation and analytics, \ac{IoT} gateway and cloud management, dynamic service delivery, among others \cite{crIoT_2,cIoT}. The capabilities offered by \ac{IoT} are countless and find vast applications to improve the economic and social well-being of humans such as smart home, smart lighting systems, smart healthcare, assisted driving, environmental monitoring, mobile ticketing, etc. \ac{IoT} enables interconnection of various heterogeneous devices which communicate with each other without human intervention in what is known as \ac{M2M} communication \cite{crIoTsurvey}. The limitless possibilities of \ac{IoT} through Massive and Critical \ac{IoT} will influence several aspects of everyday life. Massive \ac{IoT} involves the large deployment of smart devices in smart agricultural monitoring, smart grid, smart surveillance systems, smart home, etc. which require \textit{low-cost user equipment, low energy consumption, and scalability for massive deployment}. Critical \ac{IoT}, on the other hand, applies to critical operations such as remote healthcare monitoring, smart traffic surveillance, smart industrial operations which requires \textit{low latency, highly reliable and safe end-user experience}. Such large deployments as in Figure \ref{fig:IoTArch} generate an enormous amount of sensed data and requires seamless communication with each other and to the cloud. A critical consequence of such large deployments is \textit{spectrum congestion} which can hinder and prevent the seamless interconnected operation as intended for the \ac{IoT} applications. The large data generated from these devices require \textit{high-speed connection} to the cloud while interaction among the devices involving control signaling can be satisfied by \textit{low-speed wireless links}. Further, \ac{IoT} devices are resource-constrained in terms of available energy and computational resources. Consequently, a fundamental requirement for \ac{IoT} applications is the low power operation such that the deployed devices need not be replaced frequently. There have been several standardization efforts to support emerging \ac{IoT} communication. Few of these are Zigbee \cite{zigbee}, \ac{6LOWPAN} \cite{6lowpan}, RPL routing protocol, \ac{BLE} \cite{ble}, EPCGlobal \cite{epc}, WirelessHart \cite{whart}, ISA100.11a \cite{isa}, MiWi \cite{miwi}, \ac{LoRaWAN}, \ac{NB-IoT}, enhanced-Machine Type Communications, and Extended Coverage-Global System for Mobile Communications for \ac{IoT}. Among these, Zigbee, \ac{6LOWPAN}, WirelessHart, ISA100.11a and  MiWi employ IEEE 802.15.4 Physical and \ac{MAC} layers while  LoRaWAN adopts the \ac{LoRa} physical layer. The communication protocols at various layers of the \ac{IoT} protocol stack are shown in Figure \ref{fig:protocol}. These standards portray the shared interest and vision shared by standardization institutions and interest groups around the world in realizing the \ac{IoT} vision.

The Ericsson report for massive \ac{IoT} \cite{ericMIoT} project the number of connected smart devices around the world will reach 28 billion by 2021.  Such a surging number of devices pose a significant constraint on the wireless communication capacity of current and future deployments. The current static spectrum utilization policies lead to inefficient use of spectrum \cite{crIoT_1,crIoT_2}. Several research works have been conducted in this regard to demonstrate the benefits of dynamic spectrum sensing, opportunistic spectrum access, and cooperative communications \cite{oppSA_1,oppSA_2,dsa_1,dsa_2,Jagannath16GLOBECOM,Jagannath_TMC_2018}. Such studied interactions between devices with strategic spectrum access methodologies introduce \ac{CR} networks. Realizing the extent of capabilities that can be achieved with cognition, a new paradigm termed cognitive \ac{IoT} has been introduced. Such \ac{CR-IoT} systems has been studied by \cite{nbcriot,cogcom,learnCIoT,crIoT_1, crIoT_2,crnIoT,crIoTsurvey}.  There are several ongoing standardization efforts to incorporate \ac{CR} techniques for \ac{IoT} communication such as ETSI Reconfigurable Radio systems \cite{etsi}, ECMA-392 \cite{ecma}, IEEE 802.22b \cite{802_22b}, IEEE 802.11af \cite{802_11af}. They allow dynamic spectrum sensing, spectrum access, and spectrum management. ECMA-392 is a cross-layer scheme that interfaces the \ac{MAC} and physical layers and enables wireless home and business networks to dynamically use TV white spaces. The \ac{CR} aspect will have wide applications in disaster response and management, \ac{WBAN}, smart-healthcare facilities, vehicular networks, smart grid, among others. In this regard, \cite{crvanet} discusses the challenges and requirements in realizing \ac{CR-VANET}. The authors of \cite{crvanetML} explored the applicability of ML techniques and proposed a learning architecture for CR-VANET.  The \ac{CR} based smart grid architectures has been studied in \cite{crgrid1,crgrid2,crgrid3}. The works in \cite{crban1,crban2,crban3} integrates \ac{CR} to \ac{WBAN} architectures. The work in \cite{crps} explores the potential benefits of incorporating \ac{CR} in public safety and emergency response communications. The cognitive-\ac{IoT} aspect must address several key issues to allow efficient communication between the devices, viz., 1. Resource-constrained \ac{IoT} devices, 2. Communication between heterogeneous hardware, 3. Dense deployments in confined space, 4. Interference between the devices, 5. Heterogeneous connectivity requirements, 6. Communication privacy and security, and 7. Large data management. 

The authors of \cite{cogcom} presented the COGNICOM+ concept, a hybrid architecture that jointly use \ac{CE} and \ac{SC} to allow optimal use of local gateways and cloud computing. The authors present the software and hardware architecture required to support the COGNICOM+ concept. The envisioned \ac{IoT} hybrid architecture houses the \ac{CE} and \ac{SC} in a \ac{SAG} that is local to the connected devices. The \ac{CE} is envisioned to employ compressed \ac{DL} and game theory built on a \ac{CNN} \ac{ASIC} accelerators. The authors introduce \ac{SAG} to perform local computing unlike cloud and fog computing aiming to reduce latency and costs while improving capacity, scalability, privacy, and security. The SC module collects spectrum sensing data from the deployed devices which are relayed to the CE. The CE gathers the collaborative spectrum sensing data to detect unoccupied spectrum bands and dynamically access them. Such collaborative spectrum sensing and accessing maximize spectrum utility. The CE applies reasoning to make a strategic decision to maximize certain user-defined objective. 

In game-theoretic sense, the authors consider each SAG as an agent in a multi-agent \ac{NSG}. Let the set of players (\acp{SAG}) be denoted as $\mathbb{P}$ and the strategy of player ($i$) be $\mathfrak{s}_i \in \mathfrak{S}_i$. Let $\mathfrak{s}_{-i}$ denote the strategy of all players except $i$. The strategy in the COGNICOM+ aspect refers to decisions on transmit power, data rate, accessible frequency bands, and interference to primary users. Each player has an associated utility $ U_i\left(\mathfrak{s}_i,\mathfrak{s}_{-i}\right)$ which is resultant of their own strategy and strategies of other players. The \ac{NSG} can be expressed as $\mathcal{G}=\left\{ \mathbb{P}, \left\{ \mathfrak{S}_i\right\}_{i\in\mathbb{P}}, \left\{  U_i\left(\mathfrak{s}_i,\mathfrak{s}_{-i}\right) \right\}_{i \in \mathbb{P}} \right\}$. Now, if the players are operating in a greedy fashion, each player searches for the optimum strategy $\mathfrak{s}_i^*$ that maximizes their utility such that

\begin{equation}
  \underset{\mathfrak{s}_i \in \mathfrak{S}_i}{\max\;}U_i\left(\mathfrak{s}_i,\mathfrak{s}_{-i}\right).  
\end{equation}

A fundamental concept in \ac{NSG} is \ac{NE} \cite{gametheory} where each player adopts their best possible strategy while being fully aware of the strategies of other players. In NE, neither player gains a unilateral incentive by deviating from the strategy. The authors propose to adopt a distributed optimization strategy in a more cooperative manner where each player optimizes its strategy to maximize their modified utility function,

\begin{equation}
\widetilde{U}_i\left(\mathfrak{s}_i,\mathfrak{s}_{-i} \right) \overset{\Delta}{=} w_i U_i\left(\mathfrak{s}_i,\mathfrak{s}_{-i}\right) – p_i \mathcal{I}_i\left(\mathfrak{s}_i,\mathfrak{s}_{-i}\right),
\end{equation}
where $w_i$ represent the weights of player $i$ and $p_i$ is the penalty for inducing interference $\mathcal{I}_i\left(\mathfrak{s}_i,\mathfrak{s}_{-i}\right)$ to other players. In this way, the collaborative operation of SAGs imparts a balance between the greedy maximization of self utility and the interference caused to other players.

The authors propose to use compressed \ac{DCNN} in the \ac{CE}. The \ac{CNN} compression is achieved by weight/activation compression, model compression, and \ac{CNN} computation acceleration in convolutional layers. Weight compression can be achieved by quantizing the pre-trained weights or quantizing during training process which significantly reduces the memory requirements. Additionally, input feature maps can be compressed by converting floating point to fixed point resulting in significant power and computational gains. Model compression is achieved by pruning less significant connection from the \ac{CNN}. A similar strategy is employed in SqueezeNet \cite{squeezenet} whereby smaller convolution filters ($1 \times 1, 3\times 3$) are employed resulting in microarchitectures called Fire modules. The Fire modules are reconfigured by choosing between $1\times 1$ and $3\times 3$ filters forming larger \ac{CNN}  macroarchitectures. SqueezeNet has been shown to achieve accuracy comparable to AlexNet \cite{alexnet} with $50\times$ fewer samples and less than $0.5$ MB model size ($\equiv 510\times$ smaller than AlexNet). Finally, \ac{CNN} computation acceleration can be attained by compressing each convolutional layer by an equivalent low-rank approximations and adapting the upper layers until desired prediction performance is met.

The authors of \cite{IoTDSA} presented end-to-end dynamic spectrum management facilitated by \ac{IoT} big data and ML algorithms. The authors propose the ML enabled \ac{IoT} spectrum management system comprised of spectrum sensing and measurement collector, deep analytics for spectrum activity learning, and spectral reasoning and decision-making. The authors implemented the proposed spectrum management framework on the testbed \cite{IoTTB} which connects to distributed sensors via \ac{IoT} service covering frequencies $70$ MHz to $6$ GHz. The sensor management, data storage, ML decision-making are performed in the cloud. The \ac{LMR} band which ranges from $70$ MHz to $1$ GHz is considered in their experiment. The \ac{LMR} band spans the very high frequency, ultra high frequency, and public safety channels. The incoming spectrum access requests could either be \ac{LMR} service type or M2M applications. The spectrum sensing data from the deployed sensors contain measured energy levels in the \ac{LMR} bands. The energy level above a preset level identifies as an occupied channel. The spectrum sensing data is processed in conjunction with the license database information to generate a channel occupancy time series for each channel every hour. The incoming spectrum sensing data is passed through usage characterization module which along with the candidate channel feature forecast module \cite{forecast} generates candidate and training channels. The candidate channels have unused spectrum bands that can be shared with other users. The training channels represent all spectrum occupancy patterns of the incoming request. The candidate and training channels form the spectrum-sharing training dataset comprised of features and sharing labels. The sharing labels represent the sharing performance of the users such as the delay incurred by users and the channel loading with respect to the channel capacity. The channel is deemed to be overloaded if the loading label exceeds 1 and available to share otherwise. A sharing predictor is trained using the training dataset which assigns candidate channel and label (sharing performance). The authors used K-means clustering prior to training the predictor with the gradient boosting tree (XGBoost) \cite{xgboost} algorithm. The candidate channels are ranked for spectrum sharing for each incoming request as per the predicted sharing labels. Subsequently, the refining process improves the accuracy and robustness of the sharing label prediction of the ranked candidate channels before predicting the final match. The authors compared the predictor performance trained with XGBoost, random forest and \ac{SVM} algorithms and demonstrated faster training speed and improved accuracy with XGBoost.

The work in \cite{learnCIoT} proposed a \ac{CR} network architecture that employs multi-stage online learning techniques to perform spectrum assignment to \ac{IoT} devices with an aim to improve their throughput and energy efficiency. The authors considered a \ac{IoT} network with \acp{PU} and \acp{SU} . The \acp{PU} are the licensed users who have the primary right of accessing the channels while \acp{SU} can opportunistically access the channels as it becomes available. The \acp{PU} are categorized into idle and active states depending on whether they are actively transmitting. A collision can occur if a \ac{SU} sensed the channel idle and starts transmission while the \ac{PU} moves into an active state and starts transmission simultaneously. This happens as the \ac{PU} has the exclusive right to the channel and will use it without sensing its availability. If a collision occurs the \ac{PU} retransmits the data until successful transmission prior to switching back to an idle state. The authors considered a central node that has access to all the \ac{IoT} devices in the network which will perform the channel assignment based on the channel sensing data from them. The \ac{PU} traffic model is considered to be either generalized Pareto or hyper-exponential while the \ac{SU} traffic could be either event-driven, periodic, or high volume payload exchange. The proposed approach comprises a channel order selection for sensing, and OFF time prediction for each channel. The OFF time prediction allows the \acp{SU} to access the channels without sensing. The authors exploit the fact that the central node has access to all the \ac{IoT}s in the network and can gather the channel sensing data from them to model the traffic characteristics. The central node assigns one channel at a time to save energy consumed in sensing all channels. If the sensed channel is deemed available by the central node, the \ac{SU} will access and proceed to transmission. If the transmission was a success, the corresponding throughput is returned to the central node else if a collision occurs it will inform the central node and switch to wait state. A value table ($V_{c,d}$) for each channel ($c$) and device ($d$) is maintained at the central node with their corresponding throughput ($\mathfrak{T}$). The value table is updated as, 

\begin{equation}
    V_{c,d} \leftarrow \eta \mathfrak{T} + \left( 1-\eta \right) V_{c,d},
\end{equation}
where $\eta$ is the learning rate that affects the priority given to the latest and past observations. The value table signifies the quality of each channel to each device. The authors adopt a hill climbing strategy to randomly swap some entries in the value table and recalculate the value of the resultant configuration. If the new channel-device configurations offer better quality compared to the previous, the new configuration is saved while discarding the previous. This swapping continues until there are no new configurations available that could improve the quality. The hill climbing will work only if the value table maintained at the central node is correctly estimated. However, this knowledge is unavailable initially and requires an exploration strategy to build the value table. Accordingly, an $\epsilon-$ greedy strategy is adopted to randomly explore different configurations for a fraction of time. Further, to predict the OFF time of the channels, the central node is required to learn the \ac{PU} traffic distribution. Accordingly, a non-parametric Bayesian learning method is employed to perform online learning of the \ac{PU} traffic distribution. Subsequently, a function $\mathcal{C}\left( \epsilon , \omega \right)$ representing the number of observed collisions which is dependent on the exploration factor $\epsilon$ and other factors $\omega$ is used. The objective is to achieve a value close to a predetermined threshold level ($C^{*}$) for $\mathcal{C}\left( \epsilon , \omega \right)$.  In order to achieve this objective, a loss function $L\left(\epsilon\right) = loss(C^*,\mathcal{C}\left( \epsilon , \omega \right))$ is optimized using \ac{SGD}. The gradient loss function with respect to $\epsilon$ is expressed as,

\begin{equation}
    \frac{\partial L\left(\epsilon\right) }{\partial \epsilon} = \frac{\partial \; loss(C^*,\mathcal{C}\left( \epsilon , \omega \right))}{\partial \epsilon} \frac{\partial \mathcal{C}\left( \epsilon , \omega \right)}{\partial \epsilon}.
\end{equation}
However, the functional relationship between $ \mathcal{C}\left( \epsilon , \omega \right)$ and the parameters $\epsilon, \omega$ are unknown. In order to circumvent this, the authors adopted a Simultaneous Perturbation Stochastic Approximation \cite{spsa} that allows performing \ac{SGD} while the functional relationship is unknown. The predicted OFF time allows the central node to assign skip period to the \ac{IoT} devices enabling them to use the channel directly without sensing. The authors demonstrated using simulations that the proposed approach requires less channel sensing and achieve comparable throughput while not exceeding the collision threshold $C^{*}$.

\subsection{Adaptive Physical Layer for Cognitive \ac{IoT} frameworks}
The signal processing techniques that enable the physical layer functionalities have a direct impact on the data rate and sensitivity of the radio. With the increasing number of \ac{IoT} devices that communicate over networks, some of which stream multimedia data, there is a growing need for high speed, low latency, and higher capacity systems. \ac{IoT} devices are often deployed densely with several devices interconnected and communicating in the same spectrum posing severe constraints on bandwidth. To enable communication in such dense \ac{IoT} networks, several challenges such as interference, power and bandwidth constraints come into play. Adaptive signal processing is a well-researched topic aimed around suppressing interference and noise from received attenuated signal samples by estimating the interference plus noise covariance from the received samples and suppressing their effect to improve the spectral efficiency of the system \citep{Marko_InterfSupp,Ajagan,PadosKarystinosAV}. Another well-known approach to increase spectral efficiency is to adjust the modulation and coding scheme \textit{on-the-fly} based on instantaneous channel conditions. The promising capabilities of \ac{MIMO} systems to increase channel capacity has led to their adoption in wireless communication standards. Significant performance gain can be achieved by learning and estimating the varying channel dynamics and nullifying the channel' effect from the received signal samples to estimate the actual transmitted bits, in what is commonly known as adaptive channel equalization.
Research surrounding the physical layer has historically been aimed at pushing the boundaries against the norm to provide increased agility to the radios, subsequently enhancing their performance. Enabling the radios with cognitive skills at the physical layer can revolutionize the wireless communication capability of the \ac{IoT} devices. The \ac{ML} based solutions can transform the \ac{IoT} framework into cognitive \ac{IoT} that can adaptively decide which actions are necessary to achieve a certain objective based on parameters learned by the system. This section will explore the various aspects of signal processing at the physical layer and how \ac{ML} based solutions can offer a better alternative.


 
\subsubsection{Adaptive Rate and Power control}
\ac{RL} based solutions have been extensively used in wireless communications to estimate the dynamic system model on the fly \citep{RL1,RL2,Rl3,RL4,RL5,RL6,RL7,RL8,RL9}. In the context of the physical layer, \ac{RL} based solutions can extensively improve the system data rate, bit error rate, goodput (\textit{i.e.}, the amount of useful information that successfully arrived at the destination over the time-varying channel) and energy efficiency \citep{linkadaptfde,linkadaptSVR,adaptRP,nmastron}. Adaptive rate control can serve as a useful tool to selectively adapt the data rate depending on the instantaneous channel conditions. Such flexibility aids the system in leveraging the channel statistics to its benefit, essentially maximizing the channel utilization. \ac{IoT} devices are often battery powered and hence constrained in power. Each layer of the protocol stack must be designed to reduce the energy consumption and prolong the device' lifetime. Therefore, adaptive power control at the physical layer is imperative to the longevity of the device and consequently the \ac{IoT} network lifetime.

In \cite{Li}, an adaptive rate control strategy based on \ac{RL} is proposed to learn the dynamically varying channel conditions. The time-varying fading channel is modeled as a finite state Markov chain, whose channel state transition probabilities are unknown but the instantaneous channel gains can be estimated. Now the optimization problem forms a \ac{MDP} which can be solved in \ac{DP}. However, the \ac{DP} approach is suited best for static systems and hence would not be suitable for a dynamic system where the channel statistics vary with time. 

In this work, the authors propose to use Q($\lambda$)-learning \citep{SuttonBarto98} to track the varying environmental changes in pursuit of the optimal control policy online. Q($\lambda$)-learning is a popular \ac{RL} based algorithm used to solve \ac{MDP} when the system's state transition probabilities are unknown. 
The Q($\lambda$)-learning algorithm is similar to the standard Q-learning except that it updates the learning rate based on the Q value of the state-action pair. The incremental learning process involves the learning agent transitioning from system state of one block to another at the next block by choosing an action. For each chosen action, the agent observes the reward and modifies its control policy with an aim to maximize the expected reward for future actions. This foresighted iterative learning process will repeat at each block and the agent will eventually converge at the optimal policy.

In this context, the objective is to find a rate-control scheme that maximizes the system throughput subject to a fixed \ac{BER} constraint and long-term average power constraint. The system state is characterized by the instantaneous channel gain and buffer occupancy as $s_n = \{g_n, b_n\}$. The receiver estimates the channel gain and feeds back to the transmitter. In a practical system, this could be accomplished by having the receiver and transmitter exchange estimated statistics via control packets using a separate control channel. Consider the transmission buffer is a \ac{FIFO} queue that can hold a maximum of $N$ packets each of size $B$ bits. The packet arrival process to the buffer follows a Poisson distribution $P_a = \frac{\nu^a e^{-\nu}}{a!}$, where $a$ is the number of packets that arrived at the buffer and $\nu$ is the expected number of packets that will arrive in one block. The number of packets dropped from buffer in the $n^{\text{th}}$ block can be expressed as $d_n = \max\left[b_{n-1} - p_{n-1} + a_n-  N, 0\right]$ , where $p_n$ is the number of packets leaving the buffer in the $n^{\text{th}}$ block.

Consider an \ac{M-QAM} system which, based on the learning agent's rate-control policy, can change the number of bits per symbol ($\log_2(M)$). There are numerous ways to change a system's transmission rate; (i) vary coding rate, (ii) vary modulation scheme, i.e., constellation size, and (iii) careful combination of both. Let us denote the bits per symbol in the $n^{\text{th}}$ block as $m_n =\{1,2,3,...,K\}$ and the number of symbols in a block as $N_{sym}$. Then, the number of packets that can be transmitted in the $n^{\text{th}}$ block is $p_n = \frac{m_nN_{sym}}{B}$, referred to herein as rate. For a $W$ bandlimited system operating in an additive white gaussian noise environment with noise spectral density $N_0$, the minimum transmission power required to maintain an acceptable \ac{BER} ($\epsilon_*$) in the $n^{\text{th}}$ transmission block is, 

\begin{equation}
P_{n} \ge \frac{WN_0}{g_n}\frac{(-\log{5\epsilon_*})(2^{p_nB/N_{sym}}-1)}{1.5}.
\end{equation}
Now, the long-term average power consumption can be expressed as,

\begin{equation}
    \bar{P} = \lim_{n\xrightarrow{}\infty}\frac{1}{n}\sum_{i=0}^{n}P_i.
\end{equation}
The rate control scheme must now aim to maximize the system throughput ($\mathfrak{T} = \nu(1-\mathfrak{P}_d)$) subject to the \ac{BER} and average power constraints. Here, $\mathfrak{P}_d$ is the packet drop probability. This escalates to a dual objective optimization; maximizing system throughput and minimizing average power. This multi-objective optimization will be solved to arrive at a Pareto-optimal solution (rate control policy).  Q($\lambda$)-learning aims to find the optimal control policy by estimating an action-value function for each state-action pair. The action-value function is the long-term discounted reward if the system starts at state $s_n$ taking an action $p_n$. The reward per block for taking an action/transmission rate $p_n$ at a state $s_n$ has a Lagrangian form which essentially implies the system gets a larger reward if the packet drops and transmission power is lower. The negative cost (reward) per block can be expressed as,

\begin{equation}
    r_{n+1} = -\left[ \textit{E}(d_{n+1}) + \lambda P_n \right].
\end{equation}
The Q($\lambda$)-learning can be solved in a way similar to the standard Q-learning except here the learning rate ($\rho$) is updated based on the state-action pair which is kept a constant in the standard Q-learning. 

\begin{equation}
    \rho = r_{n+1} + \gamma Q(s_{n+1},p^*_{s_{n+1}}) - Q(s_n,p_n),
\end{equation}
where $\gamma$ is the discount factor and $p^*_{s_{n+1}}$ is the action which maximizes the action-value function $Q(s_{n+1},p^*_{s_{n+1}})$.
The Q($\lambda$)-learning demonstrates faster convergence compared to the standard Q-learning. The authors demonstrated the ability of learning agent to acclimate to the varying wireless channel to learn and adapt the rate control policy best suited for the channel conditions.

The authors of \cite{linkadaptfde} attempt to solve the link adaptation problem of \ac{SC-FDE} systems. \ac{SC-FDE} systems use cyclically prefixed \ac{M-QAM} to allow frequency domain equalization at the receiver. The authors approach the problem from a classification perspective such that the optimum modulation and coding scheme that would deliver the highest goodput for the current channel conditions would correspond to the best classification of the multidimensional data. The feature vectors considered include estimated post-processing \ac{SNR}, estimated channel coefficients, and noise variance. \ac{PCA} is used for dimensionality reduction such that an orthogonal transformation maps the features from a higher dimensional space to lower dimension. The kNN algorithm is used to classify the reduced dimensional feature vectors. A significant drawback of using kNN algorithm is that it requires storing the previously observed values which is memory intensive and computationally expensive. For a low power wireless device, such an algorithm is a poor choice for real-time operations. \cite{linkadaptSVR} tackle this problem to perform real-time link adaptation of \ac{MIMO}-\ac{OFDM} systems by using online kernelized \ac{SVR}. \ac{SVR} attempts to minimize the generalization error bound to achieve generalized performance rather than minimizing training error like \ac{SVM}. \ac{SVR} requires minimal memory and computational power and was demonstrated to adapt quickly to varying channel conditions in their simulations. For every packet, the receiver observes the packet failure/success, channel measurements and the modulation and coding scheme corresponding to that packet. To prevent memory explosion, the authors use a sparsification algorithm \citep{sparse} such that only linearly independent samples are preserved in the dictionary. The \ac{SVR} algorithm finds the linear regression function that corresponds to the minimum mean squared loss function. The authors compared the performance of online kNN versus online \ac{SVR} to demonstrate the monotonically increasing memory and time consumption with online kNN while it remained constant for online \ac{SVR}.

A \ac{RL} based solution is proposed in \cite{adaptRP} to achieve adaptive rate and power control for point-to-point communication and extend it to a multi-node scenario. The receiver is assumed to feedback channel gain and packet success/fail status (\ac{ACK}/\ac{NACK}) to the transmitter allowing it to choose the modulation and transmitter power based on the obtained information. Accordingly, the authors formulate the objective to maximize the throughput per total consumed energy considering the channel conditions, queue backlog, modulation and transmit power. The authors incorporate buffer processing cost/energy into the total energy consumption cost such that there is a cost incurred for buffer overflows. Imposing buffer processing cost can be viewed as a \ac{QoS} factor. The formulated \ac{MDP} is solved using the \ac{AC} algorithm \citep{SuttonBarto98} which involves two parts: actor and critic. The actor decides the action and the critic estimates the state-value function and the error which criticizes the actor's action. The actor selects the action based on Gibbs \textit{softmax} method \citep{SuttonBarto98} such that the action corresponding to the highest conditional probability of state-action is chosen. The authors demonstrated the throughput achieved with \ac{AC} algorithm is twice that of a simple policy where the highest modulation order that maintains a predefined link \ac{SNR} is chosen.

Another notable application of \ac{ML} in improving real-time video streaming is presented in \cite{qarc}. The authors propose \ac{QARC}, a \ac{DL} based adaptive rate control scheme to achieve high video quality and low latency. The complex challenge posed by the varying video quality and dynamic channel conditions is solved by two \ac{RL} based models; a \ac{VQPN} and a \ac{VQRL}. The \ac{VQPN} predicts future video quality based on previous video frames whereas the \ac{VQRL} algorithm adopts the \ac{A3C} \ac{RL} \citep{a3c} method to train the neural network. \ac{VQRL} accepts historic network status and video quality predictions from \ac{VQPN} as inputs. 

The authors use a neural network in this video streaming application motivated by the effectiveness of the neural network in the prediction of time sequence data. The \ac{VQPN} model adopts \ac{CNN} layers to perform feature extraction of the input video frames to obtain spatial information. The \ac{CNN} layers are followed by a two layered \ac{RNN} which extracts temporal characteristics of the video frames in the past $k$ sequences. The output of the \ac{VQPN} is the prediction of video quality for the next time slot. The weights are updated based on the mean squared error loss function between the actual video quality score and the estimated video quality score. Specifically, the \ac{VQPN} has $5$ layers to perform feature extraction; a convolution layer with $64$ filters each of size $5$ with stride $1$, a $3 \times 3$ average pooling layer, a second convolution layer with 64 filters of size 3 with stride 1, a $2 \times 2$ max-pooling layer and a hidden layer with 64 neurons. The output of the feature maps represents time series data which is fed into the \ac{RNN}. The \ac{RNN} comprises a \ac{GRU} layer with 64 hidden units which then connects to another \ac{GRU} layer of 64 hidden units. The hidden layer connects to the hidden output of the last \ac{GRU} layer resulting in a 5-dimensional vector output corresponding to the video quality scores for the bit rates $\mathrm{\left[300, 500, 800, 1100, 1400\right]~kbps}$. The authors use Adam gradient optimizer to train the \ac{VQPN} with a learning rate of $10^{-4}$. The \ac{VQPN} was realized using the open source \ac{ML} library, TensorFlow \cite{tensorflow2015}.

In modeling the \ac{VQRL}, the neural network must be trained to learn the relationship between the video quality and bit rate. The sender serves as a learning agent who observes the future video quality and previous network status in the state space. The network status in the state space is comprised of sender's video transmission bit rate, received bit rate of past $k$ sequences, delay gradient, and packet loss ratio of previous $k$ sequences. The action taken refers to the video bit rate selected for the next time slot. Since in this case the states will be represented by continuous numbers which leads to a fairly large state space, it is unable to store them in a tabular form. As Q-learning cannot be effective in solving large state space problems, the authors have combined \ac{RL} with a neural network.  The authors solve this \ac{RL} problem using A3C \ac{RL} algorithm \cite{a3c} whereby the policy training is achieved by means of a policy gradient algorithm. The authors further propose a multiple-training agent version to accelerate the training process. The multiple agents comprise of a central agent and several forward propagation agents. The forward propagation agent only decides with policy and critic via state inputs and neural network model received by the central agent for each step. The central agent uses the actor-critic algorithm to compute gradients and then updates its neural network model which is then pushed to the forward propagation agent. This can happen asynchronously among all agents with no thread locking between agents. The \ac{VQPN} is trained and tested on two video datasets; VideoSet (a large scale compressed video quality dataset) and self-collected video sets (live concerts, music videos, short movies). To train \ac{VQRL}, the authors use packet-level, chunk-level, and synthetic network traces. The \ac{QoE} metric is defined as a weighted function of video quality, sender's bit rate and delay gradient measured by the receiver at a time instant $n$. The \ac{QARC} algorithm was tested for its efficacy in real-world operating conditions by video-streaming on three different networks (public WiFi, Verizon cellular network and a wide area network between Shanghai and Boston) at a local coffee shop. The client was running on a MacBook Pro laptop connected to a server running on a Desktop machine located in Boston. The authors demonstrated the \ac{QARC} algorithm outperform Skype and WebRTC in terms of the \ac{QoE} metric.

Future \ac{IoT} systems will involve a variety of traffic types ranging from bursty small packets, emergency low-latency transmissions, and high-rate multimedia traffic. Adaptive rate control strategies which intelligently respond to link quality, as well as traffic type, will be imperative for such systems.

\subsubsection{Adaptive Channel Equalization}

Another key area where \ac{ML} algorithms, and more specifically neural network models, have been successfully employed to enhance the physical layer is adaptive channel equalization \citep{Ceq1,Ceq2,Ceq3,Ceq4,Ceq5,Ceq6,Ceq7,DLchannel,Chaneq2001,deepDec}. \ac{IoT} networks are usually dense comprising of several devices attempting to communicate simultaneously. Such dense deployment with multiple transmissions results in a harsh communication environment. Channel equalization techniques must be employed at the receiver for efficient signal demodulation. \cite{Chaneq2001} employs \acp{MLP} to perform non-linear channel equalization of a 16-QAM system. The use of non-linear power amplifiers result in non-linear amplitude and phase distortion resulting in a non-linear channel model as expressed by the following relation,

\begin{equation}
    \mathfrak{r}(t) = \mathcal{A}(\mathfrak{a}(t))e^{j[\;\phi(t) + \mathcal{P}(\mathfrak{a}(t)) \;]} + g(t),
\end{equation}
such that $\mathcal{A}(x) = \frac{\alpha_a x}{1 + \beta_a x^2}$ and $\mathcal{P}(x) = \frac{\alpha_\phi x}{1 + \beta_\phi x^2}$ are the non-linear amplitude and phase distortions and $g(t)$ is the \ac{AWGN}.

\begin{figure}
    \centering
    \includegraphics[width=4 in]{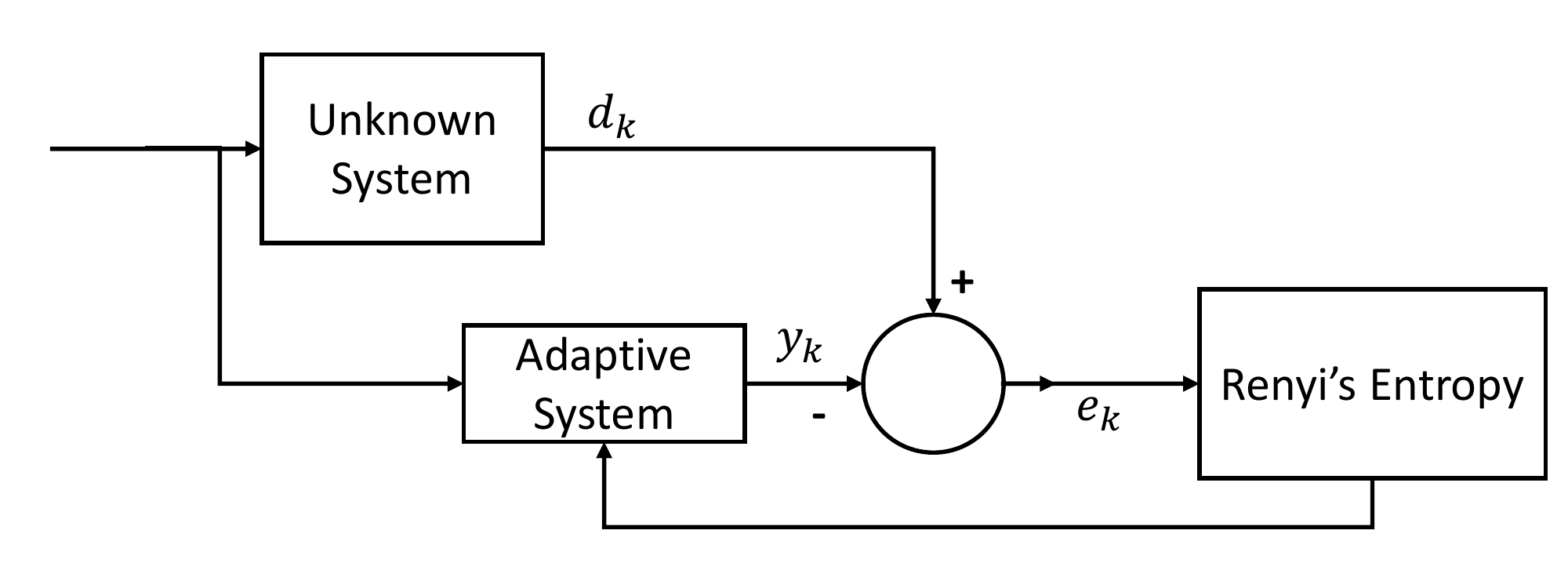}
    \caption{Adaptive System}
    \label{fig:adapt}
\end{figure}

The goal of non-linear channel equalization is to estimate the transmitted symbol from the received distorted symbols. The \ac{MLP} is trained following a minimum error entropy criterion \citep{entropy}. The adaptive system training aims to minimize/maximize the information potential based on the Renyi's entropy order. Figure \ref{fig:adapt} shows an adaptive system learning to update its weights such that the difference ($e_i$) between the output ($y_i$) and desired response ($d_i$) is minimized. 
The weights ($w$) are trained based on the gradient of the information potential ($I_\rho$) as
 
 \begin{equation}
     \frac{\partial I_\rho}{\partial w} = \frac{\rho-1}{N^\rho}\sum_j \left( \left[ \sum_i \mathfrak{G}_\sigma(e_j - e_i) \right]^{\rho-2}\sum_i \mathfrak{G'}_\sigma(e_j - e_i) \frac{\partial y_i}{\partial w}\frac{\partial y_j}{\partial w} \right),
 \end{equation}
where $\mathfrak{G}_\sigma(.)$ denotes the Gaussian kernel with standard deviation $\sigma$. The gradients of the outputs with respect to the weights can be computed using the standard \ac{BP} algorithm. The proposed adaptive equalizer is composed of two \acp{MLP} operating in parallel, say MLP1 and MLP2. MLP1 is trained to learn the mapping of the transmitted signal amplitude $\mathfrak{a}_i, i= 1, 2, ..., N$ to the received signal amplitude $|\mathfrak{r}_i|$. Since for a 16-QAM, the transmitted signal amplitude can only have three different amplitude levels, the output of the MLP1 corresponding to the transmitted signal amplitude is compared to the measured $|\mathfrak{r}_i|$. The output that gives the closest estimate to the possible values is chosen as the estimate for transmitted signal amplitude. The MLP2 is trained to learn the mapping from received signal amplitude to the non-linear phase distortion. From the estimated amplitude and phase from MLP1 and MLP2, the in-phase and quadrature components of the transmitted symbol are determined. In this work, authors train the system for an entropy order ($\rho = 3$), steepest ascent for information potential, a Gaussian kernel with $\sigma = 1$ and a dynamic step size. The training initially starts with unitary step size which then updates depending on the weight update such that the value increases when the update yields better performance and vice-versa. The authors demonstrated in simulations that the information potential maximization approach converges in fewer iterations than the mean squared error technique.
 
The authors of \cite{deepDec} explore the capabilities of \ac{DL} for joint channel equalization and decoding. The \ac{DL} model comprises of an increased number of hidden layers to improve the representation capability of the neural network. Similar to the \citep{Chaneq2001}, the channel is assumed to introduce non-linear distortion to the transmitted symbols. The authors train the network to minimize the mean squared error loss ($L = \frac{1}{N}\sum_i\left(\mathfrak{r}_i - m_i \right)^2$) between the transmitted symbol ($m_i$) and received symbol ($\mathfrak{r}_i$) as presented to the network in the training phase. The neuron weights in each layer can be updated using any gradient descent algorithms such as to minimize the loss function. The activation functions could be \textit{ReLU} or \textit{sigmoid} functions. The authors demonstrated the performance of the proposed \ac{DL} for joint channel equalization and decoding with a six layer neural network comprising of 16, 256, 128, 64, 32 and 8 neurons in each layer. The modulation used is \ac{BPSK} for a $(16,8)$ polar code and \textit{ReLU} activation function.

In \cite{CNNdec}, three \ac{DL} models for channel decoding are proposed. In this book, we will cite the \ac{CNN} model for the channel decoder. The \ac{CNN} employs a convolution operation which significantly reduces the number of parameters allowing the network to be deeper with fewer parameters. The hidden layers use either convolution or pooling. The input to the \ac{CNN} is batch-normalized \cite{batchnormCNN} such that any layer that previously received input $x$ will receive $BatchNorm(x)$ which is a normalized, scaled and shifted version of original input with respect to a mini-batch.
 
\begin{equation}
     BatchNorm(x) = \theta_1 \hat{x} + \theta_2,
\end{equation}
where $\hat{x} = \frac{x - \mu_\chi}{\sqrt{var_\chi + \epsilon}}$ is the normalized $x$ over the mini-batch $\chi$ with mean ($\mu_\chi$) and variance ($var_\chi$), $\theta_1$ and $\theta_2$ are the parameters to be learned. 

The batch-normalized \ac{CNN} is trained with mini-batch \ac{SGD} to minimize the mean-squared error loss function. The authors observed the \ac{CNN} decoder offered better performance compared to a \ac{MLP} but at the cost of increased computational time.
 
The applicability of \ac{DL} in channel estimation and signal detection in \ac{OFDM} systems is demonstrated in \cite{DLchannel}. The DL model is trained offline with simulated data to learn the channel distortions and reconstruct the transmitted symbols. Let $m(n)$ be the baseband \ac{OFDM} modulated symbols transmitted over an $N$-path multipath channel $\{h(n)\}_{n=0}^{N-1}$ with \ac{AWGN} $g(n)$ as shown by,

\begin{equation}
    \mathfrak{r}\left(n\right) = m\left(n\right)h\left(n\right) + g\left(n\right).
\end{equation}
After removing cyclic prefix and converting back to frequency domain the signal representation translates to,
\begin{equation}
    R\left(k\right) = M\left(k\right)H\left(k\right) + G\left(k\right).
\end{equation}
The pilot symbols are transmitted in the first \ac{OFDM} block followed by user data in the subsequent blocks. The received pilot block and one data block are fed as input to the \ac{DL} model. During the offline training stage, the model is trained with various received \ac{OFDM} symbols generated with varying channel conditions under certain statistical profiles. The trained model when deployed for online signal detection, would estimate the signals without explicit channel estimation. The received signal and original transmitted symbols are supplied to the model to train it such that the difference between the model output and the original transmitted data are minimized. The model consists of five layers, three of which are hidden. Each layer comprises 256, 500, 250, 120 and 16 neurons respectively. The \textit{ReLU} function is used as the activation function in all layers to map the input to the outputs of each layer except the last layer where \textit{sigmoid} function is used to map to the interval $[0,1]$.

A \ac{DL} based method to improve the \ac{BEP} algorithm for decoding linear codes is proposed in \cite{DLdecoder}. \ac{BEP} also known as Sum-Product algorithm is a message passing algorithm to derive statistical inferences from graphical models such as Bayesian networks. \ac{BEP} is a form of Maximum A Posteriori (MAP) decoding of linear codes. \ac{BEP} was first used in information theory by Gallager's iterative decoder for LDPC \citep{Gallager} which was a generalized case of belief propagation. Tanner graph forms a Bayesian network on which \ac{BEP} operates. The \ac{DL} model is trained with a single codeword. Conventional \ac{BEP} decoder is constructed from the Tanner graph which is a graphical representation of the parity check matrix that describes the code \citep{TannerGraph}. The messages are transmitted over edges such that each edge calculates the outgoing message based on messages received over all its edges except for the transmitting edge.

To enable \ac{DL} for a \ac{BEP} decoder, the authors propose an alternative trellis representation where nodes in the hidden layer represent edges in the Tanner graph. If $N$ denote the code block length, the number of neurons in the input layer is a vector of size $N$. The subsequent layers except for the final output layer i.e., the hidden layers have size $E$ implying the number of edges in the Tanner graph. Each neuron in the hidden element corresponds to the message transmitted over some edge in the Tanner graph. The output layer has a size $N$ that outputs the final decoded codeword. Let $e=(v,c)$ denote the neuron in the hidden layer $i, i\in {1,2,...,2L}$, $l_v$ is the log-likelihood ratio of the variable node $v$  and $y_{i,e}$ represent the output message from the neuron after $\lfloor \frac{i-1}{2} \rfloor$ iterations. To allow the \ac{DL} model, to learn based on the inputs, they are assigned weights which will be updated using the \ac{SGD} method. The output of a neuron in the hidden layer, for an odd $i$ is expressed as,

\begin{equation}
    y_{i,e=(v,c)} = \tanh(\frac{1}{2}(w_{i,v}l_v \;+\sum_{e'=(v,c'),c'\neq c} w_{i,e,e'}y_{i-1,e'} ))
\end{equation}
and for an even $i$,

\begin{equation}
    y_{i,e=(v,c)} = 2\tanh^{-1}(\prod_{e'=(v',c),v'\neq v}y_{i-1,e'})
\end{equation}
and the final $v$th output of the network is expressed as

\begin{equation}
    z_v = \sigma\left[ w_{2L+1,v}l_v +\sum_{e'=(v,c')} w_{2L+1,v,e'}y_{2L,e'},  \right]
\end{equation}
where $\sigma(x) = (1+e^{-x})^{-1}$ is the \textit{sigmoid} function to map the final output codeword in the range $[0,1]$. The goal is to train the weights $\lbrace w_{i,v},w_{i,e,e'},w_{i,v,e'} \rbrace$ to achieve an $N-$dimensional output codeword.

The computationally constrained \ac{IoT} systems desire low complexity channel equalization approaches. The trained neural networks will be able to perform channel equalization without requiring channel estimation, hence rendering them suitable for the \ac{IoT} systems.

\subsubsection{Adaptive Array Processing}
\ac{MIMO} systems are a trending physical layer solution to meet the increasing demand for high-speed, high-multiuser capacity communication systems. MIMO systems, due to their antenna arrays, can exploit spatial and temporal diversity to increase the communication data rate and spectral efficiency. Systems with adaptive antenna arrays can perform smart signal processing to combine the signals received at each array and nullify the interference and/or transmit the signals to steer the beam in an intended direction.  Multi-user MIMO \citep{LTEA} is already adopted in the developed and evolving communication standards like \ac{3GPP} \ac{LTE}, and \ac{LTE-A}. Another emerging \ac{MIMO} technology, Massive \ac{MIMO}, is the physical layer technology of choice for the latest \ac{5G} technology \citep{5GAkyi}. Massive-MIMO can revolutionize the \ac{5G} communication by providing reliable faster communication to more number of users simultaneously. Emerging \ac{5G} wireless networks promise ubiquitous connectivity, high data rates, energy efficiency, and spectrum availability. The dense, diverse and heterogenous nature of \ac{IoT} networks can be fulfilled by the disruptive \ac{5G} technologies such as massive MIMO, \ac{NOMA}, \ac{M2M}, etc. Beamforming is a prominent MIMO solution to enable communication to the desired device allowing to coexist with the other devices in the dense network. An essential step that enables beamforming is \ac{DoA} estimation that allows the transmitter/receiver to learn the direction to/from which the signal should be directed/arrived. In this section, we will discuss a few prominent adaptive array techniques and how \ac{ML} solutions can improvise them. 

Several works \citep{doa1,doa2,doa3,doa4,doa5,doa6} address the problem of \ac{DoA} estimation in array signal processing using \acp{ANN}. Let us look at each of these solutions. In \cite{doa2}, authors propose the use of a three-layer \ac{RBFNN} that can learn multiple source-direction findings of a six-element linear antenna array. The \ac{RBFNN} does not require training with all possible combinations of training sets. The network will generalize when trained with an expected range of input data. In this case, authors trained the network with input data whose \ac{DoA} is uniformly distributed in the range $-90^{\circ}$ to $90^{\circ}$. The performance is compared to the conventional \ac{MUSIC} algorithm for \ac{DoA} estimation of correlated and uncorrelated signals. The linear antenna array performs the mapping from the angle space to the sensor output space such that

\begin{equation}
    o_i = \sum_{k=1}^{K}a_ke^{j2\pi f_0d\sin{\theta_k} + \alpha_k},
\end{equation}
where $i = 1, 2, ..., M$ and $k$ denote the respective antenna element and incident signal respectively, $f_0$ is the frequency of incident signal, $d$ is the inter-element spacing, $\theta_k$ is the angle of arrival of $k$-th signal and $\phi_k$ is the initial phase of $k-$th incident signal. The \ac{RBFNN} is trained with $N$ patterns to perform reverse mapping of received array data ($o_i$) to the angle space ($\theta_k$). The incident array vectors are preprocessed prior to  feeding them to the \ac{RBFNN}. To train the neural network, the antenna array output vectors are generated ($\mathbf{o}(n), n= 1,2,...,N$). Each of the array output vector is further transformed to the spatial correlation matrix $\mathbf{R}(n)$. Since the diagonal elements of the correlation matrix does not carry any angle information, i.e. $R_{mm'} = \sum_{k=1}^{K}a_k$, only the cross-correlation terms are considered. These cross-correlated terms are arranged into an input vector $\mathbf{v}(n)$. 
The output node subsequently computes the weighted sum of the hidden layer outputs.

\begin{equation}\label{eqn:theta}
    z_k(j) = \sum_{i=1}^{N}w_i(k)\mathfrak{G}\left(\Vert \mathbf{o}(j) - \mathbf{o}(i) \Vert^2\right), k = 1, 2, ..., K, \;j= 1, 2, ...,N,
\end{equation}
where $w_i(k)$ represents the $i-$th weight of the network for the $k-$th incident signal and $\mathfrak{G}(.)$ is the Gaussian function performed by the hidden layer. Now, the equation (\ref{eqn:theta}) changes to 

\begin{equation}
    z_k(j) = \sum_{i=1}^{N}w_i(k)e^{-\Vert \mathbf{o}(j) - \mathbf{o}(i) \Vert^2/\sigma_g^2},
\end{equation}
where $\sigma_g$ controls the influence of each basis function. The above equation can be rewritten in matrix form as, 

\begin{equation}\label{eqn:doa}
    \mathbf{\Theta} = \mathbf{WF}.
\end{equation}
Here, $\mathbf{\Theta}$ and $\mathbf{W}$ are the $K\times L$ angle and weight matrices and $\mathbf{F}$ is the $L\times N$ hidden layer matrix. $L$ is chosen to be less than $N$ to prevent ill-conditioning arising from large matrix. The input vectors $\mathbf{v}(n)$ are normalized according to equation \ref{eqn:doa}. Using \ac{LS} approach, the weights can be obtained as

\begin{equation}
    \hat{\mathbf{W}} = \mathbf{\Theta F^\dagger},
\end{equation}
where $F^\dagger$ is the pseudo-inverse given by

\begin{equation}
    \mathbf{F^\dagger} = \mathbf{F}^T(\mathbf{F}\mathbf{F}^T)^{-1}.
\end{equation}
Now, the \ac{DoA} estimate can be obtained as

\begin{equation}
    \hat{\mathbf{\Theta}} = \hat{\mathbf{W}}\mathbf{F} = \mathbf{\Theta}^T(\mathbf{F}\mathbf{F}^T)^{-1}\mathbf{F}.
\end{equation}
The \ac{RBFNN} is trained with the Normalized cumulative delta rule \cite{haykin94} such that the weight changes are accumulated over several training presentations as specified by the Epoch. The trained \ac{RBFNN} will give the \ac{DoA} estimates when presented with the normalized input vector. The authors demonstrated the computational advantage gained by adopting the \ac{RBFNN} based \ac{DoA} estimator as opposed to the conventional \ac{MUSIC} algorithm. The estimation accuracy of the proposed \ac{DoA} estimator in addition to the computational efficiency are the key merits of the proposed solution and presents itself as a computationally efficient alternative.

In a recent work \cite{doa1}, the \ac{DoA} estimation is performed using a \ac{ANN} with three layers; input, hidden and output layers. The authors study the estimation accuracy in terms of the number of neurons in the hidden layer. Unlike the \ac{RBF} approach adopted by authors of \cite{doa2}, here the input activation function is the \textit{hyperbolic tangent sigmoid transfer function} and the output activation function is the \textit{logarithmic sigmoid}.

In yet another work \cite{zbeam}, the authors employ \ac{RBFNN} to perform adaptive beamforming. Adaptive beamforming is a method of updating weights of an adaptive antenna array such that the antenna radiation pattern will form beams such that strong beam is sent towards intended user's direction and nulls to sources of interference. The authors adopt a two-step approach to tackle this problem. First, the \ac{DoA} of desired users are determined as in \cite{doa2} and secondly, the beamformer weights are estimated to direct the beams. Similar to the \ac{DoA} estimation problem, the authors approach this using \ac{RBFNN}. For $K$ incident signals, let the signal received at an $M$ element linear antenna array at $\textit{p}^{\text{th}}$ time instant be

\begin{equation}
    \mathbf{o}(p) = \sum_{k=1}^{K}\mathbf{b}_ks_k(p) + \mathbf{g}(p) = \mathbf{Bs}(p) + \mathbf{g}(p)
\end{equation}
such that, $\mathbf{B} = [\mathbf{b}_1, \mathbf{b}_2, ..., \mathbf{b}_M]^T $   
is the array steering matrix that holds the spatial signal of the $k^{\text{th}}$ source, $\mathbf{s} = [s_1, s_2, ..., s_K]$ is the signal vector and $\mathbf{g}(p)$ is the noise vector. Here, 

\begin{equation}
\mathbf{b}_m = [1,e^{-j2\pi d\sin{\theta_k}/\lambda},...,e^{-j2\pi (M-1)d\sin{\theta_k}/\lambda} ].   
\end{equation}
\ac{RBFNN} is trained to compute the \ac{MVDR} beamformer weights such that

\begin{equation}\label{beameq}
    \Hat{\mathbf{w}}_{MVDR} = \frac{\mathbf{R}^{-1}\mathbf{b}_m}{\mathbf{b}_m^H\mathbf{R}^{-1}\mathbf{b}_m},
\end{equation}
where $\mathbf{R} = \frac{1}{P}\sum_p\mathbf{o}(p)\mathbf{o}^H(p)$ is the sample averaged covariance matrix computed from $P$ snapshots of the received signal vector. The beamformer output can be denoted as $\mathbf{y}(p) =  \Hat{\mathbf{w}}_{MVDR}^H\mathbf{o}(p)$. The beamformer vector estimation can be extended to any adaptive antenna array, the model considered in this example is a linear array for notational simplicity. The input and output layer of the \ac{RBFNN} consists of $2M$ nodes to accommodate the in-phase and quadrature components of the input vector $\mathbf{o}(p)$ and the hidden layer' outputs. Much alike the \ac{DoA} estimation problem in \cite{doa2}, the \ac{RBFNN} is trained to perform an input-output mapping from the received vector space to the beamformer weight space. The weights from the input to the hidden layer are identified using unsupervised $k$-means clustering and those from hidden to output layer follows the supervised $delta$ learning rule. During the training phase, the \ac{RBFNN} is trained with $N_t$ training array output vectors $\mathbf{x}_n(p)$ and their corresponding $\Hat{\mathbf{w}}_{MVDR}^n \forall n \in {1, 2, ..., N_t}$. The array output vector is normalized prior to computing the corresponding covariance matrices $\mathbf{R}_n$. The beamformer weights $\Hat{\mathbf{w}}_{MVDR}^n$ are then computed according to equation \ref{beameq}. The trained \ac{RBFNN} can be used to estimate the optimum \ac{MVDR} beamformer weights for a presented normalized array output vector in a computationally inexpensive manner.

In this section, we explored the various ML techniques pertinent to the physical layer that is currently proposed to enhance the \ac{IoT} framework. The integration of such \ac{ML} solutions with the \ac{IoT} devices would be a prominent step in developing cognitive \ac{IoT} architectures that can learn, adapt and behave under the varying system and environmental dynamics. Table \ref{tab:my_phy} enlists the various \ac{ML} algorithms and their corresponding physical layer objective.

\begin{table}[h!]
    \centering
    \caption{Summary of Applications of \ac{ML} in Physical layer\label{tab:my_phy}}
    \begin{tcolorbox}[tab2,tabularx={|p{4.3cm}||p{3.8cm}|X}]
      \textbf{Physical layer solution}   & \textbf{ML Algorithm}  & \textbf{Objective}\\ \hline\hline
      V. T. Nguyen et al. \cite{cogcom}   & \ac{DCNN} &  Cognitive communication architecture   \\ \hline
      Li et al. \cite{IoTDSA}   & XGBoost &  End-to-end dynamic spectrum management   \\ \hline
      T. Tholeti et al. \cite{learnCIoT}   & Non-parametric Bayesian learning &  \ac{CR} architecture for spectrum assignment   \\ \hline
      Li et al. \cite{Li}   & \ac{RL} &  Adaptive rate control   \\ \hline
      Puljiz et al. \cite{linkadaptfde}   & kNN & Adaptive rate control \\ \hline
      Yun and Caramanis \cite{linkadaptSVR}   & SVR & Adaptive rate control  \\\hline
      Li \cite{Li_2010}   & \ac{RL} & Adaptive rate and power control   \\\hline
      Huang et al. \cite{qarc}   & \ac{RL} wt \ac{CNN} and \ac{RNN} & Adaptive rate control  \\ \hline
      Erdogmus et al. \cite{Chaneq2001}   & \ac{MLP} & Non-linear channel equalization  \\\hline
      Ye and Li \cite{deepDec}   & \ac{DNN} & Non-linear channel equalization \\ \hline  
      Lyu et al. \cite{CNNdec}   & \ac{CNN}& Channel decoder  \\ \hline
      Ye et al. \cite{DLchannel}   & \ac{DNN} & Channel equalization in \ac{OFDM} systems  \\ \hline
      Nachmani et al. \cite{DLdecoder}   & \ac{DNN} & Improve \ac{BEP} algorithm for decoding linear codes  \\ \hline
      Zooghby et al. \cite{doa2}   & \ac{RBFNN} & \ac{DoA} estimation  \\ \hline
      Nleren and Yaldiz \cite{doa1}   & \ac{ANN} & \ac{DoA} estimation  \\ \hline
      Zooghby et al. \cite{zbeam}   & \ac{RBFNN} & Adaptive beamforming  \\ \hline
    \end{tcolorbox}{}
\end{table}

\subsection{Open Problems and Challenges}
We have discussed the capabilities introduced by integrating cognition into an \ac{IoT} framework. The \ac{CR-IoT} framework is truly an invaluable component to keep up with the rising \ac{IoT} device density and its tailored comeuppances. While \ac{CR} holds the key in realizing the full potential of future \ac{IoT} architectures, several open challenges exist that readers can derive motivation from for future research. 

\subsubsection{Optimizing Distributed Spectrum Utilization}

Though a few spectrum utilization techniques have been introduced for \ac{CR} \ac{IoT} frameworks, they involve cognition in the cloud/fog or in the gateway. Such centralized spectrum assignment decision-making introduces additional latency to the \ac{IoT} communications. Further, the scalability of such techniques would be speculative and intractable in terms of latency and the computational and data management load on the centralized access point (cloud/fog/gateway). A lightweight distributed spectrum decision making would be desired for the \ac{IoT} frameworks whereby each \ac{IoT} device uses its own cognitive ability to access the spectrum based on its historical and current spectrum sensing data.

Another viable approach would be to perform opportunistic spectrum access decisions based on the geographical location of the spectrum sensing history. Such that even when the \ac{IoT} device (\ac{SU}) senses a \ac{PU}/\ac{SU} traffic it can identify the radio frequency identification tags and map them to their location. Accordingly, a geolocation-based spectrum occupancy history can be built to predict the traffic and perform efficient transmit power control scheme to use the channel without interfering with the ongoing traffic. The success of such transmissions can be recorded over time to learn the collision and transmit power level records. An efficient ML technique can be used to perform online learning of the spectrum records to optimize the transmit power level to carry out interference-free spectrum sharing. 

\subsubsection{Mobility support}

Exploring efficient communication strategies for mobile \ac{IoT} applications such as moving vehicles in the connected vehicular network, drones in an \ac{UAV} network, mobile smartphone users, among others pose unforeseen challenges to the capacity, spectrum handoffs, cloud connectivity, scalability, etc. Adaptive physical layer technique such as adaptive intelligent beamforming in conjunction with opportunistic spectrum access in the mobile scenario for reliable energy-efficient communication is another area to be explored. Specifically, \ac{ML} can be exploited to learn the user mobility pattern and data traffic model to assign radio resources such as transmission rate, power and the frequency band based on link reliability, spectrum congestion, spectrum availability, among others.

\section{Machine Learning For Signal Intelligence}\label{sec:Phy_Int}

As IoT devices become more pervasive throughout society the available operational \ac{RF} environment will contain more non-cooperative signals than ever seen before. Subsequently, the ability to garner information about signals within a spectrum of interest will become ever more important and complex, motivating the use of \ac{ML} for signal intelligence in the IoT. \ac{ML} techniques for signal intelligence typically manifest themselves as solutions to discriminative tasks, and many applications specifically focus on multi-class or binary classification tasks. Problems of these types arise in the context of IoT in many ways including \ac{AMC} tasks, wireless interference classification tasks, and signal detection tasks, each of which, is relevant to signal intelligence for the IoT in their own way. 

\ac{AMC} is the task of determining what scheme was used to modulate the transmitted signal, given the raw signal observed at the receiver. Knowledge of the modulation format used by the transmitter is essential for proper demodulation of the received signal at the receiver, thus solutions to \ac{AMC} tasks are paramount in scenarios where the operational environment may distort the transmitted signal. Such is the case in the IoT, where multipath fading channels are regular in device to device communication. \ac{AMC} may also find application in next-generation intelligent, adaptive transceiver technology in which the radios rapidly switch between modulations based on the channel conditions without requiring a dedicated feedback channel. 

Solutions to wireless interference classification tasks aim primarily to associate a given received signal with an emitter from a known list of emitters. Typical implementations consider emitters that use common communication standards including WiFi, Zigbee, and Bluetooth. Other such solutions consider additional signals that are present in the environment, such as those emanating from a household microwave oven appliance, as they may play an interfering role in some operational environments. Wireless interference classification in this nature is particularly important in the IoT, as IoT devices are often deployed in the homes of users and around other devices that emit RF signals. Classification of a signals emitter can provide insight into the behavior of its use and subsequently its effect on the operability of the local IoT devices.

Problems of signal detection arise in many different areas of communications and the resulting applications of signal detection very widely. In the simplest case, signal detection can be formulated as a binary classification problem with an output corresponding to whether or not a signal is present in the locally sensed RF environment. While interesting solutions exist for the aforementioned problem formulation, within the IoT more complex detection problems often arise in the context of security. Interesting signal detection algorithms can thus be extended to classify the presence of an intruder provided characteristics of their transmission in the environment. Problems of these types are discussed later in section \ref{sec:SigIntOpenProb}.

\subsection{Modulation Classification}\label{sec:AMC}
\ac{DL} solutions to modulation classification tasks have received significant attention in recent years \citep{OShea-ieeejstsp2018,o2017introduction,wang2017deep,West-dyspan2017,Kulin-ieeeaccess2018,Karra-ieeedyspan2017}.
Several \ac{DL} models are presented in \cite{OShea-ieeejstsp2018} to address the modulation recognition problem. Hierarchical \acp{DNN} used to identify data type, modulation class, and modulation order are discussed in detail in \cite{Karra-ieeedyspan2017}. A conceptual framework for end-to-end wireless \ac{DL} is presented in \cite{Kulin-ieeeaccess2018}, followed by a comprehensive overview of the methodology for collecting spectrum data, designing wireless signal representations, forming training data and training deep neural networks for wireless signal classification tasks.

The task of \ac{AMC} is pertinent in signal intelligence applications as the modulation scheme of the received signal can provide insight into what type of communication frameworks and emitters are present in the local RF environment. 
The problem at large can be formulated as estimating the conditional distribution, $p(y|x)$, where $y$ represents the modulation structure of the signal and $x$ is the received signal.

Traditionally, \ac{AMC} techniques are broadly classified as maximum likelihood-based approaches \citep{ML_Onur, TWC, WimalajeewaMILCOM2015, Foulke14MILCOM, Jagannath15CISDA}, feature-based approaches \citep{FB_Azzouz_1, Feature_1, Feature_2} and hybrid techniques \citep{Jagannath17CCWC}. Prior to the introduction of \ac{ML}, \ac{AMC} tasks were often solved using complex hand engineered features computed from the raw signal. While these features alone can provide insight about the modulation structure of the received signal, \ac{ML} algorithms can often provide a better generalization to new unseen data sets, making their employment preferable over solely feature-based approaches. The logical remedy to the use of complex hand engineered feature-based classifiers are models that aim to learn directly from received signal data. Recent work \cite{o'sheaConvMod} show that \acp{DCNN} trained directly on complex time domain signal data outperform traditional models using cyclic moment feature-based classifiers. In \cite{pengCNN}, the authors propose a \ac{DCNN} model trained on the two-dimensional constellation plots generated from the received signal data and show that their approach outperforms other approaches using cumulant-based classifiers and \acp{SVM}.


While strictly feature-based approaches may become antiquated with the advent of the application of \ac{ML} to signal intelligence, expert feature analysis can provide useful input to ML algorithms. In \citep{Jagannath18ICC}, we compute hand engineered features directly from the raw received signal and apply a feedforward neural network classifier to the features to provide a \ac{AMC}. The discrete time complex-valued received signal can be represented as,

\begin{equation}
y(n)=h(n)x(n)+w(n),\;\;\;\;\;\; n=1,...,N 
\end{equation}
where $x(n)$ is the discrete-time transmitted signal, $h(n)$ is the complex valued channel gain that follows a Gaussian distribution and $w(n)$ is the additive complex zero-mean white Gaussian noise process at the receiver with two-sided \ac{PSD} $N_0/2$. The received signal is passed through an Automatic Gain Control prior to the computation of feature values.

The first feature value computed from the received signal is the variance of the amplitude of the signal and is given by,

\begin{equation}
    Var(|y(n)|)= \frac{\sum_{N_s} (|y(n)|-\mathbb{E}(|y(n)|))^2}{N_s}
\end{equation}
where $|y(n)|$ is the absolute value of the over-sampled signal and $\mathbb{E}(|y(n)|)$ represents the mean computed from $N_s$ samples. This feature provides information which helps distinguish \ac{FSK} modulations from the \ac{PSK} and \ac{QAM} modulation structures also considered in the classification task. The second and third features considered are the mean and variance of the maximum value of the power spectral density of the normalized centered-instantaneous amplitude, which is given as,

\begin{align}
& \gamma_{max}=\frac{max\left | FFT(a_{cn}(n)) \right |^2}{N_s},\label{eq:gamma_max}
\end{align}
where $a_{cn}(n)\triangleq\frac{a(n)}{m_a}-1 $, $m_a=\frac{1}{N_s}\sum_{n=1}^{N_s}a(n)$, and $a(n)$ is the absolute value of the complex-valued received signal. This feature provides a measure of the deviation of the \ac{PSD} from its average value. The mean and variance of this feature computed over subsets of a given training example are used as two separate entries in the feature vector input into the classification algorithm, corresponding to the second and third features, respectively.

The fourth feature used in our work was computed using higher order statistics of the received signal, namely, cumulants, which are known to be invariant to the various distortions commonly seen in random signals and are computed as follows,

\begin{equation}
C_{lk}=\sum_p^{\text{No. of partitions in }l} (-1)^{p-1}(p-1)!\prod_{j=1}^p\mathbb{E}\{y^{l_j-k_j}y^{*k_j}\},
\end{equation}
where $l$ denotes the order and $k$ denotes the number of conjugations involved in the computation of the statistic. We use the ratio, $C_{40}/C_{42}$ as the fourth feature which is computed using,

\begin{equation}
    C_{42} = \mathbb{E}(|y|^4) - |\mathbb{E}(y^2)|^2 - 2\mathbb{E}(|y|^2)^2,
\end{equation}

\begin{equation}
    C_{40} = \mathbb{E}(y^4) - 3\mathbb{E}(y^2)^2.
\end{equation}

The fifth feature used in our work is called the in-band spectral variation as it allows discrimination between the FSK modulations considered in the task. We define $Var(f)$ as,

\begin{equation}
    Var(f) = Var\Big(\mathcal{F}\big(y(t)\big)\Big),
\end{equation}
where $\mathcal{F}(y(t))=\big\{Y(f)-Y(f-F_0)\big\}_{f=-f_i}^{+f_i}/F_0$, $F_0$ is the step size, $Y(f)=FFT(y(t))$, and $[-f_i,+f_i]$ is the frequency band of interest.

The final feature used in the classifier is the variance of the deviation of the normalized signal from the unit circle, which is denoted as $Var(\Delta_o)$. It is given as,

\begin{equation}
    \Delta_o=\frac{|y(t)|}{\mathbb{E}(|y|)}-1. \\
\end{equation}
This feature helps the classifier discriminate between \ac{PSK} and \ac{QAM} modulation schemes.

The modulations considered in the work are the following: \ac{BPSK}, \ac{QPSK}, 8PSK, 16QAM, \ac{CPFSK}, \ac{GFSK}, and \ac{GMSK}, resulting in a seven class classification task using the aforementioned six features computed from each training example. To generate the data set, a total of 35,000 examples were collected: 1,000 examples for each modulation at each of the five \ac{SNR} scenarios considered in the work. Three different feedforward neural network structures were trained at each \ac{SNR} scenario using a training set consisting of $80\%$ of the data collected at the given \ac{SNR} and a test set consisting of the remaining $20\%$. The three feedforward nets differed in the number of hidden layers, ranging from one to three. Qualitatively, the feedforward network with one hidden layer outperformed the other models in all but the least favorable \ac{SNR} scenario, achieving the highest classification accuracy of $98\%$ in the most favorable \ac{SNR} scenario. The seemingly paradoxical behavior is attributed to the over-fitting of the training data when using the higher complexity models, leading to poorer generalization in the test set.

This work has been further extended to evaluate other \ac{ML} techniques using the same features. Accordingly, we found that training a random forest classifier for the same \ac{AMC} task yielded similar results to the feedforward network classifier. Additionally, the random forest classifier was found to outperform the \ac{DNN} approach in scenarios when the exact center frequency of the transmitter was not known, which was assumed to be given in the previous work. The random forest classifier was comprised of 20 \ac{CART} constructed using the gini impurity function. At each split a subset of the feature vectors with cardinality equal to 3 was considered.

An alternative approach to the previously described method is to learn the modulation of the received signal from different representations of the raw signal. \cite{Kulin-ieeeaccess2018} train \acp{DCNN} to learn the modulation of various signals using three separate representations of the raw received signal. The authors denote the raw complex valued received signal training examples as $\mathbf{r}_k \in \mathcal{C}^N$, where $k$ indexes the procured training data set and $N$ is the number of complex valued samples in each training example. We inherit this notation for presentation of their findings. The data set in the work was collected by sampling a continuous transmission for a period of time and subsequently segmenting the received samples into $N$ dimensional data vectors.

The authors train separate \acp{DCNN} on three different representations of the raw received signal and compare their results to evaluate which representation provides the best classification accuracy. The first of the three signal representations are given as a $2 \times N$ dimensional \ac{I/Q} matrix containing real-valued data vectors carrying the \ac{I/Q} information of the raw signal, denoted $\mathbf{x_i}$ and $\mathbf{x_q}$, respectively. Mathematically,

\begin{equation}
    \mathbf{x}^{IQ}_k = \begin{bmatrix}
           \mathbf{x_i}^T \\
           \mathbf{x_q}^T \\
         \end{bmatrix}
\end{equation}
where $\mathbf{x}^{IQ}_k \in \mathcal{R}^{2 \times N}$. The second representation used is a mapping from the complex values of the raw received signal into two real-valued vectors representing the phase, $\Phi$ and the magnitude, $A$,

\begin{equation}
    \mathbf{x}^{A/\Phi}_k = \begin{bmatrix}
           \mathbf{x_A}^T \\
           \mathbf{x_{\Phi}}^T \\
         \end{bmatrix}
\end{equation}
where $\mathbf{x}^{A/\Phi}_k \in \mathcal{R}^{2 \times N}$ and the phase vector $\mathbf{x_{\Phi}}^T \in \mathcal{R}^N$ and magnitude vector $\mathbf{x_A}^T \in \mathcal{R}^N$ have elements,

\begin{equation}
    x_{\Phi_n} = \arctan\left(\frac{r_{q_n}}{r_{i_n}}\right), x_{A_n} = (r_{q_n}^2 + r_{i_n}^2)^{\frac{1}{2}}
\end{equation}
respectively. The third representation is a frequency domain representation of the raw time domain complex signal. It is characterized by two real-valued data vectors, one containing the real components of the complex FFT, $\Re(X_k)$, and the other containing the imaginary components of the complex FFT, $\Im(X_k)$, giving,

\begin{equation}
    x^{F}_k = \begin{bmatrix}
           \Re(X_k)^T \\
           \Im(X_k)^T \\
         \end{bmatrix}
\end{equation}
Using these three representations of the raw signal, the authors train three \acp{DCNN} with identical structure and compare the accuracy of the resultant models to determine which representation allows for learning the best mapping from raw signal to modulation structure.

The authors use training examples comprised of $N = 128$ samples of the raw signal sampled at $\mathrm{1~MS/s}$ and consider the following 11 modulation formats: \ac{BPSK}, \ac{QPSK}, 8-\ac{PSK}, 16-\ac{QAM}, 64-\ac{QAM}, \ac{CPFSK}, \ac{GFSK}, 4-\ac{PAM}, \ac{WBFM}, \ac{AM}-\ac{DSB}, and \ac{AM}-\ac{SSB}. Thus, the training targets $\mathbf{y}_k \in \mathcal{R}^{11}$ are encoded as one-hot vectors where the index holding a 1 encodes the modulation of the signal. The authors use a total of 220,000 training examples $\mathbf{x}_k \in \mathcal{R}^{2\times128}$. Additionally, samples were acquired uniformly over different \ac{SNR} scenarios ranging from $-20dB$ to $+20dB$.

The \ac{CNN} structure used for each signal representation is the same, and consists of two convolutional layers, a fully connected layer, and a \textit{softmax} output layer trained using the negative log-likelihood loss function. The activation function used in each of the convolutional layers and the fully connected layer is the \textit{ReLU} function. The \acp{CNN} were trained using a training set comprised of $67\%$ of the total data set, with the rest of the data set used as test and validation sets. An Adam optimizer \citep{KingmaB14Adam} was used to optimize the training processes for a total of 70 epochs. The metrics used to evaluate each of the models include the precision, recall, and F1 score of each model. The authors provide a range of values for the three aforementioned metrics for the CNN models trained on different data representations for three different \ac{SNR} scenarios: high, medium, and low, corresponding to $18dB$, $0dB$, and $-8dB$, respectively. In the high \ac{SNR} scenario, the authors report that the precision, recall, and F1 score of each of the three CNN models falls in the range of $0.67-0.86$. For the medium and low \ac{SNR} scenarios, the same metrics are reported in the ranges of $0.59-0.75$ and $0.22-0.36$, respectively. The authors attribute the relatively low performance to the choice of the time-varying multipath fading channel model used when generating the data.

The authors go on to evaluate the classification accuracy of each of the three models trained using different data representations under similar \ac{SNR} conditions. Qualitatively, each of the three \ac{CNN} models performs similarly at low \ac{SNR}, while the \ac{CNN} trained on the I/Q representation of data yields a better accuracy at medium \ac{SNR} and the \ac{CNN} trained on the amplitude and phase representation yields a better accuracy at high \ac{SNR}. Interestingly, the \ac{CNN} trained on the frequency domain representation of the data performs significantly worse than the $I/Q$ and $A/\phi$ \acp{CNN} at high \ac{SNR}. The authors mention that this could potentially be due to the similar characteristics exhibited in the frequency domain representation of the PSK and QAM modulations used in the classification problem. The primary takeaway from this work is that learning to classify modulation directly from different representations of the raw signal can be an effective means of developing a solution to the \ac{AMC} task; howeve, the efficacy of the classifier is dependent on how the raw signal is represented to the learning algorithm.

The following table provides the summary of the methods for \ac{AMC} discussed in this section.

\begin{table}[h!]
    \centering
    \caption{Summary of \ac{ML} Solutions for Automatic Modulation Classification\label{tab:sigintAMC}}
    \begin{tcolorbox}[tab2,tabularx={|p{4 cm}||p{1.5 cm}|p{3.3 cm}|X}]       \textbf{Classifiers}& \textbf{Model} & \textbf{Representation}   & \textbf{Objective}  \\ \hline\hline
      Jagannath et al. \cite{Jagannath18ICC}& \ac{DNN}   & Feature-Based  & 7-Class task considering PSKs, FSKs, QAMs   \\ \hline
      Kulin et al. \cite{Kulin-ieeeaccess2018}& \ac{DCNN}   & I/Q, A/$\Phi$, FFT &  11-Class task considering PSKs, FSK, QAMs, PAM, DSB, SSB\\ \hline
      O'Shea and Corgan \cite{o'sheaConvMod} &\ac{DCNN}   & I/Q & 11-Class task considering PSKs, FSK, QAMs, PAM, DSB, SSB \\ \hline
      Shengliang Peng and Yao \cite{pengCNN}&\ac{DCNN}  & Constellation &  4-Class task considering PSKs and QAMs \\ \hline
      West and O'Shea \cite{West-dyspan2017} & \ac{DCNN}, \ac{LSTM}, \ac{RN} & I/Q & 11-Class task considering PSKs, FSK, QAMs, PAM, DSB, SSB \\ \hline
      Karra et al. \cite{Karra-ieeedyspan2017} & \ac{DCNN}, \ac{DNN} & I/Q, FFT & 11-Class task considering PSKs, FSK, QAMs, PAM, DSB, SSB \\ \hline
      
    \end{tcolorbox}{}
\end{table}

\subsection{Wireless Interference Classification}

The task of \ac{WIC} regards identifying what type of wireless emitters exist in the local \ac{RF} environment. The motivation behind such a task is that it can be immensely helpful to know what type of emitters are present (WiFi, Zigbee, Bluetooth, etc.) in the environment when attempting to avoid and coexist with interference from other emitters. Solutions to \ac{WIC} tasks are often similar in nature to \ac{AMC} techniques. For example, \cite{SchmidtBM17} employ \acp{DCNN} to classify IEEE $802.11$ b/g, IEEE $802.15.4$, and IEEE $802.15.1$ emitters using a frequency domain representation of the captured signal. \ac{WIC} tasks may also consider emitters in the environment that are not used in communication systems. In \cite{Grimaldi}, an \ac{SVM} solution is developed to classify interference in \acp{WSN} from IEEE $802.11$ signals and microwave ovens. A recent work \cite{SelimIF} shows the use of \acp{DCNN} to classify radar signals using both spectrogram and amplitude-phase representations of the received signal. In \cite{Akeret}, \ac{DCNN} models are proposed to accomplish interference classification on two-dimensional time-frequency representations of the received signal to mitigate the effects of radio interference in cosmological data. Additionally, the authors of \cite{CzechLSTM} employ \ac{DCNN} and \ac{LSTM} models to achieve a similar end.

In \cite{Kulin-ieeeaccess2018}, \acp{DCNN} are employed for the purpose of the wireless interference classification of three different wireless communication systems based on the WiFi, Zigbee, and Bluetooth standards. They look at five different channels for each of the three standards and construct a fifteen class classification task for which they obtain $225,225$ training vectors consisting of 128 samples each, collected at$\mathrm{10~MS/s}$. A flat fading channel with additive white Gaussian noise is assumed for this classification task.

Three \acp{DCNN} were trained and evaluated using the wireless interference classification data set described above. Each of the three \acp{CNN} was trained on one of the representations of the data that were presented in the previous section, namely, \ac{I/Q}, $A/\Phi$, and frequency domain representation. The \ac{CNN} architectures were also the same as presented previously in Section \ref{sec:AMC}.

Each of the three \acp{CNN} trained using different data representations was evaluated in a similar fashion to the evaluation method described in Section \ref{sec:AMC}, namely, using precision, recall, and F1 score under different \ac{SNR} scenarios. For the wireless interference classification task, the precision, recall, and F1 score of each of the three \acp{CNN} all fell in the interval from $0.98-0.99$ under the high \ac{SNR} scenario. For the medium and low \ac{SNR} scenarios, the analogous intervals were from $0.94-0.99$ and $0.81-0.90$, respectively.

Additionally, the authors provide an analysis of classification accuracy for each of the three \ac{CNN} models at varying \acp{SNR}. For the task of wireless interference classification, the \ac{CNN} model trained on the frequency domain representation of the data outperforms the other models at all \acp{SNR}, especially in lower \ac{SNR} scenarios. The authors claim that these findings are due to the fact that the wireless signals considered have more expressive features in the frequency domain as they have different bandwidth, modulation, and spreading characteristics.

The authors of \cite{youssefIF} take a different approach to the wireless interference classification task and primarily compare different types of learning models rather than different types of data representation. The models the authors propose include deep feedforward networks, deep convolutional networks, support vector machines using two different kernels, and a \ac{MST} algorithm using two different learning algorithms. The authors consider 12 different transmitters and collect 1,000 packets from each transmitter for a total of 12,000 packets which comprise the entire data set. Each transmitter transmitted the same exact 1,000 packets, which were generated using pseudo-random values injected into the modem. All of the transmitters used a proprietary \ac{OFDM} protocol with a \ac{QPSK} modulation scheme and a baseband transmitter sample rate of $\mathrm{1.92~MS/s}$. At the receiver, each packet is represented by 10,000 time domain \ac{I/Q} samples. Each of the models was trained on data sets consisting of training examples made up of 32, 64, 128, 256, 512, and 1024 samples from each packet, and their performance is compared across data sets. Given the complex-valued received signal,

\begin{equation}
    r = (r_1, r_2,....,r_N)
\end{equation}
$N$ samples were selected by skipping the first $N_0$ samples of a packet where $|\Re(r_i)| < \tau$ for some $\tau > 0$ yielding the signal vector $x$,

\begin{equation}
    x = (r_{N_0}, r_{N_0+1}, ..., r_{N_0+N-1})
\end{equation}
For the \ac{DNN}, \ac{SVM}, and \ac{MST} models each training example was constructed by concatenating the real and imaginary parts of the signal vector, yielding a vector of dimension $2N$. For the \ac{CNN} model the real and imaginary parts of the signal vector were stacked to generate $2\times N$ dimensional training vectors.

The \ac{DNN} architecture considered in the work consisted of two fully connected hidden layers, comprised of 128 \textit{ReLU} units each and an output layer consisting of logistic \textit{sigmoid} units. The network was trained using the Adam optimizer \citep{KingmaB14Adam} and a mini-batch size of 32.

The \ac{CNN} model used by the authors was composed of two convolutional layers using 64 ($8\times 2$) and 32 ($16 \times 1$) filters, respectively. Each convolutional layer was input into a max-pool layer with a pool size of $2\times 2$ and $2\times 1$, respectively. The output of the second max-pool layer was fed into a fully-connected layer consisting of 128 \textit{ReLU} units. An output layer employing logistic \textit{sigmoid} units was used on top of the fully-connected layer.

The two \ac{SVM} architectures analyzed in the work differ only in the kernel function used. The first architecture employed the polynomial kernel and the second employed the Pearson VII Universal Kernel \citep{pearsonKernel}. Both architectures used Platt's Minimization Optimization algorithm to compute the maximum-margin hyperplanes.

\begin{figure}[h!]
    \centering
    \includegraphics[width=5 in]{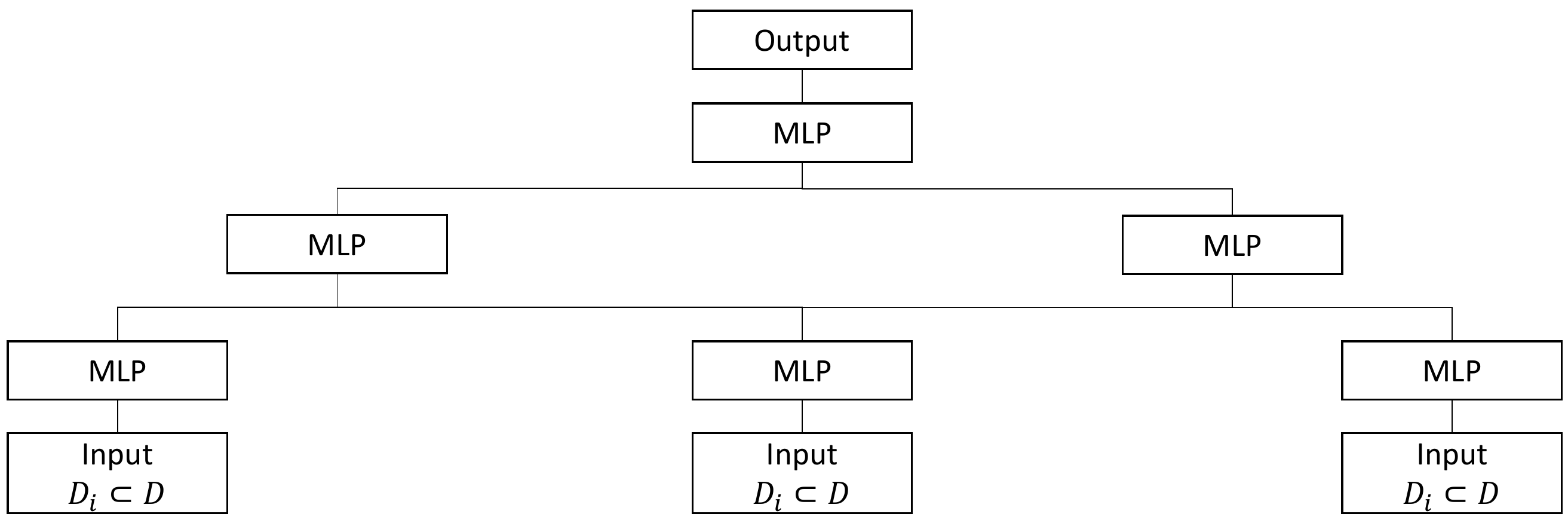}
    \caption{Adaptation of MST MLP used in \cite{youssefIF}.}
    \label{fig:MST_MLP}
\end{figure}

The authors also analyze the performance of \ac{MST} \acp{MLP} trained using first order and second order methods. A high-level description of \ac{MST} \ac{MLP} is presented here and we refer the interested reader to \cite{youseffMST} for a more rigorous derivation. The MST method to training neural networks, as presented in the work, is essentially a hierarchical way to solve an optimization problem by solving smaller constituent optimization problems. To this end, in what is called the first stage, a number of separate \acp{MLP} would be trained on different subsets of the training data set. This can be seen in the lowest layer of the hierarchical representation adapted from \cite{youssefIF}, and provided herein Figure \ref{fig:MST_MLP}. 

Once the first stage is trained, a second stage is trained by taking the concatenation of the network outputs from the first stage as input. Training can continue in this fashion for subsequent stages. One of the advantages of training networks in this way is that the many smaller \acp{MLP} comprising the larger classifier can be efficiently trained using second-order optimization methods. Second-order optimization methods such as Newton, Gauss-Newton, or Levenberg-Marquardt methods are usually intractable due to the size of typical networks but can provide better convergence when applicable. The authors train two 3-stage \ac{MST} systems, one using the first order method of SGD, and one using the second-order Accelerated Levenberg-Marquardt method \citep{acceleratedLM}. Each \ac{MST} system had the identical structure where stage 1 consisted of 60 \acp{MLP} with 2 hidden layers and 10 units in each layer. Stage 2 and 3 had the same architecture and were comprised of 30 \acp{MLP} with each \ac{MLP} consisting of 2 hidden layers made up of 15 units each. All hidden units employed the \textit{tanh} activation function and all output layers contained linear units.

All of the models described above were trained on 10 different iterations of the collected data set and their performance was compared. Five data sets were constructed using training examples made up of 32, 64, 128, 256, and 512 samples and each model was trained twice, using a training set comprised of $90\%$ and $10\%$ of the total data set, for a total of 10 different data sets for each model. In general, the \ac{MST} system trained using second-order methods on $90\%$ of the training data performed best across all sizes of training examples, yielding a classification accuracy of $100\%$ for each data set. All of the models performed better when trained using $90\%$ of the data set as opposed to $10\%$ of the training data set. Generally, each model performed better when provided with training examples that contained more samples, with the exception of the deep feedforward network model, which the authors attribute to the fact that longer sequences of samples may contain an increasing number of artifacts which the \ac{DNN} may not be robust to. A summarization of the different models presented in this section is provided in Table \ref{tab:sigintWIC}.

\begin{table}[h!]
    \centering
    \caption{Summary of \ac{ML} Solutions for Wireless Interference Classification\label{tab:sigintWIC}}
    \begin{tcolorbox}[tab2,tabularx={|p{4.0 cm}||p{1.5 cm}|p{3.3 cm}|X}]       \textbf{Classifiers}& \textbf{Model} & \textbf{Representation}   & \textbf{Objective}  \\ \hline\hline
      Kulin et al. \cite{Kulin-ieeeaccess2018}&\ac{DCNN}   & I/Q, A/$\Phi$, FFT & Classification of 15 WiFi, ZigBee, and Bluetooth Transmitters     \\ \hline
      Selim et al. \cite{SelimIF}  & \ac{DCNN}   & 2D time-frequency, A/$\Phi$ & Classification of Radar Signals \\ \hline
      Akeret et al. \cite{Akeret}& \ac{DCNN}   & 2D time-frequency & Classification of Cosmological Interference  \\ \hline
      Czech et al. \cite{CzechLSTM}& \ac{DCNN}, \ac{LSTM}   & 2D time-frequency &  Classification of Cosmological Interference  \\ \hline 
      Youssef et al. \cite{youssefIF} & \ac{DNN}, \ac{DCNN}, \ac{SVM}, \ac{MST}   & I/Q & Classification of 12 OFDM Transmitters  \\ \hline
      Schmidt et al. \cite{SchmidtBM17} & \ac{DCNN}  & FFT & Classification of IEEE 802.11 b/g, IEEE 802.15.4, IEEE 802.15.1 signals  \\ \hline
      Grimaldi et al. \cite{Grimaldi} & \ac{SVM}  & Feature Based & Classification of IEEE 802.11 and Microwave Oven signals  \\ \hline
    \end{tcolorbox}{}
\end{table}

\subsection{Open Problems}\label{sec:SigIntOpenProb}
The solutions to the tasks of wireless interference and modulation classification fixate themselves among solutions readily available to be deployed in the IoT. This distinction is primarily a result of the mutual exclusivity that these tasks exhibit with the IoT itself; these problems exist both within and outside the context of the IoT. Contrarily, there are signal intelligence tasks that arise from and are innate to the IoT, which have been studied in comparatively less detail. These tasks, along with their potential to benefit from the application of \ac{ML} techniques, are described in the rest of this section. 

\subsubsection{Intrusion Detection}
Security in the IoT is of utmost importance as the prevalence of connected devices in society and the amount of data collected from individuals increases. Detection of an intruder is often the first step in mitigating efforts from adversaries and can be performed in myriad ways across multiple layers of the protocol stack. As the problem of intruder detection moves from the internet to the IoT, the detection of the physical presence of the intruder among things becomes a salient avenue for mitigation. In \cite{Roux2017}, \ac{RSSI} information is collected from deployed security probes in an attempt to detect behaviors and communications that are illegitimate and thus identify devices that may have been compromised or may have entered the local network illegally. The authors collect \ac{RSSI} information, reception timestamps, and radio activity during a time interval from each of the probes and route them to a central security system, which processes the information using a proposed neural network algorithm, which classifies the presence of an intruder. The authors note significant changes in collected \ac{RSSI} information in their laboratory but highlight a full implementation of the proposed solution as future work. A primary advantage to \ac{RSSI}-based intrusion detection is that the proposed solution is protocol agnostic.

\subsubsection{Indoor Localization}
The problem of indoor localization remains a challenging one in the context of the IoT and elsewhere. Generally, RF localization problems arise when trying to estimate the geolocation of a receiving or transmitting radio. In outdoor environments, this is readily accomplished on-board many devices using various geolocating signals such as GPS and GNSS; however, the efficacy of these signals use in geolocation is severely diminished without \ac{LoS} between the satellites and receivers. Thus, indoor localization becomes an important problem in many applications involving the tracking and location of devices that are associated with human users, as these applications often occur indoors. Examples of applications that benefit from indoor localization capabilities include indoor robotic systems, assisted living systems, health applications, and location-based services. Additionally, in \cite{MacagnanoLoc}, indoor localization of IoT devices is motivated as one of the key enabling technologies in increasing the utilization of the IoT.

Most indoor localization approaches in the IoT aim to make use of information transmitted from the local Wi-Fi access points and employ some form of Wi-Fi fingerprinting. In \cite{ChengLoc}, a clustering based access point selection and \ac{RSSI} reconstruction algorithm is proposed to obtain the optimal feature set for input to an \ac{ML}-based localization algorithm. Simulation results are provided using \ac{ANN}, \ac{SVR}, and ensemble \ac{SVR} to obtain localization predictions from the selected \ac{RSSI} values. In \cite{AdegeLoc}, \acp{DNN} are proposed in conjunction with a linear discriminant analysis to operate on \ac{RSSI} and \ac{BSSID} information to produce both classification and regression location information. Alternatively, \cite{WangGMP16} suggest utilizing channel state information consisting of subcarrier-level measurements of \ac{OFDM} channels as opposed to \ac{RSSI} based fingerprinting and simulation results using \acp{CNN} and \acp{LSTM} trained on channel state information are provided. 

\section{Machine Learning For Higher Layers}\label{sec:Higher_Layer}

The requirement for \ac{IoT} devices to have distributed intelligence is becoming inevitable to tackle the problems emanating from the complexity, dynamic nature of its operations and to ensure scalability. This implies that part of the \ac{IoT} "smart" devices will require autonomy to react to a wide range of situations pertaining to networking, spectrum access, among others \cite{miorandi2012internet}. This is where the role of ad hoc networking becomes a crucial part of \ac{IoT}. Examples of ad hoc interaction in the context of \ac{IoT} can include \ac{VANET} that involves vehicles communicating with each other and roadside infrastructure along with the assistance of various sensors (velocity, temperature, humidity, CO2 emissions, etc.). Similarly, the ability to deploy \acp{WASN} will also play a crucial role in the overall \ac{IoT} architecture \cite{Mainetti_WSN} that is envisioned to enable smart cities as shown in Figure \ref{fig:IoTArch}. The ad hoc networking aspect of \ac{IoT} will, therefore, find applications in areas such as healthcare, infrastructure management, disaster prevention, and management, and optimizing transportation systems \cite{AdHoc_IoT, Sood2016SoftwareDefinedWN, bruzgiene2017manet}.

The advancements in the higher layers, especially the data-link and the network layers have played a significant role in enabling \ac{IoT} devices. The necessity to provide fair and efficient spectrum access has been a key motivating factor for researchers to design \ac{MAC} protocols for \ac{IoT} \citep{Cormio_MAC, al2015internet}. In contrast to centralized designs where entities like base stations control and distribute resources, nodes in ad hoc \ac{IoT} network have to coordinate resource allocation in a distributed manner. Similarly, to ensure scalability and reduce overhead, distributed designs are usually favored while designing routing algorithms for such networks. Recently, \ac{ML} has made a significant impact on the design of these layers specifically to enhance scheduling and resource allocation, mitigating attacks like \ac{DoS} in hostile environments, and efficient routing among others. In this section, we discuss in detail some of the advances made on this front. 
  
\subsection{Data Link Layer}  
  
A key functionality of the data link layer is to negotiate the access to the medium by sharing the limited spectrum resources in an ad hoc manner. Traditional \ac{MAC} protocols designed for \acp{WANET} (including \ac{IoT} networks) include \ac{CSMA/CA} \citep{Lien_CSMA,Jain_CSMA}, \ac{TDMA} \citep{Cordeiro_TDMA,Hadded_TDMA}, \ac{CDMA} \citep{Muqattash_CDMA, Kumar_survey_07} and hybrid approaches \citep{Sitanayah_Hybrid, Su_Hybrid, JithinAnu13WUWNet}. Here, we discuss some of the recent efforts to employ \ac{ML} to enhance the data link layer. 

The \ac{BSP} is a key problem studied in a \ac{TDMA}-based network to find an optimal \ac{TDMA} schedule that provides transmission time slots to of all nodes while minimizing the \ac{TDMA} frame size \citep{Commander_BSP}. Several \ac{ML}-based approaches have been proposed to solve this combinatorial optimization of \ac{BSP} using variations of neural networks. This includes the work of \cite{Salcedo_HNN_GA} proposing a combination of \ac{HNN} and \ac{GA} and \cite{Shi_SVC_NCNN} using \ac{SVC} and \ac{NCNN}. Subsequently, these solutions were shown to be outperformed by \ac{FHNN} proposed in \cite{SHEN_FHNN}. Here, we describe how \cite{SHEN_FHNN} tackles \ac{BSP}.

Consider  $N$ nodes in a network with $N_T$ time slots to share among these nodes. 
The slot assignment matrix $\mathbf{SA}$, in which each element is defined as $SA_{ij}=1$, if time slot $j$ is assigned to node $i$, otherwise $SA_{ij}=0$. 
The set of time slots to be assigned is given by set $T=\{t_1,t_2,...,t_{N_T}\}$. The fuzzy state, a degree that time slot $t_x$ is assigned to node $i$ is represented by $\mu_{xi}$ and matrix of all the fuzzy states, $\mathbf{U}$ is called a fuzzy c-partition matrix. Next, the channel utilization of node $i$ is defined as the fraction of total time slots assigned to node $j$ from the total \ac{TDMA} frame given as $\rho_j=(\sum_{j=1}^{N_T} SA_{ij}/N_T)$. Accordingly, the total channel utilization for the network can be given as \cite{wang_channelUtilization},    

\begin{equation}
    \rho=\frac{1}{N_T N}\sum_{j=1}^{N}\sum_{i=1}^{N_T} SA_{ij}
\end{equation}

The lower bound for the frame length is given by $\max_{i \in N} \deg(i) + 1$ where $\deg(i)$ is the number of edges incident to it. In the case of \ac{FHNN}, an energy function is considered as the distance between the current state of the \ac{HNN} and its solution state. The objective is to minimize the energy function by solving the optimization problem. In this case, the energy function that considers all the  constraints is defined as follows \cite{SHEN_FHNN},

\begin{align}
    E&=\frac{\alpha}{2} \sum_{x=1}^{N_T} \left( \sum_{i=1}^{N} \mu_{xi}-1 \right)^2 + \beta \sum_{x=1}^{N_T} \sum_{i=1}^{N} \left( \sum_{y=1,y \neq i }^{N_T} d_{iy}\mu_{yi} +\right. \notag \\  
    &\left. \sum_{y=1,y \neq i }^{N_T} d_{iy} \sum_{y=1,y \neq i,k \neq y }^{N_T} d_{yk} (\mu_{xi})^f \left[ t_x- \sum_{y=1}^{N_T} \frac{t_y}{\sum_{k=1}^{N_T}(\mu_{ki})^f}(\mu_{yi})^f  \right]^2\right) \label{eq:Energy}
\end{align}
where $\alpha$ and $\beta$ are assumed to be positive coefficients, $f$ is the fuzzification parameter, and $d_{iy}=1$, if there is a connectivity between $i$ and $y$. The first term in equation (\ref{eq:Energy}) ensures that $N_T$ slots can only be distributed among the $N$ classes (nodes). The second term minimizes the inter-class euclidean distance from a sample to the cluster center of all clusters. Accordingly, \ac{FHNN} aims to classify $N_T$ time slots into $N$ nodes by minimizing $E$. In simulations, the proposed \ac{FHNN} based \ac{BSP} approach outperforms both \citep{Shi_SVC_NCNN} and \citep{Salcedo_HNN_GA} in terms of average time delay. Additionally, authors also show that performance improves with larger $f$ at the expense of increased convergence time. 

\begin{figure}
    \centering
    \includegraphics[width=3.5 in]{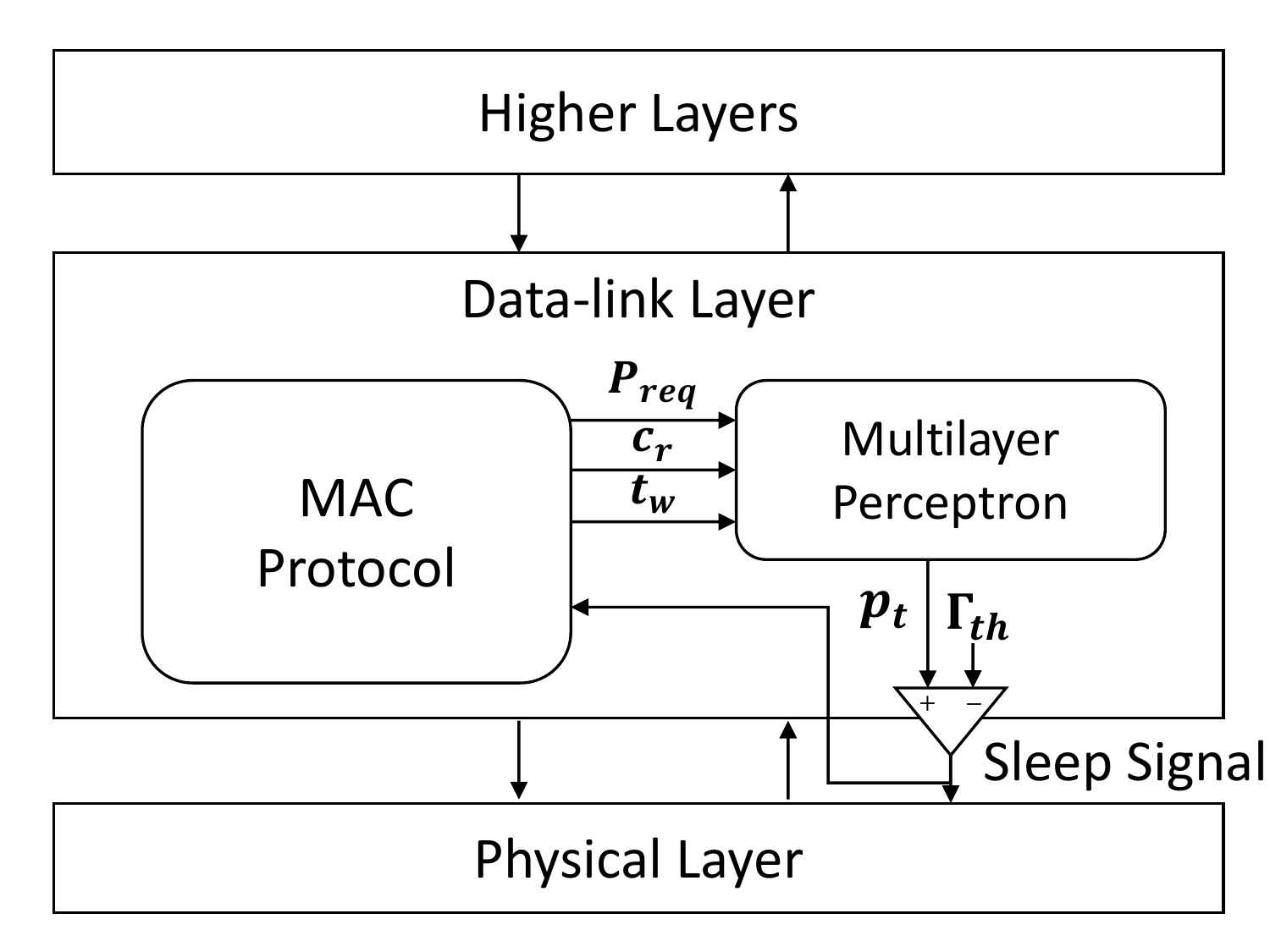}
    \caption{Block Diagram of MLP in MAC against DoS}
    \label{fig:MAC_DoS}
\end{figure}

There have also been efforts to advance the current \ac{MAC} protocols to react to different kinds of attack like \ac{DoS} that can debilitate \ac{IoT} devices. In one such case \cite{Kulkarni_MLP_MAC}, a \ac{MLP} is used to modify \ac{CSMA}-based network to identify \ac{DoS} attack and stay inactive for a duration to preserve the energy of the wireless sensor nodes. As shown in Figure \ref{fig:MAC_DoS}, the \ac{MAC} layer of each node consists of a \ac{MLP} that has been trained prior to deployment. The parameters used by \ac{MLP} include collision rate ($c_r$), packet request rate ($P_{req}$) and average packet wait time ($t_w$). The proposed solution is evaluated using both \ac{BP} and the \ac{PSO} \citep{Particle_swarm} algorithm for training. The authors show that \ac{BP} has lower computational cost compared to \ac{PSO} but provides inferior convergence point in terms of quality of the weights. The output of \ac{MLP} represents the probability that there is an active \ac{DoS} attack ($p_t$). Based on the chosen threshold $\Gamma_{th}$, the nodes decide to sleep for a predetermined period of time when $p_t>\Gamma_{th}$. The work does not discuss the optimal value for $\Gamma_{th}$ or the sleep time but provides an example of applying \ac{ML} to mitigate the effects of such attacks.

Another interesting application where \ac{ML}, specifically \ac{RL}, has been successfully applied is in the domain of \ac{DSA} for \ac{CR} which could be instrumental in enabling modern \ac{IoT} given the constrained availability of spectrum. 
An ALOHA-like scheme is developed for \acp{CR} by applying a multiagent \ac{RL} framework \cite{Li_2010}. In this work, a secondary user that is able to transmit successfully over an idle channel without collision receives a positive reward and zero otherwise. \ac{ACK} packets received after the transmission is used to ensure collision-free transmissions. Since the secondary user does not have control over the channel state, the Q-function is defined as the expected reward over a given time slot $t$. This in turn depends on the state of the overall system, $S(t)=s$ and the node $i$'s action ($a_i(t)$) at time slot $t$ to transmit on channel $h$. The expectation is taken over the randomness of other secondary user's action and the primary user's activity which can be represented as,

\begin{equation}
    Q_{ih}^{s}=E[R_i|a_i(t)=h, S(t)=s]
\end{equation}
where $R_i$ is the rewards for action. To ensure good channels are not neglected, the authors propose the use of a Boltzmann distribution for random exploration during the learning phase. Considering temperature $\mathcal{T}$, the exploration probability is given as,

\begin{equation}
    P(i \text{ chooses channel } h| \text{ state } s)= \frac{\exp(Q_{ih}^s/\mathcal{T})}{\sum_{k=1}^N \exp(Q_{ih}^s/\mathcal{T})}
\end{equation}

To accomplish this, each secondary user considers both the channel and other secondary users to update its Q-values to choose the best action. It is important to remember that this is an extreme case where no control packets are exchanged between nodes similar to traditional ALOHA. Furthermore, the authors were able to show convergence in limited circumstances even when they extend the full observations to the case of partial observations. Simulations showed how secondary users can learn to avoid collision and outperform a scheme that uses Nash equilibrium.
  
A similar case is considered in \citep{Naparstek_RL} where authors propose a distributed \ac{DSA} algorithm based on multi-agent reinforcement learning but this time employing \ac{DRL}. We have seen how Q-learning provides adequate performance when the state-action space is relatively small. As the state-action space grows exponentially for larger problems, the direct application of Q-learning becomes inefficient and impractical as discussed previously. \ac{DQN} which combines \ac{DNN} with Q-learning can overcome this challenge. The goal is to enable users to learn a policy while dealing with the large state space without online coordination or message exchanges between users. In \cite{Naparstek_RL}, the authors model their network state as partially observable for each user and the dynamics being non-Markovian and determined by the multi-user actions, they propose to use \ac{LSTM} layer that maintains an internal state and aggregate observations over time.

To ensure feasibility, the training is set to happen offline where various training experiences with changing environment and topology are considered. This ensures that the algorithm can be deployed to operate in a distributed manner with the need to be updated only if the operating conditions are significantly different from the training set. After the training phase, each user determines which channel to select and associate ``attempt probability" based on its observation. The proposed algorithm is compared against slotted-ALOHA that is assumed to have complete knowledge of the network and hence used optimal attempt probability. The proposed distributed algorithm that only used \acp{ACK} to learn outperforms slotted ALOHA by twice the channel throughput. They evaluated the network for two network utilities, (i) network rate maximization and (ii) individual rate maximization. In the case of users whose objective was to maximize the sum rate of the network, some learned to remain idle (sacrifice) incurring zero rate in order to maximize the overall network utility. In contrast, when each user aims to maximize its own rate they converged to a Pareto-optimal sharing policy.

\begin{figure}
    \centering
    \includegraphics[width=3.5 in]{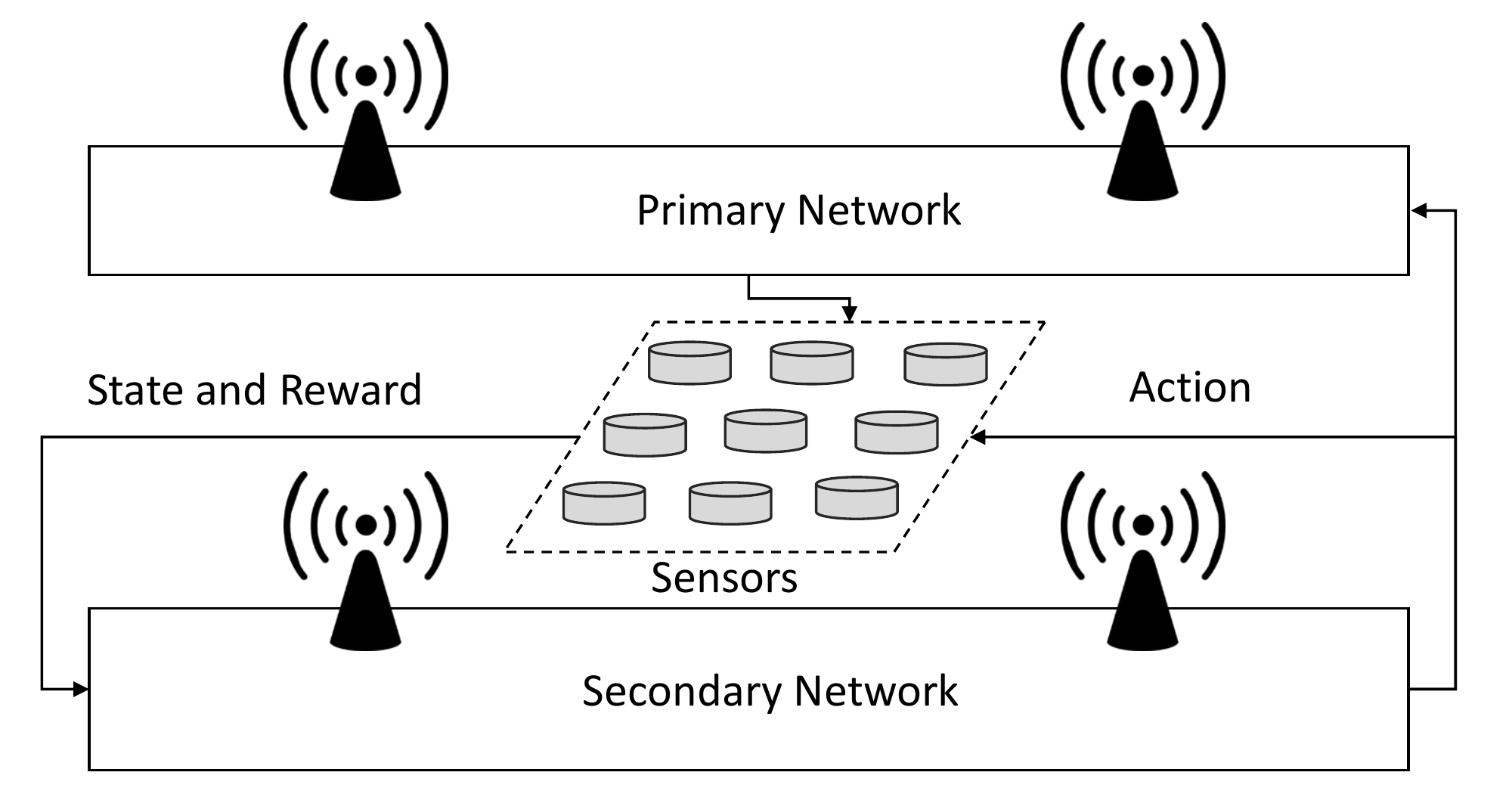}
    \caption{Framework for power control in cognitive network}
    \label{fig:MAC_LI2018}
\end{figure}

The authors of \cite{Li_2018} propose a \ac{DQN} based framework that provides an \ac{IPC} algorithm for secondary users to coexist with the primary user while ensuring \ac{QoS} for both. The overall architecture is depicted in Figure \ref{fig:MAC_LI2018}. The authors assume the presence of several sensors that are deployed to monitor and convey the \ac{RSS} to the secondary users for decision making. The infinite states associated with the continuous \ac{RSS} impose the need to employ \acp{DQN}. During the \ac{DQN}'s training phase, secondary users assume complete knowledge of whether the \ac{QoS} of every user (primary and secondary) are satisfied. The authors argue this can be achieved by overhearing the \ac{ACK} packets. Once learning is complete, only the feedback from the sensors is required to determine the optimal power level for the secondary user to access the spectrum while satisfying the \ac{QoS} constraint of both the networks. \ac{IPC} is compared against the \ac{DCPC} algorithm \cite{Grandhi1994_DCPC} which is an optimized solution. In contrast to \ac{IPC}, an optimization-based technique like \ac{DCPC} requires cooperation between both primary and secondary users. The simulation shows how \ac{IPC} converges faster compared to \ac{DCPC} while achieving a near optimal solution.



Realizing the role \ac{ML} will play in the near future in maximizing the use of scarce spectrum, \ac{DARPA} initiated a three year competition known as \ac{SC2}. The goal was for teams to propose an \ac{ML}-based spectrum sharing strategy to allow peaceful coexistence between any unknown heterogeneous wireless networks. A solution inspired from this competition is presented in \cite{Yu_2018}, which explores \ac{DLMA} for a heterogeneous wireless network consisting of various kind of networks (ALOHA, \ac{TDMA}) that coexist. To accomplish this, authors use \ac{DRL} to learn spectrum usage from a series of environmental observations and actions without actually being aware of the type of \ac{MAC} protocols being operated. The goal is to maximize the total throughput of all the coexisting networks. They exploit neural networks to employ \ac{DRL} as compared to traditional \ac{RL} to enable fast convergence and ensure robustness to non-optimal parameters. Fast convergence is essential for wireless networks as convergence time is shorter than coherence time which will give nodes an opportunity to operate using an optimal strategy rather than trying to catch up with the changing efferent every time. Similarly, the lack of knowledge of existing networks makes it difficult to obtain optimal parameters. 

The possible actions that can be taken by an agent is $a(t) \in \{wait,\; transmit\}$. The observation after taking one of these actions can be $z(t) \in \{success,\; collision, \; idleness\}$. Accordingly, the channel state at $t+1$ is given as an action-observation pair  $c(t+1) \triangleq \{a(t), z(t)\}$. Next, the environmental state at time $t + 1$ is given as $s(t+1) \triangleq \{ c(t - \mathfrak{h}+2), ..., c(t), c(t+1)\}$, where the parameter $\mathfrak{h}$ is the state history length to be tracked by the agent. The reward for transitioning from $s(t)$ to $s(t+1)$ is defined as,

\begin{equation}
r (t+1)= \begin{cases}
1 & ,\;\; \text{if}\;\; z(t)=success \\ 
0 & ,\;\; \text{if}\;\; z(t)=\{collision\; idleness\} \\
\end{cases}    
\end{equation}

In this work, a \ac{DNN} is used to approximate the state value function. Assuming $\phi$ is the parameter vector representing the weights of the \ac{DNN}, the approximation can be represented as $q(s,a;\phi) \approx Q^*(s,a)$. The authors employ ``experience replay" \cite{mnih2015humanlevel} which uses multiple experience samples $(s,a,r(t+1),s(t+1))$ in each training step using the following loss equation,

\begin{equation}
    L(\theta)= \sum_{(s,a,r,s')\in EX_t} \left( y_{r,s'}-q(s,a;\phi) \right)^2 
\end{equation}
where,

\begin{equation}
    y_{r,s'}=r+ \gamma \max_{a'} q(s',a'; \phi_t)
\end{equation}
where $EX_t$ is the set of experience samples used for training at time $t$. The authors argue the advantage of using \ac{DRL} over \ac{RL} by showing a faster convergence rate and a near-optimal strategy being achieved through simulations. They show how the network can learn and achieve near-optimal performance with respect to the sum throughput objective without the knowledge of co-existing MAC (\ac{TDMA}, ALOHA). The work is further extended \cite{Yu_2018_DLMA} by using a residual network \cite{He_ResidualNet} in place of the \ac{DNN}. The authors show how a single \ac{RN} architecture with fixed depth is suitable to ever-changing wireless network scenarios as compared to the plain \ac{DNN} which was shown to vary in performance based on the selected number of hidden layers. 

These works provide a promising direction towards solving the spectrum crunch that will be experienced with the proliferation of \ac{IoT} devices and 5G networks in the near future. We summarize the discussion of this section in Table \ref{tab:Data_Link}.

\begin{table}[h!]
    \centering
    \caption{Summary of Application of \ac{ML} in MAC protocols\label{tab:Data_Link}}
    \begin{tcolorbox}[tab2,tabularx={|p{4.8cm}||p{2.8cm}|X}]
      \textbf{MAC Protocol}   & \textbf{ML Algorithm}  & \textbf{Objective}\\ \hline\hline
      Salcedo-Sanz et al. \cite{Salcedo_HNN_GA}   & \ac{HNN} wt \ac{GA} &  Proposed to solve \ac{BSP}   \\ \hline
      Shi and Wang \cite{Shi_SVC_NCNN}   & \ac{NCNN} wt \ac{SVC} & Proposed to solve \ac{BSP} \\ \hline
      Shen and Wang \cite{SHEN_FHNN}   & \ac{FHNN} & Proposed to solve \ac{BSP}  \\ \hline
      Kilkarni and Venayagamoorthy \cite{Kulkarni_MLP_MAC}  & \ac{MLP} & Tolerance against \ac{DoS}  \\ \hline
      Li \cite{Li_2010}   & \ac{RL} & ALOHA-like spectrum access   \\ \hline
      Naparstek and Cohen \cite{Naparstek_RL}   & \ac{DQN} wt \ac{LSTM} & ALOHA-linke spectrum access  \\ \hline
      Li et al. \cite{Li_2018}   & \ac{DQN} & Intelligent power control  \\ \hline
      Yu et al. \cite{Yu_2018}   & \ac{DQN} & Non-coopertive heterogenous network \\  \hline
      Yu et al. \cite{Yu_2018_DLMA}   & \ac{DQN} wt \ac{RN} & Non-coopertive heterogenous network  \\  \hline
    \end{tcolorbox}{}
\end{table}

\subsection{Network Layer}

Routing protocols have evolved over the years to accommodate the needs of modern \ac{IoT} \acp{WANET}. The design of the routing protocols primarily depends on the context and objective of the application and can be classified in several ways. Some of these classifications include geographical location based routing \citep{GPSR, Jagannath19ADH_HELPER, Peng_VBF, Jagannath18TMC}, hierarchical \citep{Heinzelman_LEACH}, \ac{QoS}-based \citep{Akkaya_QoS,Akkaya05asurvey}, and recently cross-layer optimized routing \citep{Srivastava2005CrosslayerDA, LKuo12SECON, Jagannath16GLOBECOM, Hasan2018AnalysisOC, Xu_crosslayer, callebaut2018cross}. Similar to earlier discussions, \ac{ML} has elegantly found its way into this domain by providing a powerful tool to solve some of the problems associated with designing routing algorithms.  

One of the earliest attempts to apply \ac{ML} to routing algorithms is presented in \cite{Boyan_1993_RL-Route} in the context of a traditional wired network including \ac{LATA} telephone network. The proposed algorithm, referred to as Q-routing, uses a distributed approach which gathers estimated delay information from immediate neighbors to make the routing decision. The proposed Q-learning based routing algorithm can be represented as a variation of Bellman-Ford shortest path algorithm \citep{bellman_1,Ford_2010} that replaces hop count by delivery time and performs the relaxation step online in an asynchronous manner. In \cite{Boyan_1993_RL-Route}, the authors clearly showed how Q-routing is able to adapt to varying traffic loads after the initial inefficient learning period. When the load is low, Q-routing converges to using the shortest path and when the load increases, it is capable of handling the congestion more elegantly compared to the shortest path routing that is forced to use static routes. 

In a recent effort \cite{Mao_routing}, the need to reenvision router architectures and key routing strategies to meet the requirements of modern networks is highlighted. This was motivated by the advent of the graphical processing unit accelerated software defined routers that are capable of massive parallel computing. Accordingly, authors propose to use \ac{DL}, specifically, a \ac{DBN} based system that uses traffic patterns to determine the routes. The authors demonstrated with simulations the superiority of \acp{DBN} over \ac{OSPF} in terms of throughput and average delay per hop. This can be attributed to the reduced overhead as \acp{DBN} does not use the traditional rule-based approach. Some of these ideas are extendable to \acp{WANET} after careful consideration of the challenges and characteristics of wireless networks.

One of the key challenges that will be faced by \ac{IoT} devices operating in ad hoc mode is the reliability of routes that can get disconnected due to channel conditions or node failure. The authors of \cite{POURFAKHAR_CMAC} study this problem in the context of multicast routing and apply \ac{CMAC} \cite{Albus_CMAC}. To ensure reliability, wireless mesh networks need to have the ability to recover from link disruption due to disrupted channel or node failure. The \ac{CMAC} algorithm was first introduced around the same time that the perceptron algorithm was first introduced. While the \ac{CMAC} framework can be considered a type of neural network, it is fundamentally different from the ones previously described in this paper. The CMAC architecture can be seen as an attempt to model human associative memory and employs a sort of look-up table technique. The CMAC framework is characterized by a mapping from input space to memory address space (look-up table) and a subsequent mapping from address space to output space. The mapping from input to address space is usually denoted as $S \longrightarrow A$ where $S$ is the input space and $A$ is the address space. Typically, multiple mappings from input to address space are used such that a single input can ``activate" multiple addresses in the address space. Each address in the address space contains a weight vector, $\mathbf{w} \in A$, which is used in the subsequent mapping from address space to output space, usually denoted as $A \longrightarrow P$. The function $f: A \longrightarrow P$ is given to be the sum of the weight vectors contained in the activated memory regions. The training of the model can be conducted iteratively over training examples by updating the weight vectors used in the computation of the output by some proportion of the error observed for that training example.

In \citep{POURFAKHAR_CMAC}, authors use this concept to learn  to estimate the route disconnection expectancy between itself and \acp{AP} based on the following three parameters, (i) delay of packets in a node (i.e. sum of queuing delay and processing delay), (ii) number of node disconnections, and (iii) difference in delays between two packets that are separated by a predetermined number of packets. The proposed \ac{CMAC} uses these three parameters to estimate the \ac{NDP}. Then the \ac{NDP} estimate enables nodes to predict possible node failure and react faster enabling better throughput, higher packet delivery ratio for multicast packets and provide minimum delay without prior knowledge of the topology. 

Q-MAP is another multicast routing algorithm proposed to ensure reliable communication \cite{Sun_2002_QMAP}. The algorithm is divided into two phases; in \emph{join query forward} phase nodes use \acp{JQP} to explore all the possible routes to the multicast destination and \emph{join reply backward} phase uses \acp{JRP} to establish the optimal route that maximizes the designed Q-value. The \ac{JQP} can be considered as forwarding agents carrying the possible Q-values downstream, subsequently, \ac{JRP} packets can be considered as backward agents carrying the optimal decision information upstream to the source. 

\begin{figure}[h!]
    \centering
    \includegraphics[width=3.5 in]{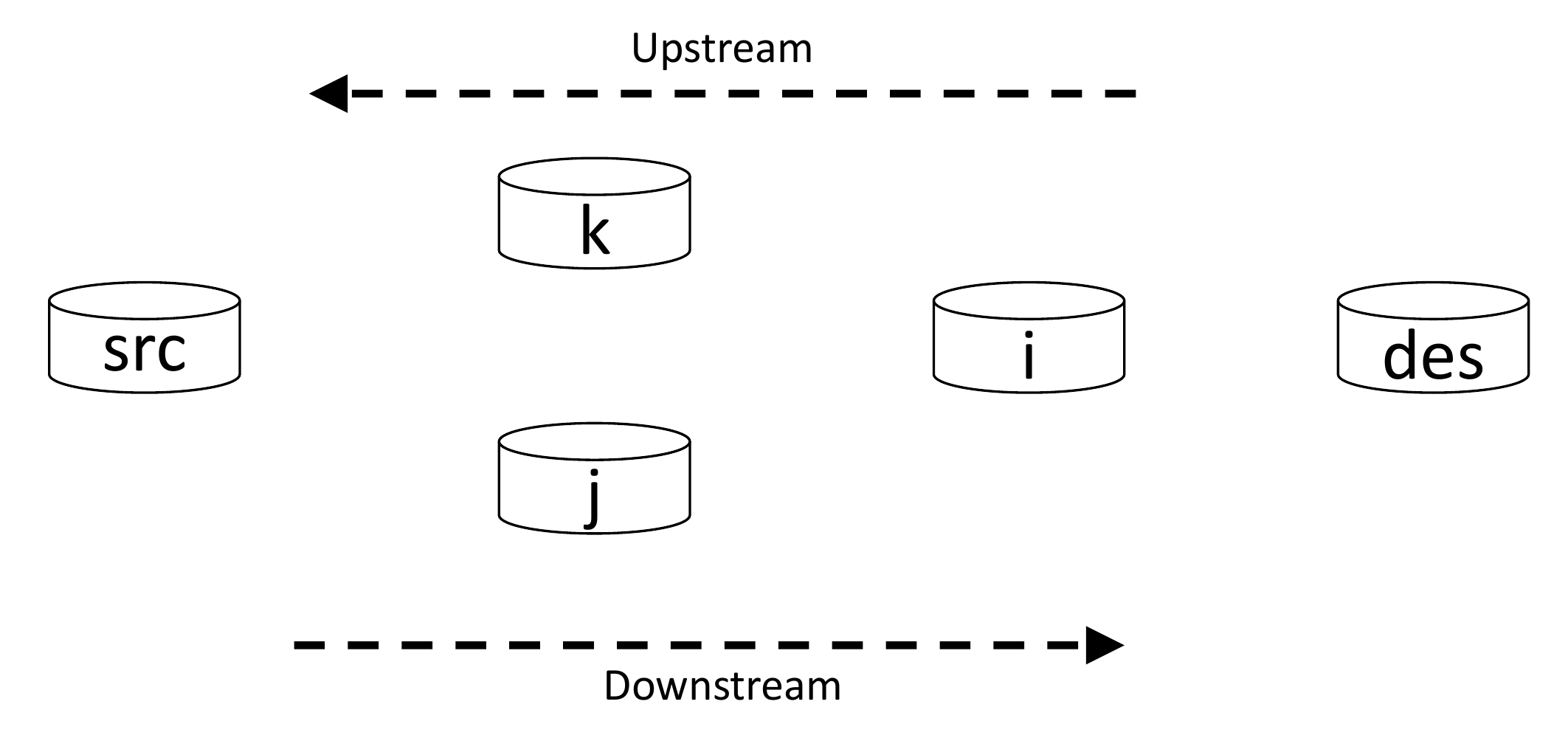}
    \caption{Framework for power control in cognitive network}
    \label{fig:Nodes_Topo}
\end{figure}

In traditional unicast routing, the information used to make route decisions (such as resource reservation information, and Q value) are derived from downstream nodes. In contrast, Q-MAP gathers information from the upstream nodes that is used to make the route selection. A simple topology is depicted in Figure \ref{fig:Nodes_Topo} where $src$ is the source node and $des$ is one of the destinations. In this example, node $i$ needs to choose between $j$ and $k$ as the upstream node. Let us consider node $i$ received a \ac{JQP} from nodes $j$ and $k$. Accordingly, node $i$ computes its reinforcement Q-function and resource reservation data. Therefore, in this case, node $i$ updates $Q(i,ux)$ for any such \ac{JQP} received from any upstream neighbor $ux$ (which in this example is $j$ and $k$) as follows,

\begin{equation}
Q_t(i,ux) \leftarrow (1-\alpha)Q_{t+1}(i,ux)+\alpha[r+\beta Q^*_t(ux) ]   
\end{equation}

Next, node $i$ configures its own \ac{JQP} and floods the packet downstream. Subsequently, when $i$ receives a \ac{JRP} from a downstream node (in this scenario $des$), node $i$ will choose an upstream forwarding node that will eventually become part of the optimal route as follows,

\begin{equation}
    Q^*_t(i)= \max_{ux} Q_t(i,ux) \;\;  \forall \;\; ux \in (i,k)
\end{equation}

Assuming that $j$ is chosen as the forwarding node, in this case, node $i$ creates its \ac{JRP} and floods it. Node $j$ receives this \ac{JRP} and configures itself as the forwarding node for this multicast group by setting its forwarding flag. Each node maintains a forwarding table consisting of a source ID, group ID, forwarding flag indication and a timer field indicating the expiry of the forwarding group. In this manner, the multicast route is selected and maintained by source periodically initiating \ac{JQP}. If any given receiver does not need to receive from a given source node, it just stops sending \ac{JRP} for the corresponding multicast group. In this work, the authors do not discuss any experimental results, rather they keep the design general stating that the reward function is designed based on the objective of the network (maximize throughput, minimize energy consumption, minimize latency, etc.) and accordingly the corresponding resource reservation decision taken at each hop can include bandwidth, power or time slot allocation.

\begin{figure}
    \centering
    \includegraphics[width=4 in]{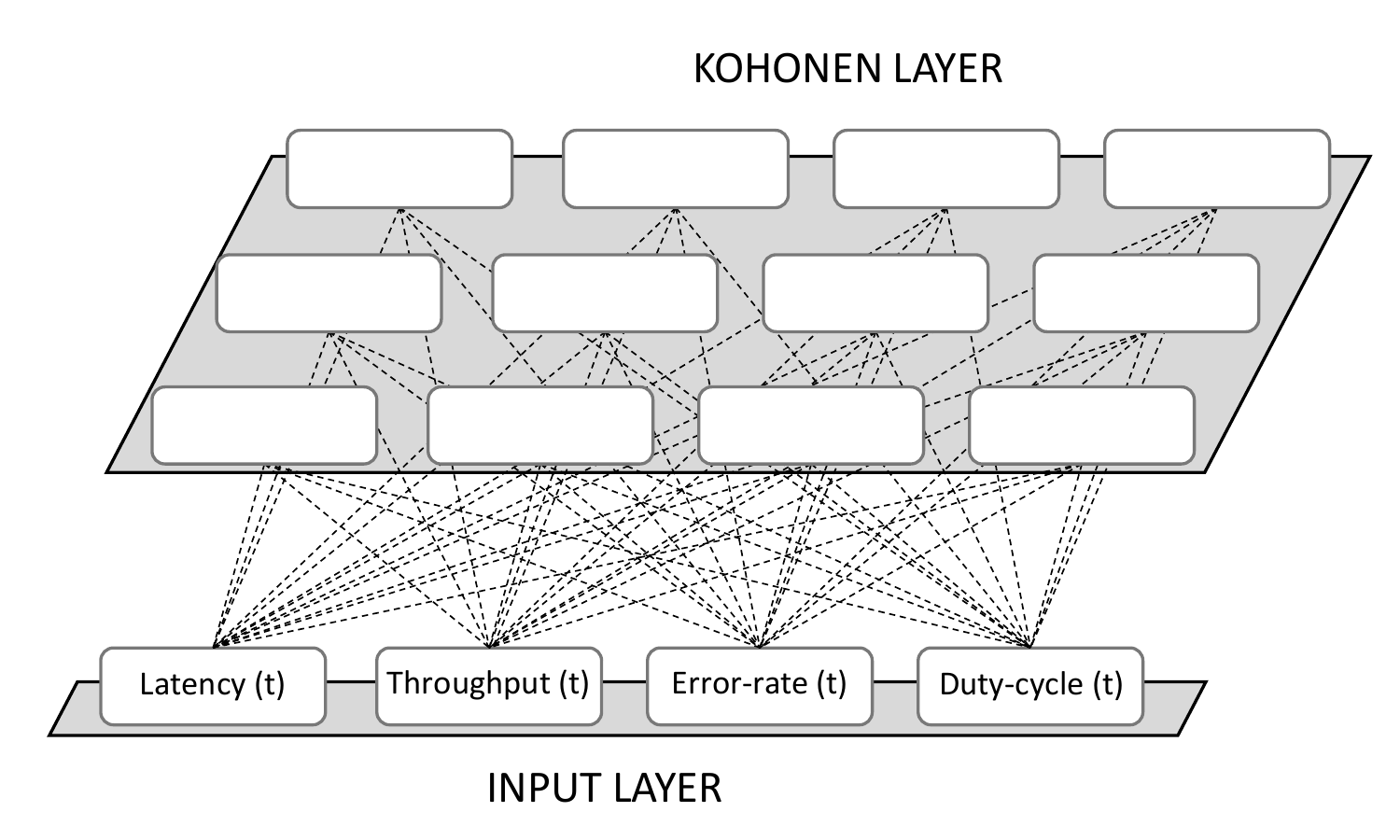}
    \caption{SOM Architecture used in SIR}
    \label{fig:SIR}
\end{figure}

An unsupervised learning based routing referred to as \ac{SIR} is proposed in \cite{Barbancho2007_SOM}. They modify the Dijkstra's algorithm utilizing \ac{SOM}. Consider a directed connectivity graph $\mathcal{G}(\mathcal{K},\mathcal{E})$, where $\mathcal{K}=\{k_0,k_1, ... , k_{N}\}$ is a finite set of nodes, and $(i,j)\in \mathcal{E}$ represents unidirectional wireless link from node $k_i$ to node $k_j$ (for simplicity, they are refer to them as node $i$ and node $j$). Each edge $(i,j)$ has a score associated with it denoted by $\gamma_{ij}$ and it is assumed that $\gamma_{ij}=\gamma_{ji}$ which depends on \ac{QoS} requirements of the network under consideration. In \cite{Barbancho2007_SOM}, the authors use latency, throughput, error-rate, and duty-cycle to represent a measure of \ac{QoS}. Accordingly, the authors use these metrics for each link to represent their input of training vectors for a two-layer \ac{SOM} architecture as shown in Fig \ref{fig:SIR}.  The input layer consists of $l=4$ neurons, for each input vector of $\mathbf{x}(t)\in \mathcal{R}^l$. The second layer consists of a rectangular grid, where each neuron has $l$ weight vectors connected from the input layer. During the learning phase, competitive learning is used such that the neuron whose vector most closely resembles the current input vector dominates. The \ac{SOM} clusters these link by assigning each cluster a \ac{QoS} rating. The learning phase is computationally intensive and hence needs to be performed offline. Meanwhile, execution can run on computational constrained sensor nodes and provide reliable performance as long as the topology and operational characteristics do not change. 

The authors compare the proposed solution to \ac{EAR} \citep{Shah_EAR} and directed diffusion \citep{Directed_Diffusion}. Directed diffusion is a data-centric routing protocol where the sink first broadcasts a request packet. This is used to set up a gradient (weighted reverse link) pointing to the sink (or the source of request). These gradients are used to find paths which are eventually pruned until the optimal path is determined. In the case of \ac{EAR},  the source maintains a set of paths chosen by means of a certain probability that is inversely proportional to the energy consumption of that given route. The goal is to distribute traffic over multiple nodes to improve the network lifetime. Simulations show that the advantage of using \ac{SIR} becomes evident only when nodes in the network start to fail. The parameters used to train \ac{SOM} enable \ac{SIR} to choose paths that are less prone to failure thereby providing better delay performance in scenarios where $40\%$ nodes are prone to failure. 


An example of \ac{RL} in geographical routing  can be seen in \cite{Dong_2007_UWB}. In  \ac{RLGR}, they proposed a distributed algorithm that utilizes residual energy $E_r$ and location of the neighbors. The \ac{MDP} is characterized by the state of the packet which is defined by the current node where the packet resides and the action represents the choice of next-hop based on the Q-value ($Q(s,a)$). In this work, the reward function is given by,

\begin{equation}
 r= \begin{cases}
\alpha \Tilde{\delta} + (1-\alpha) \Tilde{E} & ,\;\; \text{if next hop is not sink} \\ 
R_C & ,\;\; \text{if next hop is the sink} \\
-R_D & ,\;\; \text{if no next hop} \\
-R_E & ,\;\; \text{if next hop available but with low energy} 

\end{cases}   
\end{equation}
where $\Tilde{\delta}$ represents the normalized advance towards the sink, $\Tilde{E}$ is the normalized residual energy. The authors consider a constant reward, $R_C$ if the node is able to reach the sink directly. Finally, both $R_D$ and $R_E$ can be considered as the penalty suffered if no next-hop is found or if the existing next-hop has energy below the threshold. The proposed algorithm also uses $\epsilon$ to indicate the probability of exploration, i.e. how often the node will choose a random neighbor which may not be the next-hop that has the largest Q-value. For all other occasions (probability of $1-\epsilon$), each node chooses a next-hop that provides the maximum Q-value. Their simulations showed significant improvement in network lifetime comparing \ac{RLGR} to \ac{GPSR} \citep{GPSR}.

Next, we look at an example beyond \ac{RF} terrestrial networks. In \acp{UAN},  maximizing network lifetime is a key requirement. Accordingly, \cite{Hu_UAN_Routing} propose a \ac{RL} based approach that aims to distribute traffic among sensors to improve the lifetime of the network. In this work, the system state related to a packet is defined as the node that holds the packet. So $s_i$ denotes the state of a packet held by node $i$. The action taken by node $i$ to forward a packet to node $j$ is denoted as $a_j$. If this action is successful, the state transitions from $s_i$ to $s_j$ with the transition probability of $P^j_{ij}$ and stays in the same state $s_i$ with transition probability of $P^j_{ii}=1-P^j_{ij}$ if it fails. Though these transition probabilities are unknown, authors argue that this can be estimated at runtime based on history. Accordingly, the overall reward function at time instant $t$ can be defined as follows,

\begin{equation}
    r_t= P^j_{ij}R^j_{ij}+P^j_{ii}R^j_{ii} \label{eq:reward}
\end{equation}
where,
\begin{equation}
    R^j_{ij}=-c-\alpha_1 (E_i+E_j)+ \alpha_2 (D_i+D_j)
\end{equation}
where $\alpha_1$ and $\alpha_2$ are tunable weights and $c$ is the constant cost associated with consumption of resource (bandwidth, energy etc.) when a node chooses to transmit. $E_i$ is the cost function associated with residual energy ($E_i^{res}$) and inital energy ($E_i^{ini}$). The energy cost function penalizes the system as residual energy decreases and is defined as,

\begin{equation}
    E_i=1-\frac{E_i^{res}}{E_i^{ini}}
\end{equation}
Similarly, $D_i$ is defined to measure the energy distribution balance as follows,

\begin{equation}
    D_i= \frac{2}{\pi} \arctan(E_i^{res}- \Bar{E_i})
\end{equation}
where $\Bar{E_i}$ is the average residual energy of $i$ and all its direct neighbors. This parameter increases the chance of neighbors with higher residual energy being preferred.

The reward function for the case where a packet forwarding attempt fails is defined as,

\begin{equation}
    R^j_{ii}=-c-\beta_1 E_i+ \beta_2 D_i
\end{equation}
where $\beta_1$ and $\beta_2$ are again tunable weights. Authors use Q-learning at each node to enable them to learn about the environment using control packets and take action to improve network lifetime. The proposed solution is shown to outperform the vector-based forwarding protocol \cite{Peng_VBF}, a geographical routing protocol designed for \acp{UAN} by achieving $20\%$ longer lifetime. The authors claim the proposed solution can be applied for various \ac{UAN} applications by tuning the trade-off between latency and energy efficiency for network lifetime.

\ac{FROMS} \cite{Forster_FROMS_2007} is proposed  to achieve near-optimal routing from multiple source to multiple sink nodes. The goal of each node is to determine neighbor(s) for next-hop(s) towards the intended subset of sinks $SK_p \subset SK$. The state is defined as a tuple, $S = \{SK_p, H^{\mathcal{NB}}_{SK_p}\}$, where $H^{\mathcal{NB}}_{SK_p}$ is the routing information through all neighboring nodes $\mathcal{NB}$. The action is defined by a set $A_t=\{a_1,a_2,...,a_n\}$, such that $a_i=(nb_i, SK_i)$, where $SK_i \subset SK_p$. The complete action set $A$ must ensure that each sink $sk \in SK_p$ must be considered by exactly one element $a_i \in A$. The Q-value here is defined as follows,

\begin{equation}
    Q_t(a)=  \left( \sum_{i=1}^n Q_t(a_i) \right)- (n-1)
\end{equation}
where,

\begin{equation}
    Q_t(a_i)= \left( \sum_{sk \in SK_i} H^{nb_i}_{sk}  \right)- 2(|D_i|-1)
\end{equation}
where $H^{nb_i}_{sk}$ is the number of hops to the intended sink $sk \in SK_i$ through neighbor $nb_i$. $|SK_i|$ denotes the number of sinks in $SK_i$. The goal of the learning process is to decrease the Q-value as much as possible such that nodes pick the action that corresponds to the lowest Q-value. Accordingly, the reward function is defined as follows,
 
 \begin{equation}
R_t(a_i)=C+ \min_a Q(a)     
 \end{equation}
where $C$ is the cost of the action. In this manner, the Q-values propagate upstream facilitating the learning process. During the operational phase, it is assumed that nodes overhear neighbor's packets and use the information contained in the packets to update their Q-value. Eventually, the goal is to use routes that will deliver the packets to the desired subset of sinks through the least number of total hops. The authors also explore both greedy exploration and stochastic exploration techniques to avoid local minima. Simulation results validate the ability to learn shared routes to multiple sinks in an efficient manner to decrease the cost per packet compared to directed diffusion \citep{Silva_directedD}. Additionally, they show how exploration can further reduce the cost per packet albeit marginally. We summarize these routing algorithms and the \ac{ML} techniques they apply in Table \ref{tab:Route}.

 \begin{table}[h]
    \centering
    \caption{Application of \ac{ML} in routing protocols\label{tab:Route}}%
    \begin{tcolorbox}[tab2,tabularx={|p{4.8cm}||p{2.8cm}|X}]
      \textbf{Routing Protocol}   & \textbf{ML Algorithm}  & \textbf{Objective/Comments}\\ \hline\hline
      
      Boyan and Littman \cite{Boyan_1993_RL-Route}   & \ac{RL} &  Variation of Bellman-Ford proposed for wired network   \\ \hline
      Mao et al. \cite{Mao_routing}           & \ac{DBN}& Outperform \ac{OSPF} due to reduced overhead \\ \hline
      Sun et al. \cite{Sun_2002_QMAP}   & \ac{RL} & Multicast Routing Algorithm \\ \hline
      Pourfakhar and Rahmani \cite{POURFAKHAR_CMAC}   & \ac{CMAC} & Proposed to improve reliability by predicting disconnection probabilities  \\\hline
      Barbancho et al. \cite{Barbancho2007_SOM}  & \ac{SOM} & Modified version of Dijkstra's Algorithm  \\\hline
      Dong et al. \cite{Dong_2007_UWB}   & \ac{RL} & Energy efficient geographical routing   \\\hline
      Hu and Fei \cite{Hu_UAN_Routing} & \ac{RL} & Liftime-aware routing for \ac{UAN} that aims to distribute traffic load among nodes \\\hline
      Forster and Murphy \cite{Forster_FROMS_2007}   & \ac{RF} & Near-Optimal routing for multiple source to multiple sinks  \\ \hline
     
    \end{tcolorbox}
\end{table}

\subsection{Open Problems and Challenges}

In this section, we discuss some of the challenges and open problems specifically at the network and data-link layer in the context of \ac{IoT}.

\subsubsection{Scalability and Distributed Operation}

The exponentially increasing number of \ac{IoT} devices demand a scalable networking architecture to enable large scale interactions especially in the context of wireless communication. The spectrum congestion will imply more competition for limited resources. While cross-layer approaches \cite{Hasan2018AnalysisOC, Xu_crosslayer, callebaut2018cross} have been studied in the context of \ac{IoT} to enable interaction between layers and optimize the utilization of resources, the dimension of the optimization problem space is increasing drastically. This is due to the explosion in operational states (channel, residual energy, traffic level, level of \ac{QoS}, the density of the neighborhood, the priority of the entity, among others) that must be considered during decision making. This challenge is further exacerbated when a distributed operation is required to reduce the overhead and ensure scalability. In these circumstances, novel \ac{ML} approaches including \ac{DRL} needs to be explored in conjunction with network optimization techniques \cite{Minsoo_X_ML, Ajagan_x_ML}. 

\subsubsection{\ac{IoT} Data-link and Network layer Security}

Another key aspect that needs attention at the data link and network layer of the wireless \ac{IoT} network is the security threat due to various kinds of attacks \cite{Abdul-Ghani2018}. In networks like the one established by ZigBee devices, the attacker could eavesdrop and redirect traffic, launching what is known as man-in-the-middle attack \cite{Markert_ZB_attack}. In this attack, the attackers can reduce the performance of the network or even intercept and change the transmitted data. Energy efficiency is a key performance parameter of \ac{IoT} networks. Keep--Awake attack can be used to drain the battery of \ac{IoT} devices by sending control packets that constantly revive \ac{IoT} devices from their dormant sleep cycles \cite{Vidgren_ZB_sleep}. Other attacks at the network layer include selective forwarding and sinkhole (black hole) attack \cite{Mayzaud2016ATO, Pongle}. In black hole or sinkhole attack, an attacker’s node broadcasts more favorable routes attracting all traffic towards it. Due to the enormous amount of traffic handled by the \ac{IoT} network it might be challenging to identify such attacks in an efficient and effective manner. The inherent ability of \ac{ML} to use the “big data”  to its advantage can be exploited to explore solutions for these security concerns in \ac{IoT} networks.

\section{Spectrum Sensing and Hardware Implementation}\label{sec:Hardware}


One of the key challenges in enabling real-time inference from spectrum data is how to \textit{effectively} and \textit{efficiently} extract \textit{meaningful}  and \textit{actionable} knowledge out of the tens of millions of \ac{I/Q} samples received every second by wireless devices. Indeed, a single $\mathrm{20~MHz}$-wide WiFi channel generates an \ac{I/Q} stream rate of about $\mathrm{1.28~Gbit/s}$, if \ac{I/Q} samples are each stored in a 4-byte word. Moreover, the \ac{RF} channel is significantly time-varying (\textit{i.e.}, in the order of milliseconds), which imposes strict timing constraints on the \textit{validity} of the extracted \ac{RF} knowledge. If (for example) the \ac{RF} channel changes every 10ms, a knowledge extraction algorithm must run with latency (much) less than 10ms to both (i) offer an accurate RF prediction and (ii) drive an appropriate physical-layer response; for example, change in modulation/coding/beamforming vectors due to adverse channel conditions, \ac{LO} frequency due to spectrum reuse, and so on. 

 As discussed earlier, \ac{DL} has been a prominent technology of choice for solving classification problems for which no well-defined mathematical model exists. It enables the analysis of unprocessed \ac{I/Q} samples without the need of application-specific and computational-expensive feature extraction and selection algorithms \citep{OShea-ieeejstsp2018}, thus going far beyond traditional low-dimensional \ac{ML} techniques. Furthermore, \ac{DL} architectures are application-insensitive, meaning that the same architecture can be retrained for different learning problems.
 
Decision-making at the physical layer may leverage the spectrum knowledge provided by \ac{DL}. On the other hand, RF \ac{DL} algorithms must execute in \emph{real-time} (\textit{i.e.}, with static, known-a-priori latency) to achieve this goal. Traditional \ac{CPU}-based knowledge extraction algorithms \citep{abadi2016tensorflow} are unable to meet strict time constraints, as general-purpose \acp{CPU} can be interrupted at-will by concurrent processes and thus introduce additional latency to the computation. Moreover, transferring data to the \ac{CPU} from the radio interface introduces unacceptable latency for the RF domain. Finally, processing \ac{I/Q} rates in the order of $\mathrm{Gbit/s}$ would require \acp{CPU} to run continuously at maximum speed, and thus consume enormous amounts of energy.  For these reasons, RF \ac{DL} algorithms must be closely integrated into the RF signal processing chain of the embedded device.

\subsection{Existing work} 

Most of existing work is based on traditional low-dimensional machine learning \citep{Wong-isspa2001,Xu-ieeetvt2010,Pawar-ieeetifs2011,Shi-ieeetcomm2012,Ghodeshar-icscn2015}, which requires (i) extraction and careful selection of complex features from the RF waveform (\textit{i.e.}, average, median, kurtosis, skewness, high-order cyclic moments, etc.); and (ii) the establishment of tight decision bounds between classes based on the current application, which are derived either from mathematical analysis or by learning a carefully crafted dataset \citep{shalev2014understanding}. In other words, since feature-based machine learning is (a) significantly application-specific in nature; and (b) it introduces additional latency and computational burden due to feature extraction, its application to real-time hardware-based wireless spectrum analysis becomes impractical, as the wireless radio hardware should be changed according to the specific application under consideration.

Recent advances in \ac{DL} \citep{lecun2015deep} have prompted researchers to investigate whether similar techniques can be used to analyze the sheer complexity of the wireless spectrum. For a compendium of existing research on the topic, the reader can refer to \cite{Mao-ieeecomm2018}. Among other advantages, \ac{DL} is significantly amenable to be used for real-time hardware-based spectrum analysis, since different model architectures can be reused to different problems as long as weights and hyper-parameters can be changed through software. Additionally, \ac{DL} solutions to the physical layer modulation recognition task have been given much attention over recent years, as previously discussed in this work. The core issue with existing approaches is that they leverage \ac{DL} to perform offline spectrum analysis only. On the other hand, the opportunity of real-time hardware-based spectrum knowledge inference remains substantially uninvestigated.

\subsection{Background on System-on-Chip Computer Architecture}\label{sec:soc}

Due to its several advantages, we contend that one of the most appropriate computing platform for RF \ac{DL} is a \ac{SoC}. An \ac{SoC} is an integrated circuit (also known as ``IC" or ``chip") that integrates all the components of a computer, \textit{i.e.}, \ac{CPU}, \ac{RAM}, input/output (I/O) ports and secondary storage (\textit{e.g.}, SD card) -- all on a single substrate \citep{Molanes-ieeetie2018}. \acp{SoC}  have low power consumption \citep{SoCLowPower} and allow the design and implementation of \textit{customized hardware} on the \ac{FPGA} portion of the chip, also called \ac{PL}. Furthermore, \acp{SoC} bring unparalleled  flexibility, as the PL can be reprogrammed at-will according to the desired learning design. The \ac{PL} portion of the \ac{SoC} can be managed by the \ac{PS}, \textit{i.e.}, the \ac{CPU}, \ac{RAM}, and associated buses. 

SoCs use the \ac{AXI} bus specification \citep{XilinxAXI} to exchange data (i) between functional blocks inside the \ac{PL}; and (ii) between the \ac{PS} and \ac{PL}. There are three main \ac{AXI} sub-specifications: \textit{AXI-Lite}, \textit{AXI-Stream} and \textit{AXI-Full}. AXI-Lite is a lightweight, low-speed \ac{AXI} protocol for register access, and it is used to configure the circuits inside the PL. AXI-Stream is used to transport data between circuits inside the \ac{PL}.  AXI-Stream is widely used, since it provides (i) standard inter-block interfaces; and (ii) rate-insensitive design, since all the AXI-Stream interfaces share the same bus clock, the \ac{HLS} design tool will handle the handshake between \ac{DL} layers and insert \acp{FIFO} for buffering incoming/outgoing samples. AXI-Full is used to enable burst-based data transfer from \ac{PL} to \ac{PS} (and \textit{vice versa}). Along with AXI-Full, \ac{DMA} is usually used to allow PL circuits to read/write data obtained through AXI-Stream to the RAM residing in the \ac{PS}. The use of \ac{DMA} is crucial since the CPU would be fully occupied for the entire duration of the read/write operation, and thus unavailable to perform other work.

\subsection{A Design Framework for Real-time RF Deep Learning}\label{sec:dlcore}

One of the fundamental challenges to be addressed is how to transition from a software-based \ac{DL} implementation (\textit{e.g.}, developed with the Tensorflow \cite{abadi2016tensorflow} engine) to a hardware-based implementation on an \ac{SoC}. Basic notions of high-level synthesis and a hardware design framework are presented in Sections \ref{sec:hls} and \ref{sec:design_steps}, respectively.

\subsubsection{High-level Synthesis}\label{sec:hls}

\ac{HLS} is an automated design process that interprets an algorithmic description of a desired behavior (\textit{e.g.}, C/C++) and creates a model written in \ac{HDL} that can be executed by the \ac{FPGA} and implements the desired behavior \citep{winterstein2013high}. Designing digital circuits using \ac{HLS} has several advantages over traditional approaches. First, \ac{HLS} programming models can implement almost any algorithm written in C/C++. This allows the developer to spend less time on the \ac{HDL} code and focus on the algorithmic portion of the design, and at the same time avoid bugs and increase efficiency, since \ac{HLS} optimizes the circuit according to the system specifications. 
The clock speed of today's \acp{FPGA} is several orders of magnitude slower than \acp{CPU} (\textit{i.e.}, up to 200-300 MHz in the very best FPGAs). Thus, parallelizing the circuit's operations is crucial. In traditional \ac{HDL}, transforming the signal processing algorithms to fit \ac{FPGA}'s parallel architecture requires challenging programming efforts. On the other hand, an \ac{HLS} toolchain can tell how many cycles are needed for a circuit to generate all the outputs for a given input size, given a target parallelization level. This helps to reach the best trade-off between hardware complexity and latency.

\textit{Loop Pipelining:}  In high-level languages (such as C/C++), the operations in a loop are executed sequentially and the next iteration of the loop can only begin when the last operation in the current loop iteration is complete. Loop pipelining allows the operations in a loop to be implemented in a concurrent manner.

\begin{figure}[!h]
    \centering
    
\begin{lstlisting}
for (int i=0; i<2;i++) {
    Op_Read;    /* RD */
    Op_Execute; /* EX */
    Op_Write;   /* WR */
}
\end{lstlisting}
    \includegraphics[width=\textwidth]{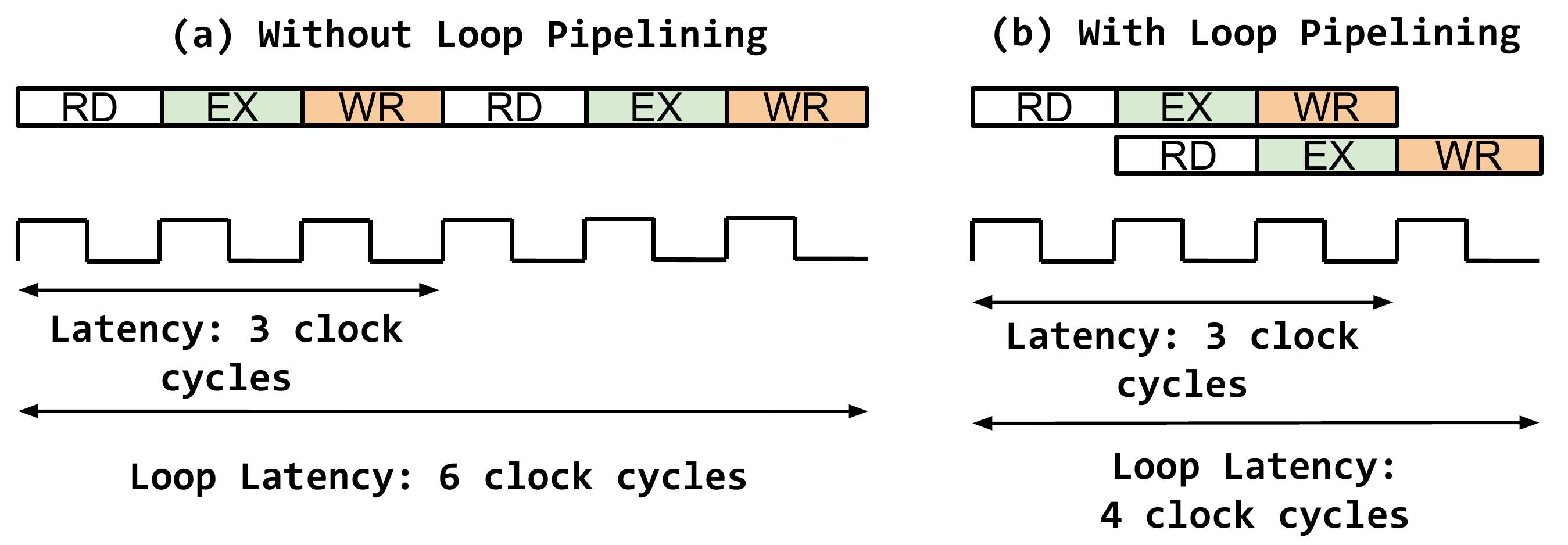}
    \caption{Loop pipelining.\vspace{-0.2cm}}
    \label{fig:pipelining}
\end{figure}

Figure \ref{fig:pipelining} shows an example of loop pipelining, where a simple loop of three operations, \textit{i.e.}, read (RD), execute (EX), and write (WR), is executed twice. For simplicity, we assume that each operation takes one clock cycle to complete. Without loop pipelining, the loop would take 6 clock cycles to complete. Conversely, with loop pipelining, the next RD operation is executed concurrently to the EX operation in the first loop iteration. This brings the total loop latency to $4$ clock cycles. If the loop length were to increase to $100$, then the latency decrease would be even more evident: $300$ versus $103$ clock cycles, corresponding to a speedup of about $65\%$. An important term for loop pipelining is called \ac{II}, which is the number of clock cycles between the start times of consecutive loop iterations. In the example of Figure \ref{fig:pipelining}, the \ac{II} is equal to one, because there is only one clock cycle between the start times of consecutive loop iterations. 
 
\textit{Loop Unrolling:} Loop unrolling creates multiple copies of the loop body and adjusts the loop iteration counter accordingly. For example, if a loop is processed with an \ac{UF} equal to $2$ (\textit{i.e.}, two subsequent operations in the same clock cycle as shown in Figure \ref{fig:unrolling}), it may reduce a loop's latency by a factor of $50\%$, since a loop will execute in half the iterations usually needed. Higher \ac{UF} and II may help achieve low latency, but at the cost of higher hardware resource consumption. Thus, the trade-off between latency and hardware consumption should be thoroughly explored.

\begin{figure}[!h]
    \centering
    \begin{lstlisting}%[caption={No loop unrolling}]
int sum = 0;
for(int i = 0; i < 10; i++) {
    sum += a[i];
}
\end{lstlisting}
\begin{lstlisting}%[caption={Loop unrolling, factor = 2}]
int sum = 0;
for(int i = 0; i < 10; i+=2) {
    sum += a[i];
    sum += a[i+1];
}
\end{lstlisting}
    \caption{Loop unrolling.}
    \label{fig:unrolling}
\end{figure}

\subsubsection{Design Steps}\label{sec:design_steps}

\begin{figure}[!h]
    \centering
    \includegraphics[width=\textwidth]{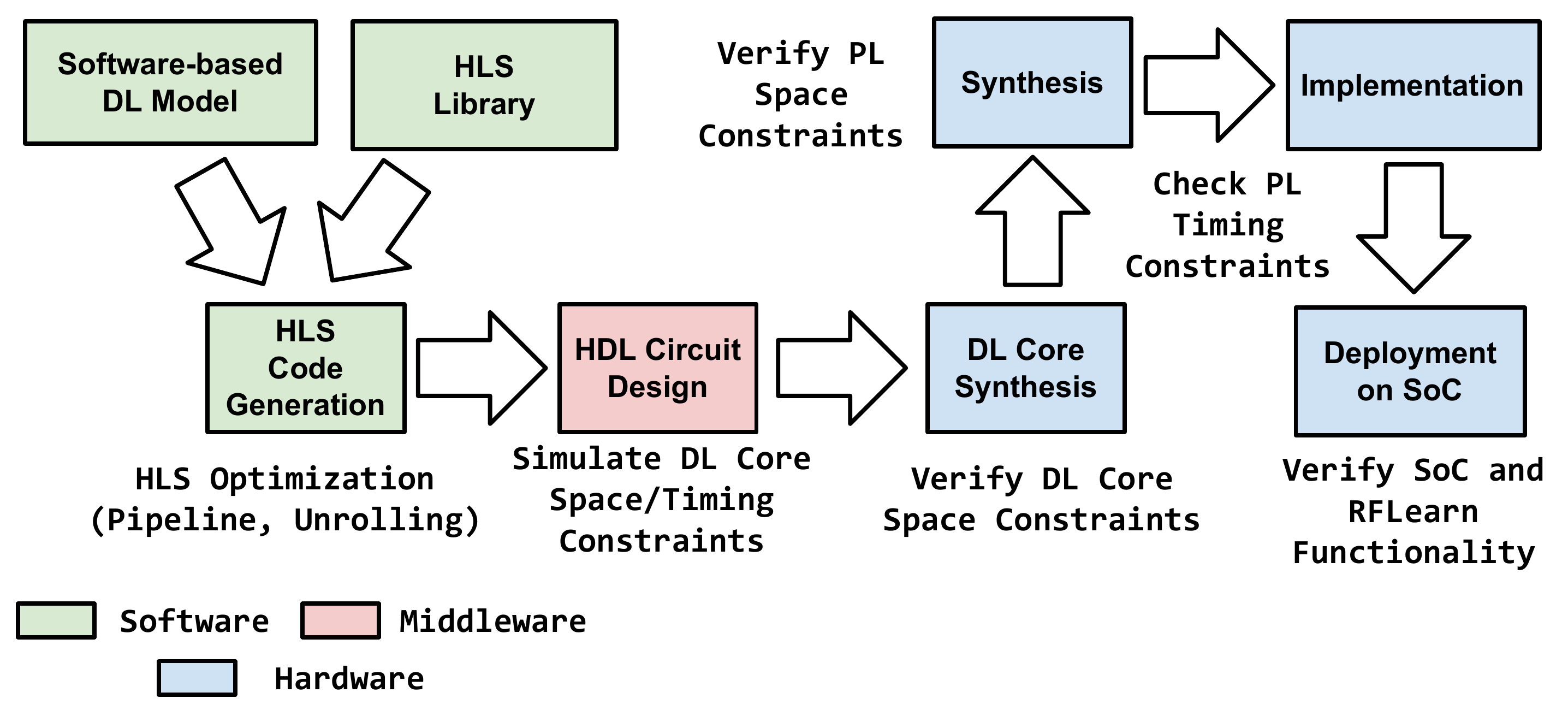}
    \caption{A Hardware Design Framework for RF Deep Learning.}
    \label{fig:rflearn_frame}
    \vspace{-0.3cm}
\end{figure}

Our framework presents several design and development steps, which are illustrated in Figure \ref{fig:rflearn_frame}. Steps that involve hardware, middleware (\textit{i.e.}, hardware description logic, or \ac{HDL}), and software have been depicted with a blue, red, and green shade, respectively. 

The first major step of the framework is to take an existing \ac{DL} model and convert the model in \ac{HLS} language, so it can be optimized and later on synthesized in hardware. Another critical challenge is how to make the hardware implementation fully reconfigurable, \textit{i.e.}, the weights of the \ac{DL} model may need to be changed by the \textit{Controller} according to the specific training. To address these issues, we distinguish between (i) the \ac{DL} model architecture, which is the set of layers and hyper-parameters that compose the model itself, and (ii) the parameters of each layer, \textit{i.e.}, the neurons' and filters' weights. 

To generate the \ac{HLS} code describing the software-based \ac{DL} model, an \emph{\ac{HLS} Library}, which provides a set of \ac{HLS} functions that parse the software-based \ac{DL} model architecture and generates the \ac{HLS} design corresponding to the desired architecture. The \textit{\ac{HLS} Library} supports the generation of convolutional, fully-connected, rectified linear unit, and pooling layers, and operated on fixed-point arithmetic for better latency and hardware resource consumption. The \ac{HLS} code is subsequently translated to \ac{HDL} code by an automated tool that takes into account optimization directives such as loop pipelining and loop unrolling. At this stage, the \ac{HDL} describing the \ac{DL} core can be simulated to (i) calculate the amount of \ac{PL} resources consumed by the circuit (\textit{i.e.}, flip-flops, BRAM blocks, etc); and (ii) estimate the circuit latency in terms of clock cycles. After a compromise between space and latency as dictated by the application has been found, the DC core can be synthesized and integrated with the other \ac{PL} components, and thus total space constraints can be verified. After implementation (\textit{i.e.}, placing/routing), the \ac{PL} timing constraints can be verified, and finally the whole system can be deployed on the \ac{SoC} and its functionality tested.

\subsection{Open Problems and Challenges}\label{Sec:Problems}

In this section, we discuss a set of open challenges overcoming which will accelerate the induction of \ac{ML} techniques to the \ac{IoT} hardware especially in the context of spectrum sensing.  

\subsubsection{Lack of Large-scale Wireless Signal Datasets}

It is well known that learning algorithms require a considerable amount of data to be able to effectively learn from a training dataset. Moreover, to compare the performance of different learning models and algorithms, it is imperative to use the same sets of data. More mature learning fields, such as computer vision and \ac{NLP} already have standardized datasets for these purposes \citep{deng2012mnist,deng2009imagenet}. However, literature still lacks large-scale datasets for RF \ac{ML}. 

This is not without a reason. Although the wireless domain allows the synthetic generation of signals having the desired characteristics (\textit{e.g.}, modulation, frequency content, and so on), problems such as \ac{RF} fingerprinting and jamming detection require data that captures the unique characteristics of devices and wireless channels. Therefore, significant research effort must be put forth to build large-scale wireless signal datasets to be shared with the research community at large.

\subsubsection{Choice of I/Q Data Representation Format}

It is still subject of debate within the research community what is the best data representation for RF deep learning applications. For example, an I/Q sample can be represented as a tuple of real numbers or a single complex number, while a set of I/Q samples can be represented as a matrix or a single set of numbers represented as a string. It is a common belief that there is no one-size-fits-all data representation solution for every learning problem, and that the right format might depend, among others, on the learning objective, choice of the loss function, and the learning problem considered \citep{OShea-ieeejstsp2018}. 

\subsubsection{Choice of Learning Model and Architecture}

While there is a direct connection between images and tensors, the same cannot be concluded for wireless signals. For example, while 3-D tensors have been proven to effectively model images (\textit{i.e.}, red, green, and blue channels), and kernels in convolutional layers are demonstrably powerful tools to detect edges and contours in a given image, it is still unclear if and how these concepts can be applied to wireless signals. Another major difference is that, while images can be considered as stationary data, RF signals are inherently stochastic, non-stationary and time-varying. This peculiar aspect poses significant issues in determining the right learning strategy in the wireless \ac{RF} domain. For example, while \ac{CNN} seems to be able to effective at solving problems such as modulation recognition \citep{West-dyspan2017,Karra-ieeedyspan2017,OShea-ieeejstsp2018}, it is still unclear if this is the case for complex problems such as \ac{RF} fingerprinting.  Moreover,  \ac{DL} has traditionally been used in static contexts \citep{krizhevsky2012imagenet,hinton2012deep}, where the model latency is usually not a concern. Another fundamental issue absent in traditional deep learning is the need to satisfy strict constraints on resource consumption. Indeed, models with a high number of neurons/layers/parameters will necessarily require additional hardware and energy consumption, which are clearly scarce resources in embedded systems. Particular care must be devoted, therefore, when designing learning architectures to solve learning problems in the RF domain.

\section{Machine Learning in IoT beyond Communication}\label{sec:Beyond_Comm}

The core objective of this work is to provide a comprehensive account of the applications of \ac{ML} for communication in \ac{IoT}. In this section, for the benefit of readers who might be exploring \ac{ML} for communication in conjunction with other \ac{IoT} related areas of research, we provide a brief discussion on how \ac{ML} has been applied to areas like security (beyond communication surfaces) and big data analysis. This is not intended to be as comprehensive as the earlier sections of this survey but provides the adequate resources for readers to understand the broad nature of ML being applied in these areas by pointing them to the relevant resources. 

\subsection{Security in IoT}

Due to the complex and integrative arrangement of IoT devices, it can be prone to a wide range of attacks. Limited computation and power resources, a wide range of accessibility, and a large amount of data being handled leads to challenging circumstances to defend IoT devices from security threats. The interdependent and interconnected environment in which IoT devices operate leads to vast numbers of attack surfaces to monitor and manage.

ML has been leveraged as a powerful tool that can monitor the vast number of IoT devices to detect and alert operators of imminent security threats \cite{Security_Survey}. One of the key security concern in an IoT network is the presence of intruders that may induce malicious behavior. Several of the \ac{ML} techniques have been used to detect these forms of attacks. In one of the earliest works \cite{Hu03robustsupport}, the author proposed a robust \ac{SVM}-based solution to intruder detection. This involved analyzing 1998 \ac{DARPA} Basic Security Module data set collected at MIT’s Lincoln Labs. Recently, a \ac{SVM}-based hybrid detection method that integrates the misuse detection model and an anomaly detection model has been proposed in \cite{KIM_RSVM}. The solution is shown to be computationally efficient, and capable of providing better detection rate for both known and unknown attacks while maintaining a low probability of false alarm. \ac{RNN}, specifically \ac{LSTM} have been proposed as an effective tool to detect malicious activity \cite{Torres_RNN} especially for time-series based threats. 

Since \ac{IoT} devices are often connected to Android-based mobile devices in order to enable remote control and configuration, there has been a growth in malware developers. Malware enables developers to control compromised devices to extract private user information or constructing botnets. Several \ac{ML} techniques have been applied to detect such malware attack. Few examples include a \ac{SVM}-based malware detection to ensure reliable \ac{IoT} services \cite{Ham_Malware}, malware detection using \ac{CNN} \cite{CNN_Malware} and an autoencoder-based approach \cite{Azar_AE}. 

\subsection{Big Data Analytics} 

The large amount of data generated and/or flowing through IoT devices have been referred to as smart data \cite{MAHDAVINEJAD_Survey, Sheth} and have been used to feed various \ac{ML} tools to enable several applications in traffic, energy management, health, environment, homes, agriculture, among others. The analysis of data can happen data centers (cloud computing) \cite{cloud}, edge devices (edge computing) \cite{Edge} or edge servers (fog computing) \cite{fog} based on the computation requirement, acceptable latency, among other factors.

To identify regular traffic patterns, authors of \cite{Ma_transit} employs \ac{DBSCAN} algorithm to analyze various trips using the operator's smart card to detect regular travel patterns and then use K-Means algorithm to classify these travel patterns. This information can then be utilized for city planning and identify the optimal use of budget to add critical infrastructure. In \cite{Derguech_predict}, an example of using IoT data in predictive analysis for enhanced decision making has been provided. In this particular example, the authors' goal was to predict energy usage of a building using four \ac{ML} models used by the WEKA data mining software \cite{WEKA} which includes \ac{SVM} for regression, two \ac{ANN} architectures, and linear regression. 

Another example of \ac{ML} being applied to classify big data is provided in \cite{Khan2014ANL}. Here, the authors provide a hybrid (unsupervised and supervised learning) solution to classify the multi-variate time series sensor data that includes environmental variables viz. temperature, humidity, light, and voltage. The authors first apply \ac{SAX} representation to the data in order to reduce its dimensions. Next, clustering techniques are applied to learn the target classes and \ac{SVM} was used thereafter to perform classification. 

Management of a large number of IoT devices is also becoming a challenging task taking into consideration the limited resources each of these devices house. Operational indicators of IoT devices that represent the reliability, \ac{QoS}, productivity, etc. are received from the management protocols. The set of these values are referred to as the state of the \ac{IoT} device in \cite{State_predict}. To enable better management and mitigate the problems arising from inefficient use of limited resources, the authors propose an ANN-based framework that enables prediction of IoT device state enabling higher efficiency in their decision making process for a wide variety of applications. These are just a subset of applications where data analytics has been exploited using \ac{ML}. Big data analytics will also find its application in health care, education, smart grid as well as other components forming a smart city.    

\subsection{Open Problems and challenges}

\subsubsection{Data Analytics}

The quality of data is a key factor affecting the efficacy of the \ac{ML} techniques applied for data analytics. The quality of sensors, the environmental condition, protocols and hardware employed along with several other factors may affect the quality of the data generated by IoT devices. This along with the fact that this data is produced in high volume, high velocity and its nature vary based on devices, application, and protocols used by these devices. It becomes an extremely challenging task to assess the quality of incoming data. The computational load required to analyze the data for quality, pre-process to enhance the data and subsequently perform application-specific data analysis in real-time will continue to be a daunting problem as the IoT revolution grows exponentially.

Beyond the quality and computational requirement of handling the data, the overarching legal and ethical concerns of handling data emanating from various IoT devices will also have to be explored in greater depth. It will be challenging to reach consensus in defining the optimal procedure/methodology of handling critical data regarding health, law enforcement or national security among others. The same data that is collected in a different context may directly impact is accessibility and sensitivity. In certain cases, the location where the data is stored (data centers) for computation can be critical to the application of the agency that generates the data. This may induce further constraints on the computational requirements. Since there is a never-ending struggle between the need of applying a centralized form of secure data handling while requiring more scalable, distributed, and low overhead operations, there will always be open challenges to determine the Pareto-optimal solutions to handle the vast amount of IoT data.

\subsubsection{Security}

Most \ac{ML} techniques employed to enhance the security of IoT that rely on supervised learning are predominantly trained using simulated or emulated data. This is due to the fact that it is very challenging to gather training data that has been obtained during real-world attack scenarios. An important research direction is to cooperatively obtain crowd-sourced data set from IoT deployed by different commercial, government and academic entities. This again will be a challenge due to the privacy, propriety and other regulatory, proprietary concerns discussed in the earlier section. Assuming this will be a difficult task in the near future, significant research will be required to design ML techniques that can provide adequate real-world protection even when trained on emulated/simulated data sets.    

The next-generation of ML-based solution needs to be able to adapt to the ever-changing landscape of the attack methodologies. Signature-based malware detection may be unable to detect zero-day attack or malware that evolve continuously as in case of metamorphic and polymorphic malware. A new emerging threat that was previously unknown to the malware detector is referred to as Zero-day attack. Though there have been recent efforts to tackle these problems \cite{Comar2013CombiningSA, Suthaharan}, there is a significant opportunity to employ \ac{ML} to mitigate or eradicate the damages caused by ever-evolving security threats.

\section{Conclusion}\label{sec:conclusion}

This paper provides a comprehensive account of advances in IoT wireless communication made possible by the application of \ac{ML}. To accomplish this, we first provide readers with a detailed overview of some of the most prevalent \ac{ML} techniques that are employed in wireless communication networks. We have done so with the hope that by elucidating the inner workings of some of the \ac{ML} algorithms relevant to communication in the IoT, we have not only enabled the reader to understand the subsequent text at a deeper level but inspired other researchers to apply the techniques discussed to their own problems in IoT communication. To a lesser degree, we have written the overview with the intent of providing a light foray into \ac{ML} for the unfamiliar reader. While it is not an all-encompassing field guide to \ac{ML}, the overview covers many of the popular algorithms from the different sub-fields of \ac{ML} and aims to provide an intuition surrounding their use. 

Next, we presented an overview of the current state-of-the-art of \ac{IoT} communication, the standardization efforts, challenges and how \ac{CR} aspect along with \ac{ML} approaches are exploited to address some of these. \ac{CR} along with \ac{ML} is a powerful tool that can take the \ac{IoT} technologies a step forward in mitigating the myriad problems that arise from large deployments.  We provided a glimpse of the subset of works proposed in realizing the \ac{IoT} vision for the foreseeable future dense, large scale \ac{IoT} deployment. The  COGNICOM+ framework is inspiring but has a long development road ahead to realize the plethora of approaches presented from designing \ac{ASIC}-based \ac{CNN} accelerators to developing the fully realized COGNICOM+. Recent works have taken algorithmic designs from simulations to real testbed implementations. More such works are essential to realize the challenges and pave way for future \ac{CR-IoT}. However, the scalability of such a centralized solution might be challenging for a large and dense deployment. The big data analytics and management will need to be addressed for such centralized approaches when applied to dense deployment. Instilling \ac{ML} techniques for future \ac{CR-IoT} enables intelligent resource management such as radio resource optimization via intelligent beamforming, channel equalization, adaptive power and rate control, spectrum allocation and management. Conventional techniques involve optimization techniques performed in an offline/semi-offline manner but ML enables such optimization to be performed in an online fashion in real-time. \ac{ML} approaches continue to learn and adapt to the varying parameters improving the cognition of the system. Such intelligent online decision making will best fit the future \ac{CR-IoT}.

Following the discussion of the application of \ac{ML} to problems in the physical layer, we introduce the use of \ac{ML} techniques for signal intelligence tasks in the realm of the IoT. We describe how \ac{ML}, and often \acp{DNN}, can be used to enhance the efficacy of the discriminative classification tasks of \ac{AMC} and wireless interference classification. The common narrative underlying the presentation of these tasks and their respective solutions is that hand-crafted feature-based classifiers of old are outperformed by their \ac{DNN} counterparts. Not only do the \ac{ML} and \ac{DL} solutions presented in this section improve upon classification accuracies, but they also allow for a model to be learned directly on the raw signal representation. The advantages of such a result are two-fold. First, learning a model that operates directly on the raw signal reduces the need for preprocessing of the data, in turn reducing latency and computational load, both of which often have stringent constraints in IoT networks. Second, the use of hand-crafted signal features limits the model's ability to adapt to new input, thus reducing the applicability of the learned model to new data sets. The raw signal representation is the most information-rich representation of the signal and thus reducing it to a set of hand-crafted features reduces the information content. The crux of \ac{DL} is to allow the algorithm to determine what aspects of and interactions between the data are important for a given task, and thus providing the algorithm with more information (raw signal) allows for a more versatile model. This is important with respect to the IoT as the wireless networks, communication protocols, and RF signals that arise in the IoT are not uniform, placing a premium on solutions that are easily adaptable to new scenarios and problem formulations. Such is the reason motivating the use of \ac{ML} in signal intelligence problems within the IoT.

Thereafter, we detail the increasing relevance of these techniques in the higher layers of the protocol stack enabling optimized utilization of limited resources which will be key to support the rapid growth of IoT devices. Deploying a dense \ac{IoT} network may rely on \ac{TDMA} to broadcast information to each other. In these scenarios, the \ac{BSP} is essentially a \ac{TDMA} cycle minimization problem which is known to be NP-complete. In this work, we have seen how \ac{ML} techniques have been successfully applied to these NP-complete problems which otherwise is challenging to overcome. While some of the solutions designed to overcome \ac{BSP} provided acceptable results they unfortunately required long computational time to reach the solution. By Applying \ac{FHNN} to solve \ac{BSP}, the problem was formulated as one that aims at minimizing the energy function associated with \ac{FHNN}. This approach outperformed the existing methodologies in terms of convergence rate. This is one example where \ac{ML} is applied to an intractable problem of wireless communication which in this case was to determine the non-conflicting transmission schedule that maximizes the utilization of the channel. Another key application of \ac{ML} during medium access is its ability to sense the spectrum and provide insight into the \ac{IoT} devices regarding possible active attacks. This is then leveraged by the decision engine to determine appropriate responses to mitigate the attack or alert the presence of a malicious entity in the spectrum of interest.

\ac{RL} becomes an excellent candidate to enable \ac{DSA} and other cognitive radio solutions because of the inherent nature of the problem that can be modeled as \ac{MDP}. Thereafter, Q-learning or even \ac{DQN} (for large state-action space) can be used to determine optimal action for a given state of the agent (transceiver). These models can be used by the data link layer for power control, negotiating spectrum access and to determining optimal transmission strategies. Similarly, at the network layer, Q-learning is used in varying traffic loads to handle congestion and \ac{QoS} requirements, optimize network parameters like delay, throughput, fairness, and energy efficiency. In contrast to traditional approaches, ML has also been used to predict route failures enabling more rapid recovery process which can be critical to large distributed \ac{IoT} networks. A key point to remember in the context of feasibility is that in many cases the learning phase might be computationally intensive and is performed offline. On the other hand, the execution itself can be light-weight thereby making \ac{ML} based approaches more feasible for \ac{IoT} devices. Realizing the importance of extending these techniques to hardware implementation, we discuss some steps that can be taken in those directions to ensure a rapid transition of these techniques to commercial hardware. 

Finally, we have also looked at a couple of key areas beyond communication where ML is being leveraged as an effective tool in the realm of IoT. Various supervised ML techniques are being employed to detect intruders and malicious behaviors which can be a key application given the risk of such attack on IoT devices. This is usually possible by analyzing the large amount of data associated with IoT. Furthermore, we have also presented some recent efforts of where big data analytics has been performed using \ac{ML} as it is a significant emerging and motivating factor in the current surge of IoT. The overarching goal of this paper is to enable researchers with the fundamental tool to understand the application of \ac{ML} in context of wireless communication in the IoT and apprise them of the latest advancements that will, in turn, motivate new and exciting works.
\small
\section*{References}
\bibliographystyle{ieeetr}
\bibliography{wiley_JJ}
\end{document}